\documentclass{elsarticle} 
\usepackage[dvipsnames]{xcolor}
\usepackage[utf8]{inputenc}

\makeatletter
\def\ps@pprintTitle{%
 \let\@oddhead\@empty
 \let\@evenhead\@empty
 \def\@oddfoot{}%
 \let\@evenfoot\@oddfoot}
\makeatother

\usepackage{moreverb}

\usepackage[colorlinks,bookmarksopen,bookmarksnumbered,citecolor=red,urlcolor=red]{hyperref}
\usepackage{orcidlink}

\usepackage{geometry}

\newcommand{\tdprod}%
     {\,{\scriptscriptstyle \stackrel{3}{\bullet}}\,}



\makeatletter
\DeclareFontFamily{OMX}{MnSymbolE}{}
\DeclareSymbolFont{myLargesymbols}  {OMX}{MnSymbolE}{m}{n}
\SetSymbolFont{myLargesymbols}{bold}{OMX}{MnSymbolE}{b}{n}
\DeclareFontShape{OMX}{MnSymbolE}{m}{n}{
    <-6>  MnSymbolE5
   <6-7>  MnSymbolE6
   <7-8>  MnSymbolE7
   <8-9>  MnSymbolE8
   <9-10> MnSymbolE9
  <10-12> MnSymbolE10
  <12->   MnSymbolE12}{}
\DeclareFontShape{OMX}{MnSymbolE}{b}{n}{
    <-6>  MnSymbolE-Bold5
   <6-7>  MnSymbolE-Bold6
   <7-8>  MnSymbolE-Bold7
   <8-9>  MnSymbolE-Bold8
   <9-10> MnSymbolE-Bold9
  <10-12> MnSymbolE-Bold10
  <12->   MnSymbolE-Bold12}{}
  
\DeclareMathSymbol{\downbrace}    {\mathord}{myLargesymbols}{'251}
\DeclareMathSymbol{\downbraceg}   {\mathord}{myLargesymbols}{'252}
\DeclareMathSymbol{\downbracegg}  {\mathord}{myLargesymbols}{'253}
\DeclareMathSymbol{\downbraceggg} {\mathord}{myLargesymbols}{'254}
\DeclareMathSymbol{\downbracegggg}{\mathord}{myLargesymbols}{'255}
\DeclareMathSymbol{\upbrace}      {\mathord}{myLargesymbols}{'256}
\DeclareMathSymbol{\upbraceg}     {\mathord}{myLargesymbols}{'257}
\DeclareMathSymbol{\upbracegg}    {\mathord}{myLargesymbols}{'260}
\DeclareMathSymbol{\upbraceggg}   {\mathord}{myLargesymbols}{'261}
\DeclareMathSymbol{\upbracegggg}  {\mathord}{myLargesymbols}{'262}
\DeclareMathSymbol{\braceld}      {\mathord}{myLargesymbols}{'263}
\DeclareMathSymbol{\bracelu}      {\mathord}{myLargesymbols}{'264}
\DeclareMathSymbol{\bracerd}      {\mathord}{myLargesymbols}{'265}
\DeclareMathSymbol{\braceru}      {\mathord}{myLargesymbols}{'266}
\DeclareMathSymbol{\bracemd}      {\mathord}{myLargesymbols}{'267}
\DeclareMathSymbol{\bracemu}      {\mathord}{myLargesymbols}{'270}
\DeclareMathSymbol{\bracemid}     {\mathord}{myLargesymbols}{'271}

\def\horiz@expandable#1#2#3#4#5#6#7#8{%
  \@mathmeasure\z@#7{#8}%
  \@tempdima=\wd\z@
  \@mathmeasure\z@#7{#1}%
  \ifdim\noexpand\wd\z@>\@tempdima
    $\m@th#7#1$%
  \else
    \@mathmeasure\z@#7{#2}%
    \ifdim\noexpand\wd\z@>\@tempdima
      $\m@th#7#2$%
    \else
      \@mathmeasure\z@#7{#3}%
      \ifdim\noexpand\wd\z@>\@tempdima
        $\m@th#7#3$%
      \else
        \@mathmeasure\z@#7{#4}%
        \ifdim\noexpand\wd\z@>\@tempdima
          $\m@th#7#4$%
        \else
          \@mathmeasure\z@#7{#5}%
          \ifdim\noexpand\wd\z@>\@tempdima
            $\m@th#7#5$%
          \else
           #6#7%
          \fi
        \fi
      \fi
    \fi
  \fi}

\def\overbrace@expandable#1#2#3{\vbox{\m@th\ialign{##\crcr
  #1#2{#3}\crcr\noalign{\kern2\p@\nointerlineskip}%
  $\m@th\hfil#2#3\hfil$\crcr}}}
\def\overbrace@#1#2#3{\vbox{\m@th\ialign{##\crcr
  #1#2\crcr\noalign{\kern2\p@\nointerlineskip}%
  $\m@th\hfil#2#3\hfil$\crcr}}}
  
\def\underbrace@expandable#1#2#3{\vtop{\m@th\ialign{##\crcr
  $\m@th\hfil#2#3\hfil$\crcr
  \noalign{\kern2\p@\nointerlineskip}%
  #1#2{#3}\crcr}}}
\def\underbrace@#1#2#3{\vtop{\m@th\ialign{##\crcr
  $\m@th\hfil#2#3\hfil$\crcr
  \noalign{\kern2\p@\nointerlineskip}%
  #1#2\crcr}}}

\def\bracefill@#1#2#3#4#5{$\m@th#5#1\leaders\hbox{$#4$}\hfill#2\leaders\hbox{$#4$}\hfill#3$}

\def\downbracefill@{\bracefill@\braceld\bracemd\bracerd\bracemid}
\DeclareRobustCommand{\downbracefill}{\downbracefill@\textstyle}

\def\upbracefill@{\bracefill@\bracelu\bracemu\braceru\bracemid}
\DeclareRobustCommand{\upbracefill}{\upbracefill@\textstyle}

\def\upbrace@expandable{%
  \horiz@expandable
    \upbrace
    \upbraceg
    \upbracegg
    \upbraceggg
    \upbracegggg
    \upbracefill@}
\def\downbrace@expandable{%
  \horiz@expandable
    \downbrace
    \downbraceg
    \downbracegg
    \downbraceggg
    \downbracegggg
    \downbracefill@}
    
\DeclareRobustCommand{\overbrace}[1]{\mathop{\mathpalette{\overbrace@expandable\downbrace@expandable}{#1}}\limits}
\DeclareRobustCommand{\underbrace}[1]{\mathop{\mathpalette{\underbrace@expandable\upbrace@expandable}{#1}}\limits}
\makeatother

\font\bigtenrm=cmr12 scaled 1200
\newcommand{\eexp}[1]{{\hbox{$\textfont1=\bigtenrm e$}}^{\raise3pt\hbox{$#1$}}}

\usepackage{xspace}

\newcommand{\exSmall}{tiny\xspace}

\newcommand{\SGS}{subgrid-scale\xspace}
\newcommand{\PN}{P\ReviewerTwo{\'e}clet number\xspace} 
\newcommand{\PNs}{P\ReviewerTwo{\'e}clet numbers\xspace}
\newcommand{\SN}{Sherwood number\xspace}
\newcommand{\SNs}{Sherwood numbers\xspace}
\newcommand{\CN}{Courant number\xspace}
\newcommand{\TS}{timestep\xspace} 
\newcommand{\BLT}{boundary layer thickness\xspace}

\newcommand{\ICs}{interfacial conditions\xspace}

\newcommand{\CapitalizeFirst}[1]{%
  \begingroup
    \edef\tempstring{#1}
    \StrLeft{\tempstring}{1}[\firstletter]%
    \StrGobbleLeft{\tempstring}{1}[\restofstring]%
    \MakeUppercase{\firstletter}\restofstring
  \endgroup
}

\newcommand{\capSGS}{\CapitalizeFirst{\SGS}\xspace} 

\newcommand{\parameter}[1]{\lstinline{#1}} 

\newcommand{\vel}{\mathbf{v}}
\newcommand{\flux}{\boldsymbol{j}} 

\newcommand{\NSigma}{{\mathbf{n}_{\Sigma}}}
\newcommand{\jump}[1]{[\![ #1 ]\!]}

\newcommand{\FullDomain}{{\Omega}}  
\newcommand{\PlusDomain}{{\FullDomain^+(t)}}
\newcommand{\MinusDomain}{{\FullDomain^-(t)}}
\newcommand{\PMDomain}{{\FullDomain^\pm}}

\newcommand{\cell}{\ensuremath{\Omega_k}}

\newcommand{\R}{\ensuremath{\mathbb{R}}}

\newcommand{\Pecl}{\operatorname{\mathit{P\kern-.08em e}}} 
\newcommand{\Sher}{\operatorname{\mathit{S\kern-.07em h}}}
\newcommand{\Cour}{\operatorname{\mathit{C\kern-.07em o}}}

\renewcommand{\S}{\ensuremath{\mathbf{S}}}
\newcommand{\vect}[1]{\boldsymbol{#1}}
\newcommand{\STensor}{\vect{S}}
\graphicspath{{figures/}}
\usepackage{subcaption}
\usepackage{algorithm}
\usepackage{algpseudocode}
\usepackage[numbers]{natbib} 
\usepackage{import} 
\usepackage{graphicx} 
\usepackage{siunitx}
\usepackage[textsize=scriptsize]{todonotes}
\usepackage{amssymb}
\usepackage{amsthm}
\usepackage{mathtools} 
\usepackage{amsmath}
\usepackage{booktabs}
\usepackage{longtable}
\usepackage{cleveref}
\usepackage{lineno}
\usepackage{txfonts} 
\usepackage{listings} 
\usepackage{lmodern} 
\usepackage[makeroom]{cancel}
\usepackage{csquotes}
\usepackage{xspace} 
\usepackage{xstring}
\usepackage{relsize}

\usepackage{soul}

\colorlet{Reviewer1}{black} 
\newcommand{\ReviewerOne}[1]{\textcolor{Reviewer1}{#1}}
\colorlet{Reviewer2}{black}
\newcommand{\ReviewerTwo}[1]{\textcolor{Reviewer2}{#1}}
\colorlet{ReviewerBoth}{black}
\newcommand{\Reviewers}[1]{\textcolor{ReviewerBoth}{#1}}
\newcommand{\otherChange}[1]{\textcolor{black}{#1}}

\begin{document}

\begin{frontmatter}

\title{{A two-sided \SGS model\\ 
for mass transfer across fluid interfaces
}} 

\affiliation[addrTUDa]{organization={Mathematical Modeling and Analysis, Mathematics department, TU Darmstadt}, country={Germany}}

\affiliation[addrUniZ]{organization={Faculty of Mechanical Engineering and Naval Architecture, University of Zagreb}, 
country={Croatia}}

\author[addrTUDa]{Moritz~Schwarzmeier\,\orcidlink{0000-0001-8992-6245}\,}
\ead{schwarzmeier@mma.tu-darmstadt.de} 

\author[addrTUDa]{Tomislav~Mari\'{c}\,\orcidlink{0000-0001-8970-1185}\,\corref{corr}} 
\ead{maric@mma.tu-darmstadt.de}
\cortext[corr]{Corresponding author}

\author[addrUniZ]{\v{Z}eljko~Tukovi\'c\,\orcidlink{0000-0001-8719-0983}\,} 
\ead{zeljko.tukovic@fsb.unizg.hr}

\author[addrTUDa]{Dieter~Bothe\,\orcidlink{0000-0003-1691-8257}\,} 
\ead{bothe@mma.tu-darmstadt.de}


\pagenumbering{arabic}
\frenchspacing


\begin{abstract}

The occurrence of extremely thin concentration boundary layers at fluid interfaces for high local \PNs is a severe obstacle for efficient and accurate numerical simulation of mass transfer processes in two-phase fluid systems.
Especially challenging are liquid-liquid systems, in which thin concentration boundary layers can appear on both sides of the fluid interface under convection-dominated conditions.
In those cases, the one-sided species concentrations at the interface are \textit{a-priori} not even known approximately, but are determined by a conjugate mass transfer problem governed by interfacial jump conditions.

\otherChange{To the best of the authors' knowledge we for the first time} introduce a two-sided \SGS~(SGS) boundary layer model for conjugate mass transfer at fluid interfaces. 
\otherChange{It} accurately computes the local mass transfer rates on moderate or coarse mesh resolutions even when very high concentration gradients in interface vicinity occur.
For this purpose, SGS modeling is applied on both sides of an interface transmissive to passive scalars, such as the \otherChange{interface in a two-phase fluid system}, enabling the accurate capture of conjugate mass transfer across thin boundary layer on one or on both sides of the interface.
We implement our approach in the unstructured Finite-Volume Arbitrary Lagrangian / Eulerian Interface-Tracking~(ALE-IT) OpenFOAM module \textit{twoPhaseInterTrackFoam}.
We have made \textit{twoPhaseInterTrackFoam} publicly available in our previous publication~(Schwarzmeier et al., 2025).

\end{abstract}




\end{frontmatter}

\section{Introduction}
\label{sec:intro}

\subsection{Motivation}

Mass transfer across fluid interfaces plays a crucial role in many chemical processes \otherChange{and engineering applications} like gas absorption, liquid–liquid extraction, chemical transformations like halogenations, polymerizations and nitrations~\cite{kashid_gasliquid_2011}.\label{l_test}
Mass transfer can not only be enhanced by increasing the interface area, a strategy extensively used in heat exchangers and also deployed for multiphase microstructured reactors \cite{kashid_gasliquid_2011}, but also by creating conditions under which thin species boundary layers exist, increasing the mass transfer per interface area.
Generally, there are different phenomena that occur in connection to thin boundary layers, which are
\emph{instationarity}~{\cite{nernst_theorie_1904,kashid_gasliquid_2011}\cite[Fig.~1c]{farsoiya_bubble-mediated_2021},}
\emph{advection-dominated transport} (high \PN~$\Pecl$)~\cite{weiner_advanced_2017,pesci_SGS_risBubb_surfactants_2018,schwarzmeier_twophaseintertrackfoam_2024},
\emph{increased mixing} due to
secondary flow profile in laminar flow, e.g. Dean or slug flow (Taylor) vortices~\cite{kurt_liquidliquid_2016} or
turbulent flow conditions~\cite{fortescue_gas_1967,liss_flux_1974,kashid_gasliquid_2011,farsoiya_bubble-mediated_2021},
and \emph{fast chemical reactions} involv\otherChange{ing} the transferred species~\cite{nernst_theorie_1904,liss_flux_1974,grunding_reactive_SGS_2016}.


There exists a variety of technical equipment used in chemical reaction engineering, where mass transfer plays a critical role: stirred tanks, centrifugal/plate/columns loop reactors, straight or coiled tubular reactors, static mixer reactors, film reactors and various microstructured reactors~\cite{kashid_gasliquid_2011}.

Chemical processes in which very thin boundary layers on both sides of an interface play a crucial role have been subject to research for more than a century~\cite{nernst_theorie_1904,whitman_twoFilm_1962}.
Liquid-liquid extraction systems and dilute gas phase gas-liquid systems, such as those, where a rapid reaction of the transfer species occurs, are examples of such systems~\cite{kashid_gasliquid_2011}.
An overview of different engineering mass transfer models is given in~\cite{wen_review_2021}.
A review of CFD technology for falling film heat and mass exchangers is given in~\cite{wen_fundamentals_2020}.
Such exchangers are used for refrigeration, carbon dioxide capture/chemical absorption, distillation/desalination, and evaporation/condensation~\cite{wen_fundamentals_2020}.

Our \otherChange{\SGS} model is able to handle cases where the concentration profiles are steep on both sides of an interface, the \BLT{}es are small and the mass transfer resistance is of similar magnitude on both sides~\cite{liss_flux_1974}.
It can be applied to cases with uneven gradients/mass transfer resistance as well.
For many gases at the air-sea interface, the mass transfer resistance on the liquid side prevails, as shown in~\cite[Table~1]{liss_flux_1974}.
An exception is sulphur dioxide~$SO_2$ because of its
high solubility in combination with a very fast hydration reaction~\cite{liss_flux_1974}.
\otherChange{However, not only turbulence, as in~\cite{liss_flux_1974}, affects gas transfer at the sea surface, but breaking waves increase it significantly by introducing instationarity, increasing the interface area and trapping air bubbles in the water phase, c.f.\ a most recent numerical study with low Schmidt number by~\cite{di_giorgio_air_2025}.}

\subsection{Approach}

We use Arbitrary Lagrangian / Eulerian Interface Tracking~(ALE-IT)~\citep{Tukovi2012} together with \SGS~(SGS) modeling.
The SGS modeling approach to mass transfer computation was introduced in~\cite{Alke_DNS-VoF_2010} for Volume of Fluid~(VoF) simulations with directional splitting, extended to general 3D SGS computations within VoF in~\cite{bothe_fleckenstein_VoF-SGS_2013} and significantly improved in~\cite{weiner_advanced_2017}.
It has been adapted to ALE-IT in~\cite{Weber2017,pesci_SGS_risBubb_surfactants_2018,schwarzmeier_twophaseintertrackfoam_2024} and applied to accurately compute diffusivity limited surfactant transport in~\cite{pesci_SGS_risBubb_surfactants_2018,schwarzmeier_twophaseintertrackfoam_2024}.\footnote{\otherChange{
Note that in this paper, we refer to surface active agents~(\textit{surfactants}) as passive scalars, even though they can have a high impact on the overall flow.
For our mathematical modeling and analysis we define \textit{passive scalars} as scalar values, which do not directly affect the flow field in the bulk, i.e.\ they neither modify the equations nor the parameters in the Navier-Stokes equations inside the flow domain, as given in eqs.~\ref{eq:localContiMomentumBulk}.
}}
The ALE-IT method in OpenFOAM~\citep{Tukovi2012,Tukovi2018,schwarzmeier_twophaseintertrackfoam_2024} is a method for the numerical simulation of multiphase flows.
Its fundamental idea involves assigning each phase its own sub-domain, which deforms according to the flow.

The incorporation of SGS modeling constitutes a distinguishing feature of our ALE-IT OpenFOAM module \textit{twoPhaseInterTrackFoam}~\cite{schwarzmeier_twophaseintertrackfoam_2024,GitlabRepositoryV2}, 
offering a significant enhancement.
The SGS modeling approach is essential for simulating the transport of passive scalars in regimes characterized by high Schmidt (high Prandtl in case of heat transfer) and high \PNs, where the associated scalar field locally exhibits length scales \otherChange{being} orders of magnitude smaller than those governing the flow field.
Traditional Finite Volume Method~(FVM) discretization would necessitate resolving these fine scales using local mesh refinement.
However, the strongly different length scales of the hydrodynamical and mass transfer boundary layers render numerical simulations of these processes computationally infeasible.
The SGS model circumvents this limitation by leveraging an analytical solution to represent thin concentration or, more generally, passive scalar boundary layers at fluid interfaces, significantly reducing the need for (otherwise excessive) local mesh refinement.
This can save multiple (local) mesh refinement levels, speeding up the simulation 
or even enabling otherwise computationally prohibitive simulations of relevant mass transfer
phenomena, while generating very accurate results, as in~\cite{weiner_advanced_2017,pesci_SGS_risBubb_surfactants_2018,schwarzmeier_twophaseintertrackfoam_2024}.

Until now, SGS modeling has been limited to cases where a steep scalar gradient\footnote{We use the term \textit{steep gradient} in this context to express a typical boundary layer characteristic:
a boundary-normal gradient, that is high at a boundary and rapidly decreases towards zero within a short distance, the \textit{\BLT~$\delta$}, in boundary-normal direction.} exists solely on one side of the interface.
Often the value of the scalar field \otherChange{at the interface} can sufficiently accurate\ReviewerTwo{ly} 
be prescribed in these cases.
In the present paper, we introduce a novel extension to the SGS framework that facilitates the computation of conjugate mass transfer across an interface, where steep gradients may occur on both sides and the one-sided limits of the scalar field are not known \textit{a~priori}.

To enforce the transmission conditions and thereby couple the two bulk phases adjacent to the interface, two (possibly different) boundary conditions~(BCs) on the two sides of the interface are applied.
Typical BCs are Dirichlet, Neumann or mixed (Robin) boundary conditions.
Reported combinations include Dirichlet-Neumann~(DN)~\cite{giles_stability_1997,tandis_analysis_2025}, Dirichlet-Robin~(DR)~\cite{tandis_analysis_2025}, Robin-Robin~(RR)~\cite{tandis_analysis_2025} boundary conditions or Dirichlet-Dirichlet~(DD)~\cite{kind_physical_2024}\footnote{
    Note that the authors of \cite{kind_physical_2024} classified their approach as RR coupling, whereas we classify this approach as DD coupling. See \cref{arg:kind_RR_or_DD}.}.
For pairs of unequal BCs, the way they are set in the solver can play a crucial role, e.g.\ in~\cite{giles_stability_1997} the Dirichlet~BC is recommended to be applied on the fluid and the Neumann~BC on the solid phase for fluid-solid systems.
On the other hand, as pointed out in~\cite{meng_stable_2017}, in certain regimes the opposite may be more appropriate.
In the present work we use DD-coupling.

Furthermore, we generalize the SGS method, \otherChange{eliminating the need for} 
a prescribed far-field concentration~$c_\infty$\otherChange{.} 
This is particularly relevant for scenarios in which far-field concentrations evolve dynamically, e.g.\ when the internal concentration of a droplet changes as a result of the transfer process or when the bulk concentration in the vicinity of a bubble in a chemical reactor changes as the bubble \otherChange{moves}.

Other enhancements to SGS modeling have been reported in the literature, such as the incorporation of local volume changes~\cite{fleckenstein_volume--fluid-based_2015}, of reactive species~\cite{grunding_reactive_SGS_2016} or local curvature effects~\cite{grosso_thermal_2024}.
Notably, the SGS has been applied in other methods than ALE-IT. 
It has been implemented in both the VoF method~\cite{Alke_DNS-VoF_2010,bothe_fleckenstein_VoF-SGS_2013,fleckenstein_volume--fluid-based_2015,weiner_advanced_2017,cai_sub-grid_2021} and the Front Tracking~(FT) method~\cite{grosso_thermal_2024,claassen_improved_2020}.

ALE-IT has been successfully utilized to address two-phase problems, including purely hydrodynamic cases~\cite{marschall_validation_2014}, cases with soluble surfactant~\citep{Dieter2014,Dieter2015,pesci_experimental_2017_book} and two-phase cases with interfacial mass transfer~\citep{Weber2017,pesci_SGS_risBubb_surfactants_2018,schwarzmeier_twophaseintertrackfoam_2024}.

Along with the present publication, we provide open source code~\citep{tudatalib_code_v2,GitlabRepositoryV2} and publicly available results of the reported cases~\cite{tudatalib_data_v2}.


\section{Mathematical model}
\label{sec:mathModel}

\subsection{Incompressible two-phase Navier-Stokes equations}

\begin{figure}[!htbp]
    \centering
    \includegraphics[]{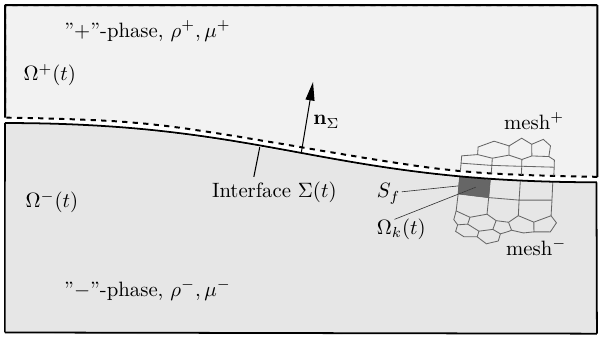}
    \caption{Definition of the solution domain for the ALE-IT method}
    \label{fig:ALE_domain}
\end{figure}

We model the incompressible and constant density, two-phase isothermal flow of immiscible Newtonian fluids with the incompressible two-phase Navier-Stokes equations, in which the two phases are separated with a sharp and mass-less interface~$\Sigma(t)$.
Since the present paper focuses on the SGS modeling \otherChange{and} basic test cases \otherChange{thereof, which have} known flow fields, we only give a brief outline.
For more details on two-phase modeling, the reader is referred to~\otherChange{\cite{bothe_sharp-interface_2022,bothe_fleckenstein_VoF-SGS_2013,slattery_interfacial_2007}}, for more details on the implementation in the ALE-IT framework to~\cite{schwarzmeier_twophaseintertrackfoam_2024}.

The whole flow domain is ${ \FullDomain \subset \mathbb{\R}^d }$ with ${ d \in \{\,1,\;2,\;3\, \} }$.
It contains two immiscible fluid phases, filling the sub-domains~$\PlusDomain$ and $\MinusDomain$, respectively, separated by the sharp interface~$\Sigma(t)$. 
The time variable~$t$ runs in~${ [\, t_0, \; t_{\rm end} \,] \subset \R }$.
The domain boundary~$\partial \, \FullDomain(t)$ may lie wholly or partially within one or both of these phases, or it may (wholly or partially) represent a solid boundary.

Within each phase, the conservation of mass and momentum is modeled by the incompressible Navier-Stokes equations
\begin{subequations} \label{eq:localContiMomentumBulk}
    \begin{align}
            \nabla \cdot \vel &= 0  &&\text{in} \ \ \Omega(t) \backslash \Sigma(t) \ \label{eq:NS_localContiBulk} \\
            \intertext{and}
            \partial_t \vel + \nabla \cdot (\vel \otimes \vel) &= \frac{1}{\rho} \nabla \cdot \S + \mathbf{g}   &&\text{in} \ \ \Omega( t) \backslash \Sigma(t) \ 
    \end{align}
\end{subequations}
\otherChange{with} $\vel$ represent\otherChange{ing} the velocity and $\mathbf{g}$ denot\otherChange{ing} the \otherChange{specific body force, typically due to} gravitational acceleration.
Furthermore, the stress tensor~$\STensor$ for a Newtonian fluid is given by
\begin{equation}
    \STensor = - p \mathbf{I} + \STensor^{\rm visc} \ ,
\end{equation}
where the viscous stress tensor is \otherChange{given} as ${ \STensor^{\rm visc} \otherChange{=} \eta (\nabla\vel + \nabla\vel^{\sf T}) }$, $p$ denotes the pressure and $\eta$ is the dynamic viscosity of the respective phase.
As usual, $\mathbf{I}$ denotes the identity tensor.

The bulk equations are complemented by interfacial transmission conditions, also known as jump conditions\otherChange{.}\footnote{Conditions coupling two fields on either side of an interface at the interface are often regarded as \textit{jump conditions}. \ReviewerTwo{Usually}\label{fn:usually} they describe the continuity of a certain quantity (or its derivation), examples being the continuity of chemical potential and mass (flux) or temperature or heat flux. Because these \textit{jump conditions} often constitute continuity, we will refer to them as \textit{interfacial conditions} for the general case in the present work.} 
We use \textit{jump brackets}~$ \jump{\Psi} $ for the difference of the one-sided limits, 
defined as
\begin{equation}
    \jump{ \Psi}(x) = \lim_{h \to 0} \left( \Psi(x+h \NSigma) - \Psi(x - h \NSigma) \right)
\end{equation}
with $\NSigma$ denoting the interface normal.
Assuming no slip at the interface and neglecting local volume effects due to mass transfer, there is continuity of velocity across the interface as \mbox{in~\cite{batchelor_introduction_2000,Tukovi2012,bothe_fleckenstein_VoF-SGS_2013,pesci_SGS_risBubb_surfactants_2018}}, i.e.
\begin{equation}
    \jump{ \vel} = 0 \qquad \qquad \text{on} \ \  \Sigma(t) \ . \label{eq:velJumCond}
\end{equation}
As a consequence of mass conservation in connection with \cref{eq:velJumCond} the normal velocity component of the interface is governed by the \textit{kinematic condition} \cite{bothe_fleckenstein_VoF-SGS_2013}, i.e.
\begin{equation}
    \rm V_{\Sigma} = \vel \cdot \NSigma \ . \label{eq:kin_BC}
\end{equation}
The quantity~$\rm V_{\Sigma}$ is also called the speed of normal displacement \otherChange{of~$\Sigma(t)$}. 

The interfacial momentum balance \otherChange{without mass transfer effect} reads as (c.f.~\cite{batchelor_introduction_2000,bothe_fleckenstein_VoF-SGS_2013,pesci_SGS_risBubb_surfactants_2018,schwarzmeier_twophaseintertrackfoam_2024}):
\begin{equation}
        \jump{ \text{p}\mathbf{I} - \STensor^{\rm visc} } = \sigma \kappa \NSigma + \nabla_{\Sigma}\sigma \qquad \qquad \text{on}  \ \ \Sigma(t) \ ,
\end{equation}

\begin{sloppypar}
\noindent where $\sigma$ is the surface tension and $\kappa$ is twice the mean curvature, defined as ${ \kappa = - \nabla_\Sigma \cdot \NSigma }$ \mbox{(c.f.~\cite{bothe_fleckenstein_VoF-SGS_2013,pesci_SGS_risBubb_surfactants_2018})} \otherChange{with}~$\nabla_\Sigma \cdot$ denot\otherChange{ing} the surface divergence.
\end{sloppypar}

\subsection{Conjugate species transport and transfer}
\label{ss:gov_eq_interface}

In the presence of a passive scalar, such as a dissolved chemical species, the mathematical model must be extended to incorporate the corresponding bulk transport equation and interfacial transmission conditions governing the transfer of chemical components, which is denoted as \textit{mass transfer}.
The interfacial transmission conditions comprise of two interfacial conditions.
These stem from mass balance and the second law of thermodynamics at the interface, ensuring physically consistent mass transfer \otherChange{modeling}. 
Other employed assumptions include dilute species and the absence of chemical reactions.
For more details we refer the reader to~\cite{bothe_fleckenstein_VoF-SGS_2013,bothe_sharp-interface_2022}.

\subsubsection*{Bulk transport}

The transport of a passive scalar within the bulk phase~$\PMDomain(t)$ is governed by the balance of species mass.
In its local form it reads as
\begin{equation} \label{eq:transport} 
    \partial_t c^\pm 
    + \nabla \cdot (c^\pm \vel^\pm + \flux^\pm)
    = \otherChange{r} \  \qquad \text{in } \PMDomain(t) 
\end{equation}
with $c^\pm$ representing the species concentration in the respective phase and the diffusive flux being modeled with Fick's law 
\begin{equation} \label{eq:ficks_law}
    \flux^\pm = -D^\pm \nabla c^\pm \ ,
\end{equation}
where $D^\pm$ denotes the diffusion coefficient~\cite{pesci_SGS_risBubb_surfactants_2018}.
\otherChange{In the present work we consider the case without chemical reaction~(${ r=0 }$), which can be included in a straightforward manner.}

\subsubsection*{Mass flux continuity at the interface}

The first interfacial condition stems from the balance of species mass at the interface and enforces the continuity of mass flux through the interface.
Ignoring local volume effects due to transfer of species mass, this balance reads as
\begin{alignat}{2}
    \jump{ \flux^\pm } \cdot \NSigma &= 0 \qquad \qquad &\text{at} \ \ \Sigma(t) \ .
\intertext{Applying Fick’s law \cref{eq:ficks_law} leads to}
    D^+ \, \partial_\NSigma c^+ &= D^- \, \partial_\NSigma c^- \qquad \qquad &\text{at} \ \ \Sigma(t) \ , \label{eq:mass_balance}
\end{alignat}
where $ \partial_\NSigma c^\pm $ denotes the directional 
derivative of the respective concentration field in the direction~$\NSigma$.

\subsubsection*{Interfacial concentration jump condition}

We employ the standard assumption of continuous chemical potentials 
across the interface, written as ${ \mu^+ = \mu^- \text{ at } \Sigma(t)}$, where $\mu$ is the chemical potential, \ReviewerOne{defined as ${ \mu_k = \partial(\rho \psi ) / \partial \rho_k }$ with the Helmholtz free energy density~${ \rho \psi(T, \rho_1,...,\rho_n) }$, where the indices denote the species in a mixture according to~\cite{bothe_fleckenstein_VoF-SGS_2013}, where a detailed derivation can be found}.
\ReviewerOne{The continuity of chemical potentials at~$\Sigma(t)$} is a reasonable \otherChange{approximation} for many practical applications, see~\citep{bothe_fleckenstein_VoF-SGS_2013}.
\ReviewerOne{It} implies a concentration jump, which with \textit{Henry's constant}~$H$ can be written as
\begin{equation}\label{eq:henry_law}
c^+ = c^- \, / \, H \qquad \qquad \text{at} \ \ \Sigma(t) \ .
\end{equation}
This is a simple way to write Henry's law, which elsewhere is formulated using par\otherChange{ti}al pressure; see e.g.~\cite{kashid_gasliquid_2011}.
\ReviewerOne{A generalization of relation~(\ref{eq:henry_law}), for example to incorporate non-constant (Henry-)coefficients for the concentration jump or incorporating curvature effects as in~\cite[eq.~(22)]{bothe_fleckenstein_VoF-SGS_2013} or temperature and pressure jump as in~\cite[eq.~(46)]{bothe_fleckenstein_VoF-SGS_2013} would be possible but the SGS modeling assumptions should be revisited for validity.
Depending on the validity of the SGS modeling assumptions, the SGS model might be applicable with or without adaptations.} 

For the case of heat transfer with the temperature~$T$ as the passive scalar, the latter field is usually assumed to be continuous at~$\Sigma(t)$.
Hence ${ T^+ = T^- }$ at~$\Sigma(t)$, or equivalently, ${ \jump{T} = 0 }$ at~$\Sigma(t)$.
This corresponds to the case of ${H=1}$ in Henry's law\otherChange{, i.e.\ in \cref{eq:henry_law}}.
But note that for very fast condensation processes temperature jumps have been reported, e.g.\ in~\cite{ward_interfacial_2001}.
Moreover, temperature jumps have been prescribed for modeling purposes. 
This is done e.g.\ for microlayers in nucleate boiling, where the temperature jump can be prescribed between solid and liquid or between liquid and gas, \mbox{c.f.~\cite{long_direct_2025,bures_comprehensive_2022}}.
This temperature jump accounts for an Interfacial Heat-Transfer Resistance~(IHTR)~\cite{long_direct_2025,bures_comprehensive_2022}.
Between two solids in contact with each other there \otherChange{usually} is a temperature jump across the contact area, which is caused by \textit{non-perfect thermal contact}, as tiny insulating gaps between two non-ideal surfaces exist\otherChange{, and where the temperature jump can be described by the \textit{contact conductance} for the interface or its reciprocal value, the \textit{thermal contact resistance}}~\mbox{\cite[p.~22f]{hahn_heat_2012}}.
If modeling temperature, additional considerations, such as the material property heat capacity, must be accounted for.

\section{Numerical method}
\label{sec:num-method}

\subsection{Unstructured Finite Volume ALE Interface Tracking Method}
\label{ss:uFVM-ALEIT}

The ALE-IT method splits the computational domain into two sub-domains with the fluid interface as the boundary between them. Each sub-domain is filled with one of the immiscible fluids or phases, as illustrated in \cref{fig:ALE_domain}.
The sub-domains are discretized with an unstructured Finite Volume mesh of non-overlapping control volumes.
As the sub-domains deform by fluid flow, their meshes deform.

The fluid interface represents a common boundary of two sub-domains, so the corresponding \ICs can be directly imposed at the interface.
In addition, the interface mesh very accurately approximates the mean curvature, which is crucial for modeling surface tension forces.

A detailed description of the ALE-IT method is provided in~\citep{Tukovi2012,Tukovi2018,schwarzmeier_twophaseintertrackfoam_2024}.

\subsection{Unstructured Finite Volume Conjugate Mass Transfer}
\label{ss:thinFilm}

\subsubsection{Discretization of the mass transport equation}

To numerically solve \cref{eq:transport,eq:ficks_law,eq:mass_balance,eq:henry_law} given in \cref{ss:gov_eq_interface}, the unstructured Finite Volume Method~(UFVM)~\citep{jasak:PhD,juretic_error_2010,moukalled_finite_2016,Maric2014} is employed.
We integrate \cref{eq:transport,eq:ficks_law} over a moving control volume~$\Omega_k(t)$ which is entirely inside one of the phases~$\Omega^\pm(t)$, and whose boundary~$\partial \, \Omega_k(t)$ is moving with some velocity~$\vel_b$, different from the flow velocity~$\vel$ everywhere except at the fluid interface~$\Sigma(t)$, where ${ \mathbf{v}_b(\mathbf{x}) = \mathbf{v}(\mathbf{x}) \otherChange{\text{ for }} \mathbf{x} \in \Sigma(t) }$.
Admitting such control volume boundary velocity~$\vel_b$ makes it possible for the ALE-IT method to deform the domains~$\Omega^\pm(t)$ so as to follow the deformation of the fluid interface~$\Sigma(t)$.
\ReviewerOne{The local mass balance in \cref{eq:transport},} 
integrated over the moving control volume~$\cell(t)$ \ReviewerOne{with the outward-pointing normal vector~${ \mathbf{n} }$ on the control volume boundary~${ \partial \, \cell }$} becomes 
\begin{equation}
    \int_{\Omega_k(t)} \partial_t c^\pm \, dV + 
    \int_{\Reviewers{\partial \,}\Omega_k(t)}  c^\pm \mathbf{v} \cdot \mathbf{n} \, d\ReviewerTwo{S} - 
    \int_{\Reviewers{\partial \,}\Omega_k(t)}  \nabla \cdot (D^\pm \nabla c^\pm) \, d\ReviewerTwo{S} = 0 \ . 
    \label{eq:cintegral}
\end{equation}
The first term in \cref{eq:cintegral} can be expressed using the Reynolds Transport Theorem for a moving volume~$\Omega_k(t)$ whose boundary~$\partial \, \Omega_k$ is moving with the velocity~$\vel_b$, i.e.
\begin{equation}
    \frac{d}{dt} \int_{\Omega_k(t)} c^\pm \, dV = \int_{\Omega_k(t)} \partial_t c^\pm \, dV 
    + \int_{\partial \Omega_k(t)} c^\pm \vel_b \cdot \mathbf{n} dS \ .
    \label{eq:rtt}
\end{equation}
Expressing $\int_{\Omega_k(t)} \partial_t c^\pm \, dV $ from \cref{eq:rtt} and inserting it into \cref{eq:cintegral} gives
\begin{equation}
    \frac{d}{dt} \int_{\Omega_k(t)} c^\pm \, dV
    +
    \int_{\Reviewers{\partial \,}\Omega_k(t)}  c^\pm (\mathbf{v} - \mathbf{v}_b) \cdot \mathbf{n} \, d\ReviewerTwo{S} - 
    \int_{\Reviewers{\partial \,}\Omega_k(t)}  \nabla \cdot (D^\pm \nabla c^\pm) \, d\ReviewerTwo{S} = 0 \ . 
    \label{eq:cintegralrtt}
\end{equation}

It is important to emphasize that $\Omega_k(t)$ is immersed entirely in either $\Omega^+(t)$ or $\Omega^-(t)$, with $\partial \, \Omega_k(t)$ possibly containing a part of the fluid interface, i.e., it is possible that ${ \partial \, \Omega_k(t) \cap \Sigma(t) \ne \emptyset }$.
In fact, when $\Omega_k(t)$ is entirely in the bulk of $\Omega^\pm(t)$, further discretization of \cref{eq:cintegralrtt} with the UFVM does not require special treatment.
However, for $\Omega_k(t)$ with ${ \partial \, \Omega_k(t) \cap \Sigma(t) \ne \emptyset }$, \ICs are required, as described below.

The UFVM discretization of \cref{eq:cintegralrtt} gives\otherChange{,} at some $t=\tau$\otherChange{,}
\begin{equation}
    \left[\frac{d}{dt}(c^\pm_k \Omega_k) + \sum_{f \in F_k} (F_f - F_b) c^\pm_f 
    - \sum_{f \in F_k} D_f (\nabla c^\pm)_f \cdot \mathbf{S}_f \right]_{t = \tau} = 0 
    \label{eq:cdiscrspace}
\end{equation}
\ReviewerOne{with $\mathbf{S}_f$ denoting the faces of the control volume.}
\otherChange{In~\cref{eq:cdiscrspace}} ${ F_f := \mathbf{v}_f \cdot \mathbf{S}_f }$ is the volumetric flux resulting from the flow velocity and ${ F_b:= \mathbf{v}_b \cdot \mathbf{S}_f} $ is the volumetric flux resulting from the motion of the control volume boundary~${ \partial \, \Omega_k(t) }$.
We aim to discretize \cref{eq:cdiscrspace} implicitly, to ensure the numerical stability \otherChange{is maintained} for large \TS{}s.
Different implicit temporal discretizations of \cref{eq:cdiscrspace} can be constructed by discretizing $\frac{d}{dt}(c_k^\pm \Omega_k)$ using Finite Differences, by evaluating  \cref{eq:cdiscrspace} at different times, e.g.\ ${ t = t^{n-1}, t^n, t^{n+1} }$, and combining the equations.
For example evaluating \cref{eq:cdiscrspace} at ${ t = t^{n+1} }$ and approximating ${ \frac{d}{dt}(c_k^\pm \Omega_k)(t^{n+1}) }$ by the implicit Backward Differencing formula with second-order accuracy~(BDS2) and with a constant \TS~$\Delta t$, gives
\begin{equation}
    \label{eq:discr_conc}
    \frac{3 c_k^{n+1} \Omega_k^{n+1} - 4 c_k^n \Omega_k^n + c_k^{n-1} \Omega_k^{n-1}}{\Delta t} + \sum_{f \in F_k} (F_f - F_b)^{n+1} c_f^{n+1} = \sum_f D_f (\nabla c)_f^{n+1} \cdot \boldsymbol{S}^{n+1}_f \ . 
\end{equation}
In \cref{eq:discr_conc}, UFVM expresses $c_f^{n+1}$ and $(\nabla c)_f^{n+1}$ from cell-centered quantities~$c_k^{n+1}$ defined in cells sharing the face~$\mathbf{S}_f$, respectively, using interpolation and gradient discretization. 

If the boundary of the control volume~$\Omega_k(t)$ contains a part of the boundary of the flow domain~$\Omega(t)$, there will be some index~${}_f$ in \cref{eq:discr_conc} for which either $c_f^{n+1}$ or $(\nabla c)_f^{n+1}$ is required at~$\partial \, \Omega(t)$, corresponding to classical Dirichlet, Neumann or mixed~(Robin) boundary conditions.
Dirichlet boundary conditions applied at ${ S_f \subset \partial \, \Omega(t) }$ result in the source term of \cref{eq:discr_conc} - as known values.
The UFVM uses linear extrapolation 
\begin{equation}
    c_k^{\pm, n+1} = c_f^{\pm, n+1} + (\nabla c^\pm)_f^{n+1} \cdot \mathbf{d}_f + O(|\mathbf{d}_f|_2^2) \ ,
    \label{eq:taylor}
\end{equation}
where $\mathbf{d}_f$ is the vector from the location of $c_k$ (the cell center) to $c_f$ (the cell face) and $O(|\mathbf{d}_f|_2^2)$ represents the error estimate with $|\cdot|_2$ being the \otherChange{$L_2$-}error norm, to implicitly discretize Neumann boundary conditions.
For example, a zero Neumann condition in \cref{eq:taylor} results in ${ c_k^{\pm, n+1} = c_f^{\pm, n+1} }$, adding $1$ to the discretization coefficient in front of $c_k^{\pm, n+1}$ in \cref{eq:discr_conc}.

If the boundary of the control volume $\Omega_k(t)$ contains a part of the fluid interface~$\Sigma$, there will be some index~$\otherChange{f}$ in \cref{eq:discr_conc}, for which either $c_f^{n+1}$ or $(\nabla c)_f^{n+1}$ at~$\Sigma(t)$ are required.
Unlike classical boundary conditions, the \ICs (\cref{eq:mass_balance,eq:henry_law}) are not straightforward to evaluate in the case of conjugate mass transfer.

To uphold the \ICs (\cref{eq:mass_balance,eq:henry_law}) for conjugate mass transfer without SGS modeling, we propose an algebraic equation system, resulting directly from the unstructured Finite Volume Discretization of boundary conditions, as described in \cref{sss:thinFilm_detCSigma}.
When \otherChange{-~in contrast~-} very steep changes of the concentration~$c$ normal to the fluid interface~$\Sigma(t)$ occur, we use SGS modeling and an additional nonlinear system is emerging, as described in \cref{ss:2sSGS_conjMassTransf}.
To solve it, we extend our one-sided \SGS modeling framework~\citep{schwarzmeier_twophaseintertrackfoam_2024} to conjugate mass transfer, marking a significant improvement in UFVM discretization for conjugate mass transfer.


\subsubsection{Linear Dirichlet-Dirichlet solution algorithm for conjugate mass transfer}
\label{sss:thinFilm_detCSigma}

The one-sided limits of~$c^\pm$ at the interface appear frequently \otherChange{below} and are abbreviated as~$c_\Sigma^\pm$.

When the boundary of the control volume $\partial \, \Omega_k(t)$ overlaps with a part of $\Sigma(t)$, 
the approximation~(\ref{eq:taylor}) used in \cref{eq:discr_conc} to implement the boundary condition changes into
\begin{equation} \label{eq:taylorinterface}
    c_k^{\pm, n+1} = c_\Sigma^{\pm,n+1} + (\nabla c^{\pm})^{n+1}_{\Sigma^{n+1}} \cdot \mathbf{d}^{\pm,n+1}_\Sigma + O(|\mathbf{d}^{\pm,n+1}_\Sigma|_2^2) \ .
\end{equation}

\begin{figure}[!htbp]
    \centering
    \includegraphics[width=0.9\textwidth]{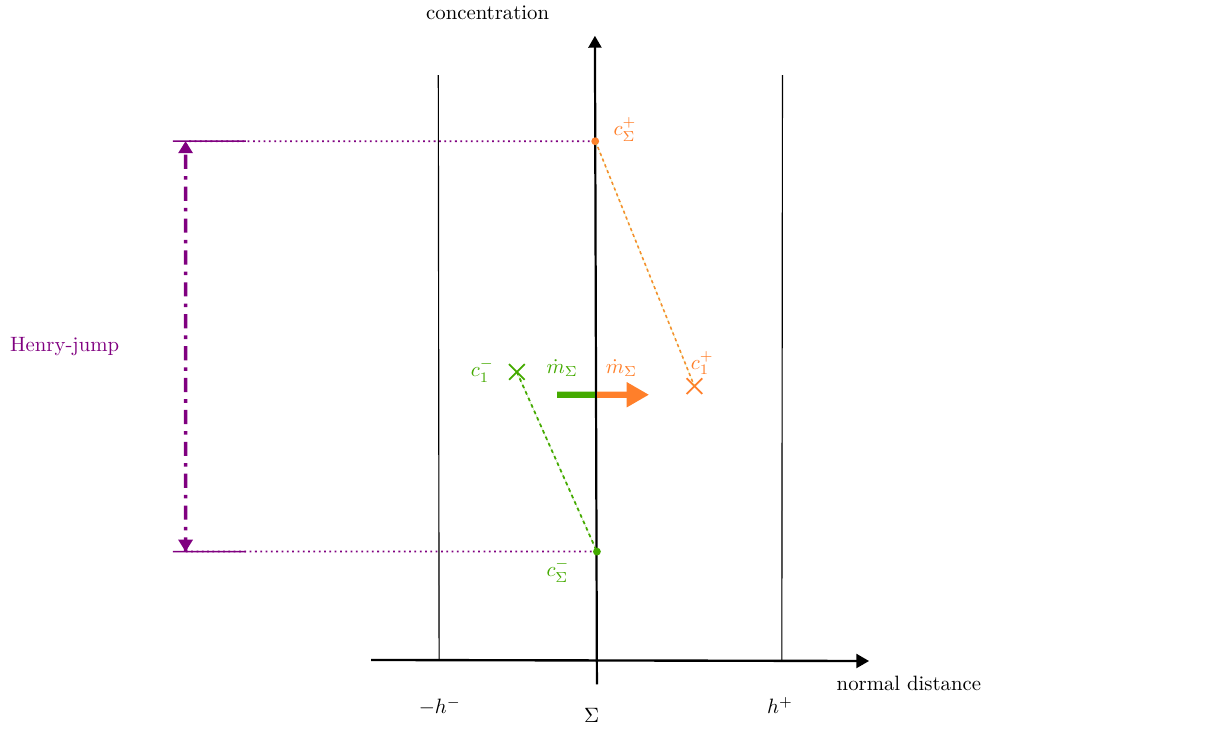}
    \caption{Concentrations at the interface for resolved simulations}
    \label{fig:thinFilm}
\end{figure}

The setup is schematically shown in \cref{fig:thinFilm}, where we associate to $c_\Sigma^-$ a volume~$\Omega_1^-$, with the cell-centered concentration~$c_1^-$, and to $c_\Sigma^+$ the volume~$\Omega_1^+$ with its cell-centered concentration~$c_1^+$.
We simplify the description by considering meshes \otherChange{orthogonal} in the interface-normal direction as shown in \cref{fig:thinFilm} without any loss of generality.

To discretize \cref{eq:discr_conc} in case of faces~$S_f$ overlapping with~$\Sigma(t)$, we require~$c_\Sigma^\pm$ in \cref{eq:taylorinterface}, but also~$(\nabla c^\pm)_\Sigma$, \otherChange{where the two~$c_\Sigma^\pm$ have to satisfy} the Henry jump~(\ref{eq:henry_law}).
\otherChange{To solve this problem we make use of two assumptions}: first, transport only occurs in interface-normal direction, and second, the gradient can be represented linearly.
\otherChange{Similar assumption can be found in~\cite{kind_physical_2024} for heat transfer.}


Concentrations on either side of the interface are coupled by the two \ICs given in \cref{eq:mass_balance,eq:henry_law}.
The discretized mass flux condition \cref{eq:mass_balance}, applied to cell-centered values in cells adjacent to the fluid interface, gives the first equation that couples the two variables~$c_\Sigma^\pm$:
\begin{equation}
    D^+ \, \frac{c_1^+ - c^+_\Sigma}{h^+ /2} = D^- \, \frac{c^-_\Sigma - c^-_1}{h^- /2} \ . 
\end{equation}

Introducing a non-dimensional parameter~${\alpha = \frac{D^-}{D^+} \frac{h^+}{h^-}}$ \otherChange{for brevity} and applying Henry's law~(\ref{eq:henry_law}), we obtain two explicit equations for the concentrations~$c^\pm_\Sigma$ at the interface:
\begin{subequations}
    \label{eq:cSigma_linear}
    \begin{align}
        c^+_\Sigma & = \frac{\alpha c^-_1 + c^+_1}{\alpha H + 1}
        \intertext{and}
        c^-_\Sigma & = H c_\Sigma^+ \ .
    \end{align}
\end{subequations}

Equation system~(\ref{eq:cSigma_linear}) makes it possible to express~$c_\Sigma^\pm$ via Dirichlet boundary conditions for \cref{eq:discr_conc} using \otherChange{existing} cell-centered quantities.
The \otherChange{value of}~$c_\Sigma^{\otherChange{+}}$ \otherChange{is} bound\otherChange{ed} within $c_1^+$ and~$c_1^- / H$ \otherChange{and, by the Henry jump condition, $c_\Sigma^-$ consequently is bounded too}.
In~\cite{kind_physical_2024} it is argued that prescribing values in such a fashion is actually a Robin-Robin boundary condition, since the fluxes are used but cancel out.
We do not follow this argumentation, as the values prescribed are fixed Dirichlet values during the solution of the bulk equations\otherChange{.}\label{arg:kind_RR_or_DD}
Indeed, they are prescribed such that -~at the moment of evaluation~- they satisfy both interfacial conditions given in \cref{eq:mass_balance,eq:henry_law}.
However, all terms in \cref{eq:discr_conc} must be evaluated at~$t^{n+1}$.

\label{sss:DD_coupling}


To ensure \cref{eq:discr_conc} is indeed solved at $t^{n+1}$,  we employ a segregated solution strategy.
We \otherChange{explicitly} compute Dirichlet boundary conditions \otherChange{from bulk values, starting with the current \TS existing~${ c_{1}^{\pm,n} }$.}
We then proceed to solve \cref{eq:discr_conc} for \otherChange{the} cell-centered concentrations, \otherChange{including~${ c_1^{\pm} }$}.
We repeat this procedure until convergence is achieved and \cref{eq:discr_conc} is satisfied for~$t^{n+1}$.
We control the convergence by monitoring residuals of the corresponding linear equation system generated by applying \cref{eq:discr_conc} in every finite volume in the discretized domain, i.e.
\begin{equation}
    \mathbf{A}\mathbf{c} = \mathbf{s}_c \ ,
    \label{eq:clinsystem}
\end{equation}
where $\mathbf{c}$ is a vector of cell-centered concentration values, $\mathbf{A}$ is the unstructured Finite Volume coefficient matrix and $\mathbf{s}_c$ is the source term vector.
Once this solution algorithm, which we call \textit{linear Dirichlet-Dirichlet~\otherChange{(LDD)} solution algorithm}, converges, when eqs.~(\ref{eq:cSigma_linear}) recover \ICs at~$t^{n+1}$, the change of the contributions of eqs.~(\ref{eq:cSigma_linear}) to the source term~$\mathbf{s}_c$ from \cref{eq:discr_conc} into \cref{eq:clinsystem} will stop being noticeable by the linear system solver.
The measure for \otherChange{convergence} is the norm of the residual of \cref{eq:clinsystem}:
\begin{equation}
    |\mathbf{r}_c|_\lambda = |\mathbf{A} \mathbf{c} - \mathbf{s}_c|_\lambda \ .
    \label{eq:residual}
\end{equation}
When the linear Dirichlet-Dirichlet solution 
\otherChange{has (sufficiently)} converge\otherChange{d}, the resulting residual, computed by inserting available~$\mathbf{c}$ into \cref{eq:residual}\otherChange{,} will be smaller than a user-prescribed tolerance \otherChange{and the iterative process can be stopped}.
The \otherChange{LDD} 
solution algorithm is summarized in \otherChange{\cref{alg:linDD}}.

\begin{algorithm}
\caption{\otherChange{Linear Dirichlet-Dirichlet algorithm:} Solve \textcolor{blue}{\textit{\ICs}} \textcolor{black}{and} \textcolor{brown}{\textit{bulk equations}} }\label{alg:linDD}
\begin{algorithmic}[1]
\For{time step \textbf{in} time steps}
    \State \texttt{...}
    \For{outer iteration $o$ \textbf{in} outer iterations}
        \State \textcolor{blue}{\textbf{Solve eqs.~(\ref{eq:cSigma_linear})} for $c_{\Sigma}^{\pm,o}$ for all faces~$f$ \Comment{for given $c_k^{\pm,o}$}}
        \State \textcolor{brown}{\textbf{Solve \cref{eq:transport}} for $c_k^{o+1}$ in $\PMDomain$ \Comment{for given $c_\Sigma^{\pm,o}$}}
        \If{\textcolor{brown}{$c_k^{\pm, o+1}$ converged acc.\ to \cref{eq:residual}}} 
            \State \textbf{break}
        \EndIf
    \EndFor
    \State \textcolor{brown}{\texttt{...}}
\EndFor
\end{algorithmic}
\end{algorithm}

\subsection{Subgrid-scale modeling} 
\label{ss:SGS}

\subsubsection{State-of-the-art model}


When the \BLT becomes small in comparison to other length scales, such as the hydrodynamic one, the numerical methods described in \cref{ss:uFVM-ALEIT,ss:thinFilm} are unable to resolve the resulting steep concentration gradients while keeping the simulation computationally tractable.
The SGS model bridges length scales and enables mass transfer simulations across the fluid interface with extremely thin boundary layers, appearing in challenging real-world mass transfer problems.

\otherChange{\capSGS modeling fits a non-linear model function to FVM data in order to capture the steep gradient's characteristic quantities.
Those quantities are the very high gradient at the interface and the very small gradient and concentration outside the boundary layer.
Fluxes in line with the SGS modeling are prescribed by scaling diffusivities and fluxes at the faces of an interface-neighbouring cell,
such that a subsequent application of the usual convective and diffusive transport step yields the result corresponding to the desired corrected flux.}


Until now SGS modeling has been employed on one side of the fluid interface, using a fixed, predefined concentration at the interface~$c_\Sigma$ as a boundary condition, c.f.~\citep{Alke_DNS-VoF_2010,bothe_fleckenstein_VoF-SGS_2013,weiner_advanced_2017,pesci_SGS_risBubb_surfactants_2018,schwarzmeier_twophaseintertrackfoam_2024}.
Alternative SGS approaches have been proposed by various researchers~\citep{Alke_DNS-VoF_2010,aboulhasanzadeh_multiscale_2012,grosso_thermal_2024,claassen_improved_2020}.
A more detailed review of most state-of-the-art methods is available in~\citep{weiner_assessment_2022}.  

In~\cite{bothe_fleckenstein_VoF-SGS_2013,weiner_advanced_2017,schwarzmeier_twophaseintertrackfoam_2024} the general solution of the simplified substitute problem, viz.
\begin{equation}
    \label{eq:SGS_anSol}
    c(x,y)=c_{\Sigma} + (c_\infty - c_{\Sigma}) \cdot {\rm erf} \left( x / \delta(y) \right)  \qquad \text{with } \delta(y)= \sqrt{4Dy/\nu}
\end{equation}
is employed, where $c_\Sigma$ is the one-sided interface concentration, $c_\infty$ is the far-field concentration and $\delta$ is the \BLT.
\ReviewerOne{The spatial coordinate~$x$ denotes the distance from the interface~$\Sigma$ in interface-normal direction and the coordinate~$y$ denotes the distance parallel to the interface, where~${ y=0 }$ marks the beginning of the solution domain in flow-wise direction.}

In state-of-the-art SGS models, \cref{eq:SGS_anSol} is fitted for each FV cell individually with prescribed~$c_\Sigma$ and~$c_\infty$.
The \BLT~$\delta$ is determined by enforcing mass conservation using the concentration value in the first cell adjacent to the interface~$c_1$:
\begin{equation}\label{species-mass_+1}
\int_0^h c (x) \,dx = h\, c_{1} \ .
\end{equation}
Following \citep{weiner_advanced_2017}, we now introduce the parameter~$\eta$ as
\begin{equation}
    \eta \left( x, y \right) := \frac{c \left( x, y \right) - c_\Sigma}{c_\infty - c_\Sigma} = {\rm erf} \left( x / \delta \left(y\right) \right) 
\end{equation}
and can then - still following \citep{weiner_advanced_2017} - state
\begin{equation} \label{eq:SGS_fittingCondition}
    \overline{\eta} = \frac{ \overline{c} - c_\Sigma}{c_\infty - c_\Sigma} 
    \overset{!}{=}
    \ReviewerTwo{ \eta^{SGS} = \ }
    \frac{1}{V} \int_V \eta \left( x / \delta \right) {\rm d} V
\end{equation}
with $ \overline{c} $ as the volume-averaged cell-centered value, as obtained from the FVM discretization\ReviewerTwo{, $ \eta^{SGS} $~as the corresponding value obtained with the SGS-profile,} and $\delta$ being the local \BLT parameter, uncoupled from the substitute problem.
To determine the \BLT~$\delta$, a Newton \otherChange{b}isection method has \otherChange{previously} been used, c.f.~\citep{pesci_SGS_risBubb_surfactants_2018}.

To implement the SGS model within ALE-IT, the diffusivity at the boundary faces~($x=0$) and in between the first cell next to an interface and its neighbour~($x=h$) is scaled \otherChange{in} such \otherChange{a way} that the gradients from the SGS model are enforced over the gradients computed by~FVM.
\Reviewers{This is done in order to maintain numerical stability instead of modifying the surface normal gradients directly.
Modifying the diffusivities has the additional advantage of circumventing modifying the solver or boundary conditions.
The diffusivity is scaled such in such a way that the mass flux is increased at the interface and decreased between the first and second cell adjacent to the interface, in accordance with the \SGS model assumptions.
As previous works~(\cite{weiner_advanced_2017,pesci_SGS_risBubb_surfactants_2018,schwarzmeier_twophaseintertrackfoam_2024}) have done, we state the face-centered diffusive concentration flux~$F_f^{c,D}$ can be calculated as follows:
\begin{subequations}
    \begin{equation}
        F_f^{c,D} = -D_f (\partial_n c)_f^{SGS} F_f \overset{!}{=} -D_f^{SGS} (\partial_n c)_f F_f \qquad \text{with } F_{\!f} := \mathbf{v}_{\!f} \cdot \mathbf{S}_f\ ,
    \end{equation}
    which can be re-arranged into
    \begin{equation}
        \label{eq:scaling_diff}
        D_f^{SGS} =  \frac{ (\partial_n c)_f}{(\partial_n c)_f^{SGS}} D_f \ . 
    \end{equation}
\end{subequations}
}

The interface-normal derivative of \cref{eq:SGS_anSol}, e.g.\ given by~\cite[eq.~(14)]{weiner_advanced_2017}, is
\begin{equation}
    \label{eq:SGS_anSol_grad}
    \partial_\NSigma c(x,y) = \frac{2}{\sqrt{\pi}} \frac{c_\infty-c_\Sigma}{\delta} e^{-(x/\delta)^2} \ .
\end{equation}
The faces in between the first cell next to an interface and its neighbours in bulk-inwards direction are also the positions\otherChange{,} where the volumetric fluxes need to be scaled accounting for the difference of the numerically computed concentration values at the respective locations to the actual concentration value 
\begin{equation}
    c\,(h,y) = c_\Sigma + (c_\infty - c_\Sigma) \, {\rm erf} \left( \frac{h}{\delta} \right) \ .
\end{equation}
The scaling approach ensures an accurate representation of interfacial transport processes\otherChange{,} while preserving boundedness.
\Reviewers{The scaling can be done according to
\begin{equation} \label{eq:advFlux}
    F_{\!f}^{c, A} := c_{\!f}^{SGS} F_{\!f} = c_{\!f} F_{\!f}^{SGS} \ ,
\end{equation}
where $ F_{\!f}^{c, A} $ is the volumetric flux of the concentration~$c$ at the face~$f$.
As before, we re-arrange the equation into
\begin{equation}
    \label{eq:scaling_adv}
    F_f^{SGS} = \frac{c_f^{SGS}}{c_f} F_f^c \ .
\end{equation}
}

\Reviewers{More information on diffusive and advective scaling, including exception handling, can be found in~\cite{weiner_advanced_2017,pesci_SGS_risBubb_surfactants_2018}.
}

\subsubsection{Robust treatment of \exSmall values of $1-\bar{\eta}$}
\label{sss:iSGS} 

\subsubsection*{Fitting algorithm}

The \BLT~$\delta$ is the root of \cref{eq:SGS_fittingCondition}: for a given cell it is found iteratively using root-finding with the cell-averaged concentration~$\overline{\eta}$ to fit the SGS-modeled concentration~$\eta(x / \delta)$ within the cell.
However, previous implementations~\cite{pesci_SGS_risBubb_surfactants_2018,schwarzmeier_twophaseintertrackfoam_2024} of SGS modeling lack robustness due to an incorrect implementation of the Newton-Bisection method, which lead to numerical inconsistencies.

The present work rectifies this issue, ensuring that the fitting procedure remains numerically stable even for \exSmall concentration values.
In situations where $1-\overline{\eta}$ approaches \exSmall values\footnote{\CapitalizeFirst{\exSmall} values can be the outcome of numerical, iterative methods and do not need to be physically correct.}, the corresponding scaling of the diffusion coefficient \ReviewerTwo{in \cref{eq:scaling_diff}} becomes excessively large, \ReviewerTwo{to account for the similarly disproportionate surface normal gradient at the interface due to the \BLT in the denominator in \cref{eq:SGS_anSol_grad}}.
In contrast, when ${ 1-\overline{\eta} = 0 }$, no scaling was applied.
Both situations can potentially destabilize the numerical solution or impede convergence:
\begin{itemize}
    \item If no \BLT~$\delta$ could be identified, because the root-finding diverges or ${1-\overline{\eta} = 0}$, no diffusion coefficient scaling occurs, leading to an underestimation of mass flux by several orders of magnitude.
    This would be equivalent to a discontinuity in \cref{fig:delta_over_eta}. 
    No scaling corresponds to a large (resolved) boundary layer.
    \item In contrast, when only a \exSmall amount of mass is present in the interface-adjacent cell, the excessively large scaled diffusion coefficient in the next iteration or \TS can result in unphysical overshoots of mass flux into the first cell.
    Note, that the overall solution still remains bounded, as the diffusion coefficient rather than the mass flux itself undergoes scaling.
\end{itemize}

\ReviewerTwo{We use a Newton method within a one-side limited search interval, where the interval boundary is a minimum.
The updated algorithm, given in \cref{alg:newtonBisection_NEW}, is simpler than the old one.}
\ReviewerTwo{The handling of the minimum boundary is special: in contrast to applications where the root lies within the search interval, the \BLT fitting to the cell-centered concentrations~$c_1$ might actually lie outside the search interval, that is $\delta$ would be smaller than a minimum value.
Because of this, instead of prescribing ${ \delta_{n+1} = (\delta_n + \delta_{min})/2 }$ in case~${ \delta_{n+1} < \delta_{min} }$, we set $\delta_{n+1}$ to~$\delta_{min}$.
If in the next iteration $\delta_{n+2}$ is again lower than~$\delta_{min}$, no solution bigger than~$\delta_{min}$ can be found because the fitted solution~$\delta$ is smaller than the limit~$\delta_{min}$.
The search is terminated and $\delta_{min}$ is used.}

\begin{figure}[!htbp]
    \begin{subfigure}{0.5\textwidth}
        \centering
        \includegraphics[width=\textwidth]{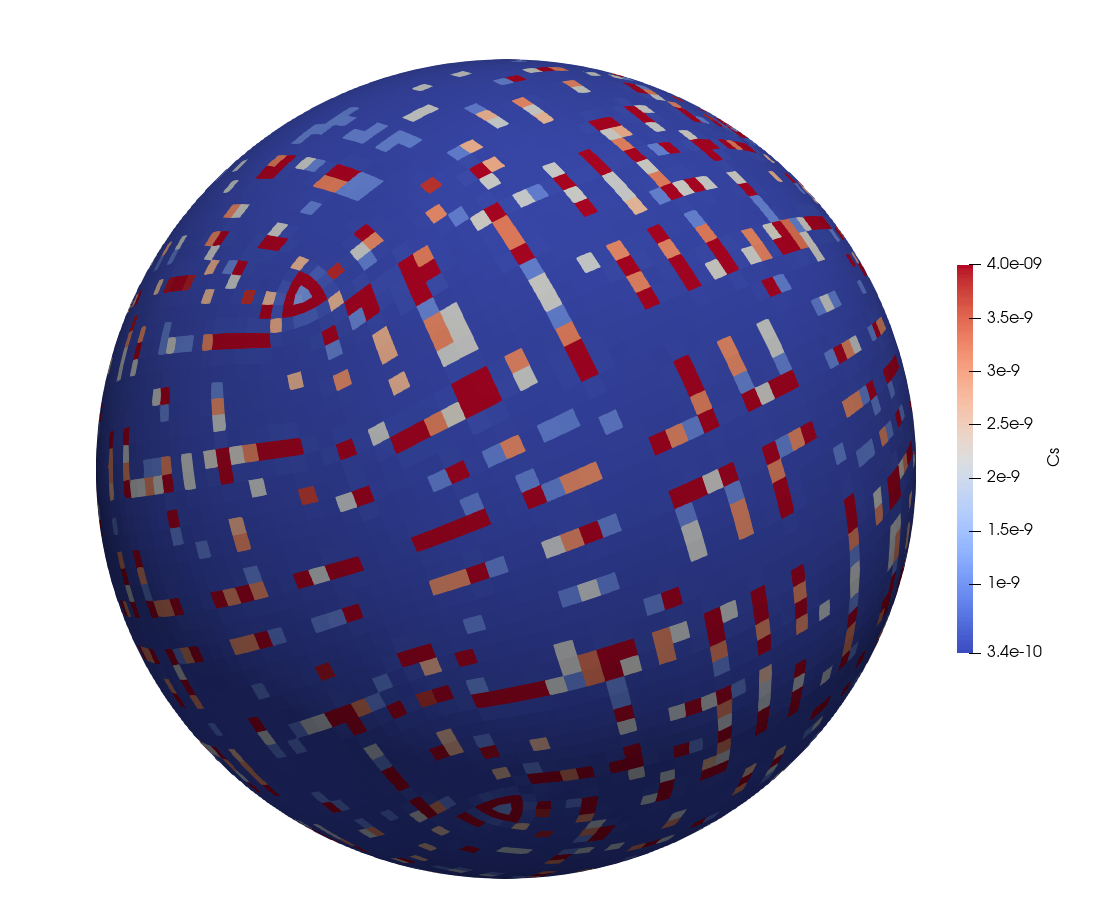}
        \caption{Old implementation}
        \label{fig:sparkled_Cs}
    \end{subfigure}
    \begin{subfigure}{0.5\textwidth}
        \centering
        \includegraphics[width=0.9\textwidth]{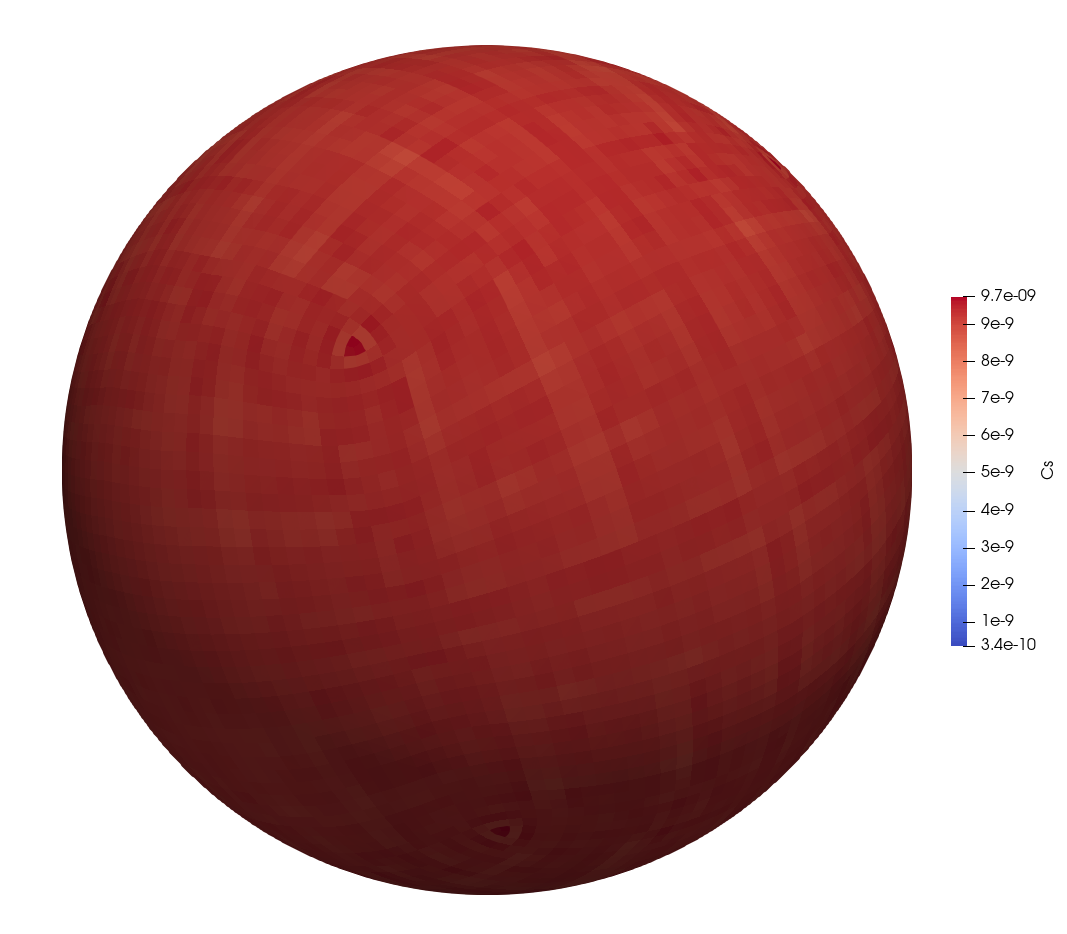}
        \caption{New implementation}
        \label{fig:new_Cs}
    \end{subfigure}
    \caption{Surfactant interface concentration at $t=\SI{5e-5}{s}$}
    \label{fig:risBubble_Cs}
\end{figure}

The case shown in \cref{fig:risBubble_Cs} was described by~\cite{pesci_SGS_risBubb_surfactants_2018,schwarzmeier_twophaseintertrackfoam_2024} and is taken from~\cite{schwarzmeier_twophaseintertrackfoam_2024}.
\Cref{fig:risBubble_Cs} shows the surfactant concentration on the interface of a rising bubble with zero initial surfactant concentration and a concentration of~\SI{5e-2}{mol/m^3} in the surrounding fluid
after $\SI{5e-5}{s}$, corresponding to 10~\TS{}s.
A long-standing but previously unnoticed consequence of the described issue is illustrated in \cref{fig:sparkled_Cs}, where we observe high differences between individual elements of the surface mesh, an observation in contrast to \cref{fig:new_Cs}.
Also, note the generally much higher concentration level in \cref{fig:new_Cs}.
This numerical artifact has remained undetected due to its transient nature, typically affecting only the initial \TS{s} of a simulation.
In previous cases, its impact on long-duration simulations was negligible.
To mitigate this \exSmall-concentration issue, an initial concentration slightly greater than~$ c_\infty $  could be prescribed, either throughout the entire computational domain or only within the cells adjacent to the interface.

\subsubsection*{Minimum \BLT}
As previously discussed, arbitrarily small boundary layer thicknesses can lead to numerical instabilities or unphysical behavior.
Introducing a minimum \BLT serves to attenuate excessive scaling effects introduced by the SGS model, thereby enhancing numerical stability.

A physically meaningful minimum \BLT can be estimated by applying boundary layer theory within the FVM framework.
This is feasible within FVM, because a spatially and temporally averaged \BLT can not become zero.

The solution to the substitute problem, as given in \cref{eq:SGS_anSol}, treats the \BLT~$\delta$ as a free parameter, which is fitted individually for each interface-adjacent computational cell.
The fitting process selects the optimal approximation from a family of error functions.
It has already been pointed out in~\cite{weiner_advanced_2017} that the substitute problem can be interpreted differently:
\textquote{Since the model problem can be interpreted as a time-dependent diffusion process with $y$ as the time coordinate and $D/\nu$ as a modified diffusivity, solutions based on Laplace transformation or similarity approach are reported in many text books.}
By exploiting this alternative formulation\otherChange{,} \cref{eq:SGS_anSol} can be rewritten in terms of time: 
\begin{equation} \label{eq:delta_of_t}
    \delta(t) = \sqrt{4Dt} \ .
\end{equation}

\begin{sloppypar}
To estimate a minimum \BLT, we assume an initial bulk concentration of zero at ${t=0}$, corresponding to a \BLT of zero.
Subsequently, we compute an averaged quantity over a single \TS~$\Delta t$.
However, rather than directly averaging the \BLT, we focus on its effect on interfacial scalar transfer.
This is achieved by averaging the surface normal concentration gradient at the interface\otherChange{, }${\partial_{\otherChange{\NSigma}} c(x,y)_{|x=0} = \frac{2}{\sqrt{\pi}} \frac{c_\infty - c_\Sigma}{\delta}}$\otherChange{,} over one \TS~${\Delta t}$, beginning at~${t=0}$:
\end{sloppypar}
\begin{equation}
    \overline{\partial_{\otherChange{\NSigma}} c_{|x=0}} 
    = \frac{1}{\Delta t} \int_0^{\Delta t} \frac{2}{\sqrt{\pi}} \frac{c_\infty - c_\Sigma}{\delta(t)} \ dt \\ 
    = \frac{2\,(c_\infty - c_\Sigma)}{\sqrt{\pi D\Delta t}} \ .
\end{equation}

Since this time-averaged concentration gradient must remain consistent with the SGS theory, we postulate
\begin{equation}
    \overline{\partial_{\otherChange{\NSigma}} c_{|x=0}} 
    \overset{!}{=}
    \frac{2}{\sqrt{\pi}} \frac{c_\infty - c_\Sigma}{\overline{\delta}} \ .
\end{equation}
Solving for $\overline{\delta}$ yields
\begin{equation} \label{eq:minBLT}
    \overline{\delta} = \sqrt{D \Delta t} \ .
\end{equation}
This expression provides a physically justified minimum \BLT for a quiescent liquid over a given (initial) \TS.
If the computed value falls below $\SI{1e-15}{}$\otherChange{\footnote{\SI{1e-15}{}, known also as parameter \parameter{SMALL}, is a common value used in software, such as OpenFOAM, to stabilize floating point arithmetic operations.
It is used in conjunction with so-called double precision, where in \textit{C++} programs floating point numbers typically are represented in the \textit{IEEE-754 binary64} format~(\cite{cppReference_fundamentalTypes}), which has a~\SI{52}{bits} \textit{trailing significand field width}~(\cite{ieee_754_2008}).
\parameter{SMALL} is bigger than~${ 2^{-52} \approx }$~\SI{2.22e-16}{}, which is also know as machine epsilon, and which is the smallest fraction, by which a floating point number represented with these\SI{52}{bits} can be in-/decreased.
}}, this threshold value is enforced.
Prescribing a lower bound for the \BLT ensures numerical stability since it limits the effect of the SGS and thus mass-transfer.

\subsubsection*{Residence time} 

For the derivation of \cref{eq:minBLT} we assumed \otherChange{the} fluid \otherChange{to be} at rest.
However, most practical scenarios involve flowing liquids, where convective transport influences boundary layer growth/thickness and the effective residence time of a fluid element within a computational cell.
To generalize this expression for flowing liquids, we again incorporate assumptions from SGS theory, particularly the parallel flow assumption, where the velocity field remains uniform within the computational cell and a negligible diffusion in the stream-wise direction, implying that advection dominates over diffusion along the primary flow direction.
To incorporate fluid motion, we evaluate the effective residence time of a fluid parcel within a FV cell.
This time depends on the \CN~$\Cour$, which quantifies how much convection happens in a given \TS within a respective FV cell with its formula ${ \Cour = \vel \Delta t / \Delta h_{c} }$, where $\Delta h_{c}$ is the characteristic mesh size.
This mesh size can be chosen as edge length for meshes with cubic elements and ${ \Delta h_{\otherChange{c}} = \sqrt[3]{V} }$ for meshes with arbitrary mesh elements.
\ReviewerTwo{Assuming parallel flow and a prismatic boundary layer mesh in later 3D~cases, we postulate that we can use the 2D~analysis leading to the SGS model in conjunction with 2D rectangular cells, which can be characterized by their cell height~$h$ and their width~$w$.
With parallel flow, in accordance to the SGS model's assumptions, we can use~$w$ in the formula given for the \CN as the characteristic mesh size~$\Delta h_c$.}

\begin{figure}[!htbp]
    \begin{subfigure}[t]{.5\textwidth}
        \centering
        \includegraphics[width=\columnwidth]{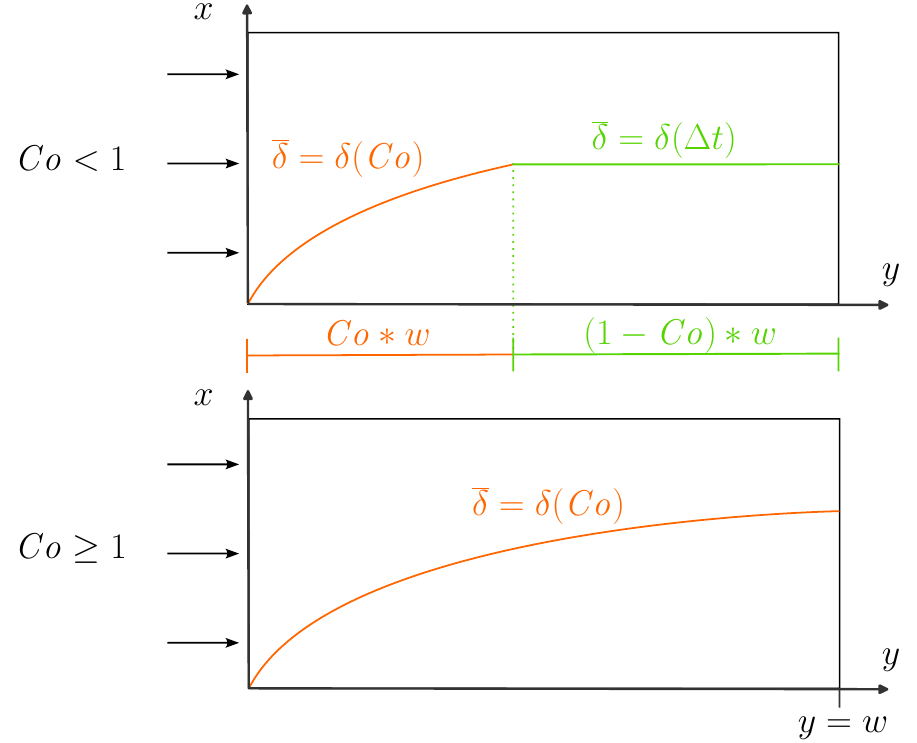}
        \caption{\otherChange{Residence} time computation (\cref{eq:tMin})}
        \label{fig:tMin_cases}
    \end{subfigure}
    \begin{subfigure}[t]{.5\textwidth}
        \centering
        \includegraphics[width=0.95\linewidth]{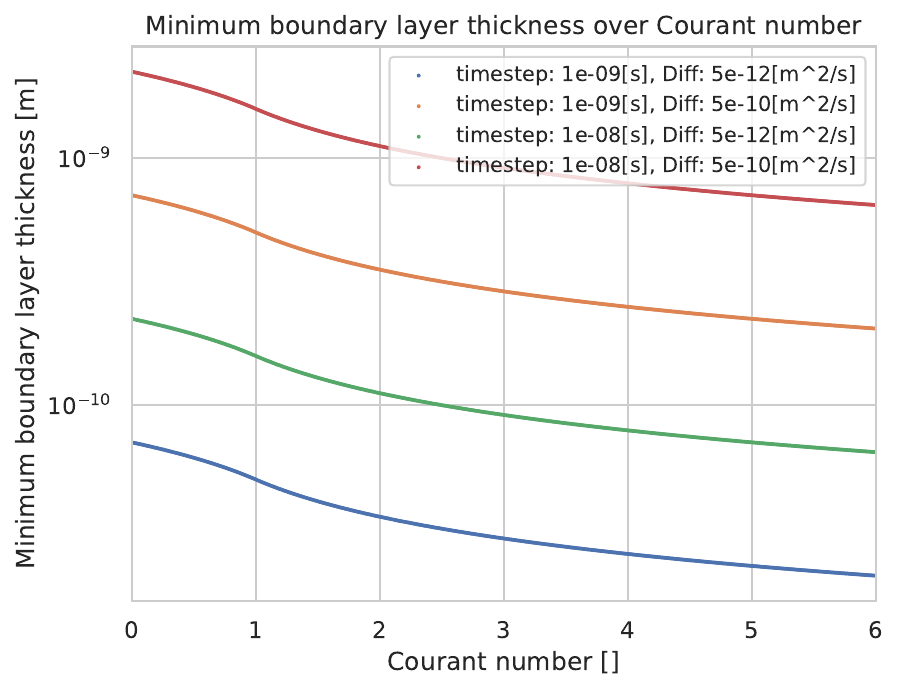}
        \caption{Parameter study over the \CN}
        \label{fig:deltaMin_PS}
    \end{subfigure}
    \caption{Minimum \BLT $\overline{\delta}$}
    \label{fig:tMin_vis}
\end{figure}

We consider two distinct cases, as depicted in \cref{fig:tMin_cases}, \ReviewerTwo{where on the horizontal axis the coordinate parallel to the boundary is depicted, which is~$y$, matching our previous notation.
The rectangle depicted represents one finite volume cell.
The cell width is denoted with~$w$.
The two cases are}:
\begin{enumerate}
    \item \textbf{Partial fluid replacement ($\Cour < 1$)}\\
    In the first case, the upper one in \cref{fig:tMin_cases}, the \CN~$\Cour$ is smaller than one.
    A fraction of the fluid in the FV cell, proportional to $(1-\Cour)$, remains within the computational cell for the full \TS~$\Delta t$.
    The remaining fraction, proportional to $\Cour$, stays within the cell for ${\Delta t / 2 / \Cour}$ on average in the present \TS.
    \item \textbf{Complete fluid replacement ($\Cour \ge 1$)}\\
    If the \CN is above one, depicted in the lower illustration in \cref{fig:tMin_cases}, the fluid is transported entirely out of the computational cell within a fraction of the \TS.
    The average time of a fluid element inside the FV cell is ${ \otherChange{(}\Delta t / 2 \otherChange{)} / \Cour}$.
\end{enumerate}
The given considerations yield these effective \TS corrections:\\
\noindent\begin{minipage}{\textwidth}
\begin{subequations} \label{eq:tMin}
    \begin{alignat}{7}
        \Delta t_{\rm eff} &= \ &\Cour  &\frac{\Delta t}{2\Cour} + (1-\Cour) \frac{\Delta t}{2} = (2-\Cour) \frac{\Delta t}{2} \qquad \qquad &&\text{for } \Cour < 1 \\
        \intertext{and}
        \Delta t_{\rm eff} &= &    &\frac{\Delta t}{2\Cour} \qquad &&\text{for } \Cour \ge 1 \ .
    \end{alignat}
\end{subequations}
\end{minipage}
For $\Cour = 0$ the fluid remains stationary and \TS $\Delta t$ is applied.

Combining \cref{eq:minBLT,eq:tMin}, we obtain the expression for the minimum \BLT in moving fluids as
\begin{equation} \label{eq:minBLT_tMin}
    \overline{\delta} = \sqrt{D \Delta t_{\rm eff}} \ . 
\end{equation}
As the \CN scales inversely to the \TS, the latter cancels out, if the formulae for $\Delta t_{\rm eff}$ and $\Cour$ are inserted in \cref{eq:minBLT_tMin}, giving an minimum \BLT irrespective of the \TS-size.
The independence of the physical \BLT from the simulation-related \TS-size not only make\otherChange{s} sense physically, it also is a necessary condition for, \textit{inter alia}, the computation of \BLT{}es in a steady-state manner.
The results of \cref{eq:minBLT_tMin} are visualized in \cref{fig:deltaMin_PS}.

\begin{algorithm}[H]
\footnotesize 
\caption{Algorithm to find the \BLT}\label{alg:newtonBisection_NEW}
\begin{algorithmic}
\State \textbf{Data:} $\text{SMALL} = \SI{1e-15}{}$, $\text{tol} = \SI{1e-9}{}$, $\text{iterations} = 10$, ${ \overline{\eta} = (\bar{c} - c_{\Sigma})/(c_\infty - c_{\Sigma}) }$
\If{($\overline{\eta} > 1$) \textbf{or} ($\overline{\eta} < \otherChange{\SI{1e-7}{}}$)} 
    \State $\delta_n = -1$
\Else
    \State $\delta_n = \ReviewerTwo{h}/(\sqrt{\pi} \, \overline{\eta} + (\pi/12) \, \overline{\eta}^3)$ 
    \State $\delta_{\text{min}} = \delta_{\text{min}}(D, \Delta t , \Cour) $ 
    
    \If{\otherChange{($\delta_n < \delta_{min}$)}}
        \State \otherChange{\textbf{return} $\delta_n = -1$}
    \EndIf
    \State $\eta^{\text{SGS}} = \text{erf}(\ReviewerTwo{h}/\delta) + (\delta / \ReviewerTwo{h})*(e^{-(\ReviewerTwo{h}/\delta)^2}-1)/\sqrt{\pi}$
    
    \State \textbf{Compute $\delta_n$:}
    \For{iteration \textbf{in} iterations}
        \State Compute $\eta^{\text{SGS}}(\delta_n)$
        \State $\text{res} = \eta^{\text{SGS}}(\delta_n) - \overline{\eta}$
        
        \If{($\text{abs}(\text{res}/(\otherChange{1-}\overline{\eta} + \text{SMALL})) < \text{tol}$)}
            \State \textbf{return} $\delta_n = -1$
        \EndIf
        
        \State $\eta' = (e^{-(\ReviewerTwo{h}/\delta)^2} - 1)/(\ReviewerTwo{h} / \sqrt{\pi})$
        \State $\delta_{n+1} = \delta_n - (\eta^{\text{SGS}}(\delta_n) - \overline{\eta})/\eta'$
        \If{($\delta_{n+1} \leq \delta_{min}$)}
            \If{($\delta_{\ReviewerTwo{n}} == \delta_{min}$)}
                \State \textbf{return} $\delta_n$
            \Else
                \State $\delta_{n+1} = \delta_{min}$
            \EndIf
        \EndIf
        \State $\delta_n = \delta_{n+1}$
        \If{($\text{iteration} \ReviewerOne{~\geq~} (\text{iterations} -1)$)}
            \State $\delta_n = -1$
        \EndIf
    \EndFor
\EndIf
\State \textbf{return} $\delta_n$
\end{algorithmic}
\end{algorithm}


\begin{figure}[!htbp]
    \begin{subfigure}[t]{0.49\textwidth}
        \centering
        \includegraphics[width=0.99\linewidth]{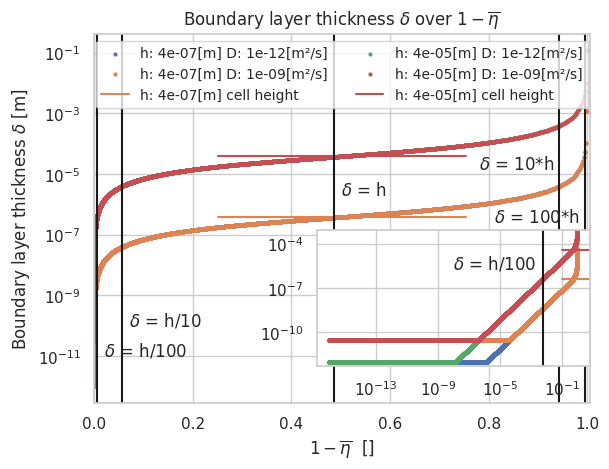}
        \caption{Calculated \BLT over $1-\overline{\eta}$}
        \label{fig:delta_over_eta}
    \end{subfigure}
    \begin{subfigure}[t]{0.49\textwidth}
        \centering
        \includegraphics[width=0.99\linewidth]{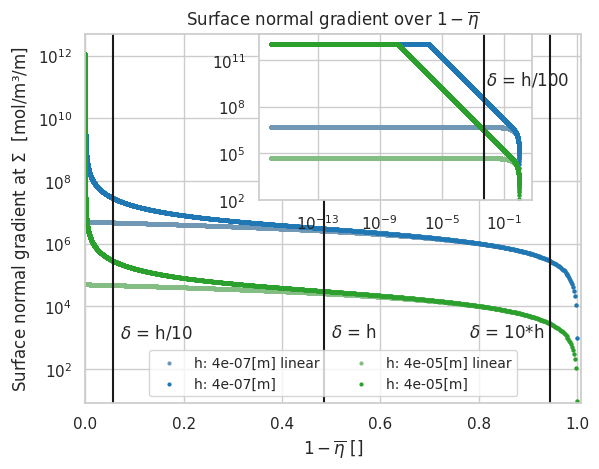}
        \caption{Surface normal gradient \otherChange{over} $1-\overline{\eta}$}
        \label{fig:limitedSGS}
    \end{subfigure}
    \caption{\otherChange{E}ffect of \otherChange{the} minimum \BLT}
    \label{fig:minDeltaViss}
\end{figure}

\Reviewers{\Cref{fig:minDeltaViss} shows the \BLT~$\delta$~(\cref{fig:delta_over_eta}) and the boundary normal derivative at the boundary~${ \partial_\NSigma c_{|x=0} }$~(\cref{fig:limitedSGS}) in dependency of~${ (1-\overline{\eta}) }$.
Later expression quantifies how close the concentration in the boundary neighbouring cell~$c_1$ is to~$c_\Sigma$ in relation to~$c_\infty$, c.f.\ \cref{eq:SGS_fittingCondition} and can be interpreted as a \textit{filling level} of the first cell.
This filling level determines the ratio~$\delta/h$ and we plot four significant levels of~$\delta/h$.
In \cref{fig:minDeltaViss} we employ the input parameters
\begin{align*}
    h = \SIlist{4e-7; 4e-5}{\ReviewerTwo{m}} \ , &\qquad
    \Delta t = \SI{1e-12}{\ReviewerTwo{s}} \ , \qquad
    \Cour = 0 \ , \qquad \\
    c_\infty = \SI{1}{\ReviewerTwo{mol/m^3}} \qquad &\text{and} \qquad
    c_\Sigma = \SI{0}{\ReviewerTwo{mol/m^3}} \ .
\end{align*}
}
\otherChange{\Cref{fig:delta_over_eta} additionally uses ${ D = \SIlist{1e-12; 1e-9}{\ReviewerTwo{m^2/s}} }$, whereas \cref{fig:limitedSGS} only presents the smaller diffusivity.}
\Cref{fig:delta_over_eta} presents the \otherChange{\BLT as a result} of \cref{alg:newtonBisection_NEW}. 
The solution exhibits two distinct states: either the minimum boundary layer constraint governs the result or the mass conservation of the fitted concentration profile can be upheld.
In the shown example, the minimum \BLT{} \otherChange{is} $\SI{1.58e-11}{m}$ and $\SI{5e-13}{m}$, \otherChange{respectively,} depending on the diffusivity given.
Its dependence on the \CN is illustrated in \cref{fig:deltaMin_PS}.
In case the \BLT is larger than the \otherChange{computed} minimum, the cell size and input parameter~${ 1-\overline{\eta} }$ determine the result.

\noindent\Cref{fig:limitedSGS} displays the (then resulting) surface normal gradient at the interface~(c.f.\ \cref{eq:SGS_anSol_grad}), \ReviewerTwo{as well as a linear approximation with~${ \partial_\NSigma c = (1-\overline{\eta} - c_\Sigma) / (h/2) }$}. 
\otherChange{The surface normal gradients computed with and without SGS show good agreement once the \BLT exceeds the cell height~$h$, that is for values of~${ (1-\overline{\eta}) \gtrapprox 0.5 }$, and diminish for values~${ (1-\overline{\eta}) \rightarrow 1 }$.}
\otherChange{\Cref{fig:limitedSGS} also shows that the maximal gradient is now given by analysis, as both shown data sets for the two cell heights collapse, determined by the physical parameters diffusion coefficient and velocity.
The (maximal) surface normal gradient is no longer a function of cell size.
For \BLT{}es bigger than the minimum value, the surface normal gradients are -~like the cell sizes~- two orders of magnitude apart, which is an expected behaviour.}

From \cref{fig:minDeltaViss} one can \otherChange{observe that} \BLT and surface normal gradient are inverse to each other.
\ReviewerTwo{One also notes how~$\delta$ and ${ \partial_\NSigma c_{|x=0} }$ inversely tend to $\infty$ or~$0$, respectively (and inversely,) close to the values~$0$ and $1$ on the abscissa.}
\Cref{fig:limitedSGS} \ReviewerTwo{illustrates}, what would happen to the surface normal \otherChange{derivative} at the interface, if no \otherChange{value for}~$\delta$ was found or it was found but very tiny.
The surface normal \otherChange{derivative} modeled by the SGS would be orders \otherChange{of} magnitude lower 
because the numerical, under-resolved value would be taken, or it would go to infinity, what impedes numerical stability and \otherChange{can lead to unphysical results, } 
as \otherChange{was} already described.

Our approach ensures that for all parameters a \BLT is found and thus an appropriate surface normal \otherChange{derivative} can be employed.
The resulting \BLT and surface normal \otherChange{derivative} are continuous and monotone.
This approach ensures robust numerical performance while preventing nonphysical concentration overshoots or unrealistic thin boundary layers in the iterative FVM framework.

\subsubsection{Novel algorithm with unknown $c_\infty$}
\label{sss:free_c_infty}

\begin{figure}[!htbp]
    \centering
    \includegraphics[width=0.7\textwidth]{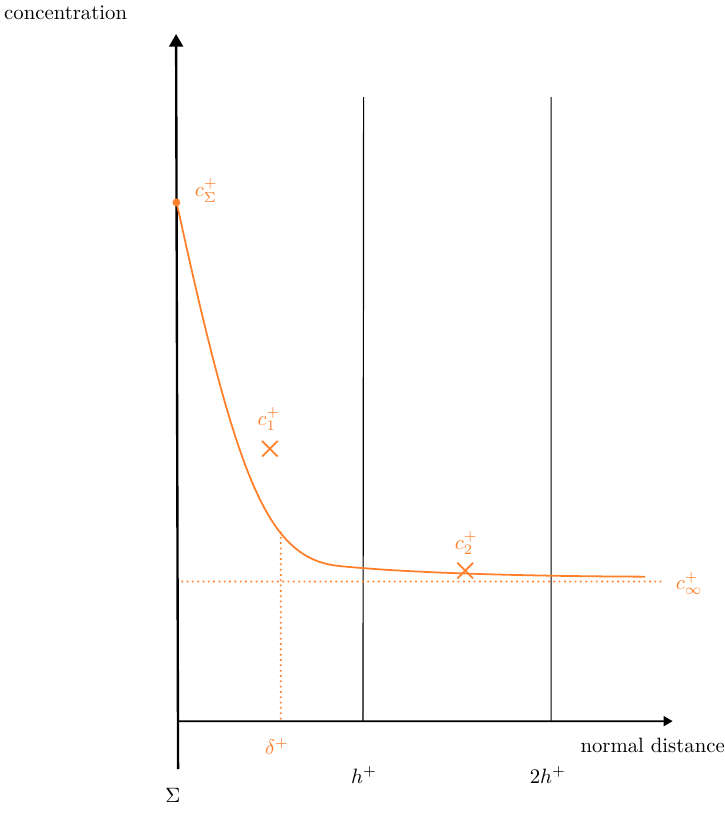}
    \caption{Concentration profile with unknown~$c_\infty$}
    \label{fig:cInfLabeling}
\end{figure}

Enhancing the SGS model with the ability to treat $c_\infty$ as an unknown parameter enables the computation of cases where the species concentration evolves dynamically.
This extension is particularly relevant in scenarios such as species accumulation inside a droplet or variable concentration in the incoming flow around a rising bubble, where the assumption of a constant far-field concentration is no longer valid.

As we now have one more free parameter we need more information to fit the concentration profile~\cref{eq:SGS_anSol} \otherChange{locally}.
\otherChange{In addition} to the concentration value in the boundary-adjacent cell \otherChange{we} take into account the concentration in the its boundary-outward neighbouring cell.
To determine the concentration profile parameters of the system\otherChange{, as} illustrated in \cref{fig:cInfLabeling}\otherChange{,} we solve the following system of equations:
\begin{equation}
    \label{eqs:freeCinf_condition}
        \int_0^{h^+} c (x) \,dx = h^+ c_{1} \qquad \qquad 
        \int_{h\ReviewerTwo{^+}}^{2{h^+}} c (x) \,dx = h^+ c_{2} \ ,
\end{equation}
where the first equation is identical to \cref{species-mass_+1} and~$h\ReviewerTwo{^+}$ denotes the cell size normal to the interface.

This modification alters the \otherChange{dependence in the computation} of the \BLT~$ \delta $, which is no longer a function ${ \delta = f(c_1) }$, but instead ${ \delta = f(c_1, c_2) }$.
Likewise, $ c_\infty $ is no longer a prescribed parameter, but instead becomes a function of the local concentration values\otherChange{, i.e.}: ${ c_\infty = f(c_1, c_2) }$.

The equation system in discretized form reads as
\begin{align} \label{eq:mass_balance_2_cells}
    \begin{split}
        c_\Sigma h^+ &+ \delta \cdot ( c_\infty - c_\Sigma ) F(\hspace{5pt}h^+/\delta \hspace{0pt}) = \hspace{12pt} c_1 h^+ \\
        2 c_\Sigma h^+ &+ \delta \cdot ( c_\infty - c_\Sigma ) F(2h^+/\delta) = (c_1 + c_2) h^+ 
    \end{split}
\end{align}
with ${ F(x)=x \, \mbox{erf} (x) + e^{-x^2}-1/\sqrt \pi \ }$.
For brevity we introduce the dimensionless parameter~${ y=h/\delta }$ and define a new parameter~$b$, allowing us to transform the equation \otherChange{system} into \otherChange{a single equation:}
\begin{equation}
    \label{eq:mass_balance_problem}
    \frac{F(2y)}{F(y)} = \frac{2 c_\Sigma - c_1 - c_2}{c_\Sigma - c_1} = 1 + \frac{c_\Sigma - c_2}{c_\Sigma - c_1} =: b \ ,
\end{equation}
\otherChange{which} can be easily solved for~$ y $.
We do so by computing $b$ from $c_1$, $c_2$ and $c_\Sigma$ and solving
\begin{equation}
    \label{eq:cInf_problem}
    f(y) := \frac{F(2y)}{F(y)} = b \ .
\end{equation}

For the SGS model to remain valid, \otherChange{the} underlying assumptions must hold.
In particular, the concentration profile must \ReviewerTwo{not only} be monotonic, \ReviewerTwo{but also the magnitude of its gradient must increase towards the interface,} which imposes
${ c_\Sigma - c_2 > c_\Sigma - c_1}$, hence ${ \ReviewerTwo{b} > 2 }$. 
On the other hand, since ${ 2 (c_\Sigma - c_1) > c_1 - c_2 }$ and ${ b = 2 + (c_1 - c_2)/(c_\Sigma - c_1) }$, it follows that $b$ must be smaller than~$4$.
Thus, the valid range for~$b$ is given by
\begin{equation}
    \label{eq:cInf_b_intervall}
    b \in \ \left] \, 2, \; 4 \, \right[ \ .
\end{equation}

\subsubsection*{Algorithm for determining unknown $\delta$ and $c_\infty$}

The function $f(y)$ is monotonically decreasing within the \otherChange{range} given \otherChange{in} \cref{eq:cInf_b_intervall}, ensuring the existence of a unique solution for~$y$.
However, numerical instabilities could arise when $b$ approaches the limiting values of $2$ or $4$, as ${ f'(x) \approx 0 }$ in these regions.
To circumvent this issue, we employ an approximate formulation~$\tilde{f}(y)$ that provides stable evaluations across the entire range:
\begin{equation}
    \label{eq:cInf_f-tilde}
    \tilde{f}(y) = 
    \begin{cases}
        2 + 2 e ^{-y^2} & \text{for } 0 \leq y \leq 0.2 \\
        f(y) & \text{for } 0.2 < y < 100 \\
        2 + \frac{0.5674}{y} & \text{for } 100 \leq y
    \end{cases} \ .
\end{equation}

This approximation introduces a relative error of less than $0.025\ \%$ for ${y \in [0, 0.2]}$ and below \mbox{\num{4e-4}~\%} for ${y \ReviewerOne{~\geq~} 100}$.
At the transition points, $b$ assumes the values~$3.9226$, respectively~$2.005676$.

Note, that $\tilde{f}(y)$ can be solved explicitly for ${b \in \ [\,3.9226,\; 4\,]}$ and ${ b \in \ ]\,2,\; 2.005676\,] }$\otherChange{, viz.}:
\begin{alignat}{3}
    y &= \sqrt{- \ln({b}/{2} - 1)} &\qquad \text{for } b \in \ &[\,3.9226, \; 4\,] \\
    \intertext{and}
    y &= \frac{0.5674}{b - 2} &\qquad \text{for } b \in \ &]\,2, \; 2.005676\,] \ .
\end{alignat}

In the other case, i.e.\ ${ b \in [2.005676, 3.9226] }$, we employ the Newton-Raphson method.
The initial guess for~$y$ is obtained by 
\begin{align}
   f_0(y) &= 
    \begin{cases}
        2.35 + 1.65 e^{-y^2} & \text{if } b > 2.5 \\
        2 + \frac{0.5674}{y} & \text{if } b \leq 2.5
    \end{cases} \ ,
\intertext{for which \cref{eq:cInf_problem} can be explicitly inverted as}
   y_0 &= 
    \begin{cases}
        \mathrlap{\sqrt{- \ln (\frac{b - 2.35}{1.65})}}\hphantom{2.35 + 1.65 e^{-y^2}} & \text{if } b > 2.5 \\
        \frac{0.5674}{b - 2} & \text{if } b \leq 2.5
    \end{cases} \ .
\end{align}

This approach ensures robust convergence of the iterative solver while maintaining numerical stability.
The overall solution procedure is summarized in \cref{alg:c_inf}.






\begin{algorithm}
\caption{Algorithm for fitting the solution with unknown $c_\infty$}\label{alg:c_inf}
\begin{algorithmic}[1]


\State Compute b
\State $b \gets 1 + (c_\Sigma - c_2)/(c_\Sigma - c_1)$

\State Compute y with $f = F(2y)/F(y)$
\If {$ b < 2$ \textbf{or} $b \ReviewerOne{~\geq~} 4$}
    \State \textbf{continue} without SGS
\ElsIf{$b = 2$}
    \State \textbf{continue} with $\delta_{min}$
\ElsIf{$b \leq 2.5$}
    \State $y_0 \gets 0.5674/(b-2)$
    \State $y \gets \text{newtonMethod}(f, y_0)$
\ElsIf{$b \leq 3.9226$} 
    \State $y_0 \gets \sqrt{-\ln((b-2.35)/1.65)}$
    \State $y \gets \text{newtonMethod}(f, y_0)$
\Else 
    \State $y \gets \sqrt{-\ln(b/2 - 1)}$    
\EndIf

\State Compute $\delta$
\State $\delta \gets \text{max}(h/y, \delta_{min})$
\end{algorithmic}
\end{algorithm}

\subsection{Subgrid-scale conjugate mass transfer} 
\label{ss:2sSGS_conjMassTransf}

\subsubsection{Governing equation system for determining $c_\Sigma^\pm$}

\begin{figure}[!htbp]
    \centering
    \includegraphics[]{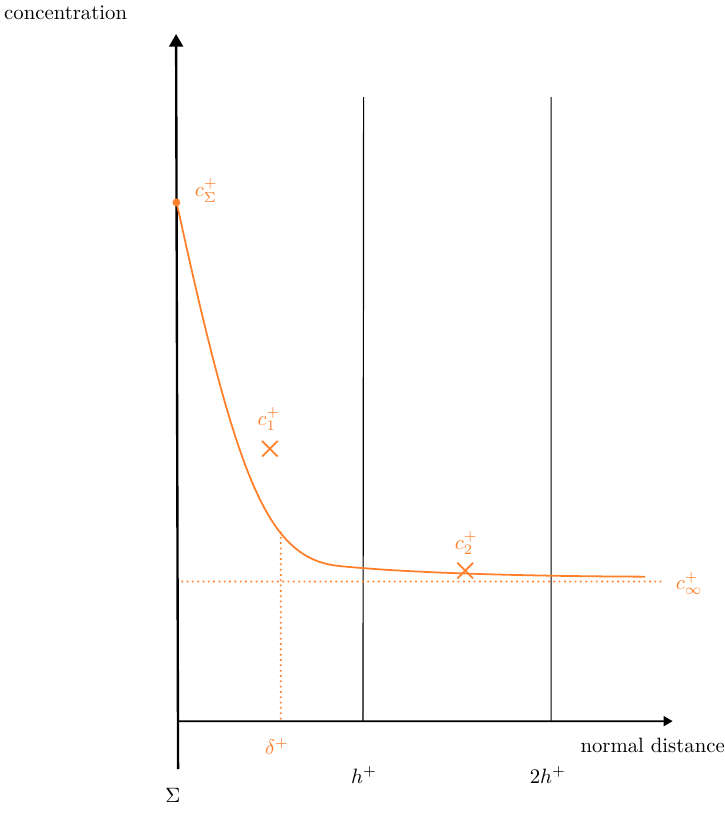}
    \caption{Concentration profiles with thin boundary layer on both sides}
    \label{fig:2sSGS_boundaryLayers}
\end{figure}

In this section, we address the scenario where thin boundary layers exist on both sides of the interface - requiring conjugate mass transfer with two-sided \SGS modeling.
The \ICs stemming from conservation equations described in \cref{ss:gov_eq_interface} are valid for conjugate mass transfer with and without thin boundary layers.
However, additional considerations must be taken into account.

Unlike the approach in \cref{ss:thinFilm}, where an explicit calculation of $c_\Sigma^\pm$ is possible from discrete cell-centered concentrations~(\ref{eq:cSigma_linear}), the presence of very thin nonlinear  boundary layers on both sides of the fluid interface introduces a coupled and highly nonlinear \otherChange{local} problem, requiring an iterative solution to determine \otherChange{the local}~$c_\Sigma^\pm$.
The resulting complex nonlinear problem is illustrated in \cref{fig:2sSGS_boundaryLayers}.
The solution to the problem is the coupling of two SGS models, one for each side of the fluid interface, fitted to the cell-centered concentrations, satisfying \ICs given by \cref{eq:mass_balance,eq:henry_law}.

In the general case, we have to determine six unknowns: $ \delta^\pm, c_{\infty}^\pm $ and~$ c_\Sigma^\pm $.
\otherChange{Depending on the simulation case in question, t}he number of required governing conditions \otherChange{changes according to} the degrees of freedom within the employed function classes\otherChange{, which need to be chosen adequately}.
Using the novel treatment of~$c_\infty$ as an unknown parameter (see \cref{sss:free_c_infty}) \otherChange{as the more general and complex case in the following part}, we  apply \otherChange{eqs.~(}\ref{eq:mass_balance_2_cells}\otherChange{)} on both sides of the interface.
This results in four equations\footnote{\otherChange{Or $2$ equations, if $c_\infty^{\pm}$ is prescribed. A mix of treatments for~$c_\infty$ is also possible and would result in $3$~equations.\\Another option is to use only one SGS and no SGS on the other side, in which case no scaling takes place, thus the gradient from the FVM method is used.}}, ensuring that the SGS-concentration profiles are consistently embedded within the FV framework.
These equations enforce mass conservation by ensuring that the species mass represented within the SGS boundary layer matches the cell-centered FV value.
To obtain the final two equations, we \otherChange{additionally} employ \otherChange{the} interfacial conditions:
\begin{enumerate}
    \item Continuity of normal mass flux across the interface, derived from \cref{eq:mass_balance} and modified for the SGS model:
        \begin{align}
            D^+ \, \partial_\NSigma^{SGS} c^+ &=
            D^- \, \partial_\NSigma^{SGS} c^- \ , \\
        \intertext{which, using the SGS formulation and already omitting the factor~$2/\sqrt{\pi}$ 
        on both sides, transforms into}
            \label{eq:2sSGS_conti}
            D^+ \, \frac{c^+_\Sigma - c^+_\infty}{\delta^+} &= 
            D^- \, \frac{c^-_\Sigma - c^-_\infty}{\delta^-} \ ,
        \end{align}
    showing the coupling between the boundary layer thicknesses~$\delta^+$ and~$\delta^-$.
    
    This formula is similar to the mass flux equation in film-theory given already in 1903 by~\cite{nernst_theorie_1904}.\\ 
    We can also draw parallels to the mass transfer coefficient in the penetration model, which is~${ 2 \sqrt{D^\pm/(2t_c)} }$, where $t_c$ denotes the exposure time~\cite{wen_fundamentals_2020}.
    \otherChange{The} above formula \otherChange{hides} this dependency on time in the parameter~$\delta^\pm$.
    
    \item Interfacial concentration jump condition, given by \cref{eq:henry_law}, which can be directly employed.
\end{enumerate}

\otherChange{Having completed} all \otherChange{the} governing equations\otherChange{, we} can rewrite and summarize the equation system.
For specific conditions coming from the adjacent bulk cells we now need to fix a class of \SGS profile functions.
Motivated by the analytical solution of the simplified local substitute problem, we employ the family of functions \otherChange{given in} \cref{eq:SGS_anSol}\otherChange{, i.e.}:
\begin{equation*} 
    c^\pm (x) = c_\Sigma^\pm \, + \, (c^\pm_\infty - c^\pm_\Sigma) \, \mbox{erf} \big( \frac{|x|}{\, \delta^\pm} \big) \ .
\end{equation*}

Insertion of these Ansatz functions into the governing equations yields, after some algebra, the following nonlinear system of six equations
for the six unknowns $c^\pm_\Sigma \, $
, $\delta^\pm \, $
, and $c^\pm_\infty \, $: 
\begin{subequations} \label{eq:6x6-eqs}
    \begin{align}
        c^+_\Sigma + (c^+_\infty - c^+_\Sigma) E(\hspace{5pt} y^+)  &= \hspace{16pt} c_1^+ \label{eq:6x6a}\\
        c^+_\Sigma + (c^+_\infty - c^+_\Sigma) E(2y^+) & = (c_1^+ + c_2^+) / 2 \label{eq:6x6b}\\
        c^-_\Sigma + (c^-_\infty - c^-_\Sigma) E(\hspace{5pt} y^-) &= \hspace{16pt} c_1^-  \label{eq:6x6c} \\
        c^-_\Sigma + (c^-_\infty - c^-_\Sigma) E(2y^-) & = (c_1^- + c_2^-) / 2 \label{eq:6x6d}\\[2pt]
        c^+_\Sigma & = \hspace{2mm} c^-_\Sigma \, / \, H   \label{eq:6x6_HJ}\\
        D^+ \frac{c^+_\Sigma - c^+_\infty}{\delta^+} & = D^- \, \frac{c^-_\Sigma - c^-_\infty}{\delta^-} \label{eq:6x6_MF} \ .
    \end{align}
\end{subequations}
Here we used the notation $ y^\pm = \frac{h^\pm}{\delta^\pm} \text{ and } E(x)=\mbox{erf} (x) +\frac{e^{-x^2}-1}{x\sqrt \pi} \, $. 

This equation system captures the nonlinear coupling between interfacial concentrations, boundary layer thicknesses, far-field species concentrations and concentration values in the first and second cells adjacent to the interface ensuring consistent conjugate mass transfer with SGS modeling.

\subsubsection{Solution strategy for the interface boundary conditions}
\label{sss:TOS-F-SGS_fitting}


The nonlinear equation system~(\ref{eq:6x6-eqs}) is solved using an iterative procedure, which fits two one-sided SGS models to cell-centered concentrations on either side, that is to $c_{1,2}^\pm \,$, while ensuring that the \ICs~(\cref{eq:mass_balance,eq:henry_law}, respectively \cref{eq:6x6_MF,eq:6x6_HJ}) are \otherChange{upheld}.
\otherChange{T}he system is initialized by currently available~$c_\Sigma^{\pm,o}$ \otherChange{from the outer iteration~$\otherChange{o}$ of the solution algorithm for conservation-law equations (e.g.~PIMPLE algorithm, c.f.~\citep{greenshields_notes_2022,nilson_open-source_2022})}.
\otherChange{In case of~$o=0$ this corresponds to the final value from the previous \TS.}
\otherChange{The coupling strategy for eqs.~(\ref{eq:transport}) and~(\ref{eq:6x6-eqs}) is described in the next \cref{sss:DD_coupling_w_SGS}.}

In the solution strategy for \otherChange{the local, nonlinear system~(\ref{eq:6x6-eqs})} - the \otherChange{inner} iterati\otherChange{ve procedure -} \otherChange{we solve eqs.}~(\ref{eq:6x6-eqs}) \otherChange{for fixed} cell-centered values~$c_{1,2}^{\pm,o}$ from the outer iteration. 
To solve the equation system, we make use of several steps:
\begin{enumerate} \label{enu:tos_f_sgs}
    \item \textbf{Fit SGS models on both sides:}\label{step:fitSGS}\\
    We fit \otherChange{the} SGS \otherChange{error function} 
    in each phase to \otherChange{currently available} boundary values~\otherChange{${ c_\Sigma^{\pm,j} }$}, 
    \ReviewerTwo{which}\label{lbl:whatWhich} corresponds to satisfying \cref{eq:6x6a,eq:6x6b,eq:6x6c,eq:6x6d}.
    \otherChange{We obtain the \BLT{}es~$\delta^{\pm,SGS,j}.$}
    
    \item \textbf{Scale the diffusion coefficients:}\label{step:scaleDiffs}\\
    \otherChange{T}he diffusion coefficients~$D^{\pm}$ are scaled to $D^{\pm,SGS,j}$ in order to mimic the \otherChange{interface normal derivative} \otherChange{at the interface} in \cref{eq:SGS_anSol_grad} on both sides. 

    \item \textbf{\otherChange{Compute} Dirichlet boundary values $c_\Sigma^{\pm,fit}$:}\\
    \otherChange{I}nterfacial concentrations $c_\Sigma^{\pm\otherChange{,j,fit}}$ are 
    \otherChange{computed} using \cref{eq:cSigma_linear}, but incorporating the scaled diffusivities $D^{\pm, SGS\otherChange{, j}}$.
    \otherChange{
    These values satisfy the \ICs~\mbox{(\ref{eq:6x6_HJ}, \ref{eq:6x6_MF})} for the scaled diffusion coefficients from step~\ref{step:scaleDiffs}.
    However, for these~$c_\Sigma^{\pm}$ the computed boundary layers from step~\ref{step:fitSGS} and thus also the~$D^{\pm,SGS,j}$ from step~\ref{step:scaleDiffs} are not valid anymore.
    }
    
    \item \textbf{Update Dirichlet boundary values $c_\Sigma^{\pm,j+1}$:}\\
    \otherChange{Using an under-relaxation approach for this nonlinear and potentially stiff problem $c_\Sigma^{\pm,j+1}$ is computed as an average of $c_\Sigma^{\pm,j}$ and~$c_\Sigma^{\pm,j,fit}$.}
    \otherChange{In essence, this step 
    facilitat\otherChange{es} the communication between the two one-sided SGS models.}
\end{enumerate}



This completes a single iteration~\otherChange{$j$}, resulting in Dirichlet boundary conditions~$c_\Sigma^{\pm,j+1}$ and scaled diffusion coefficients~$D^{\pm, SGS, j}$.
A\otherChange{n} iteration~$j$ is considered converged \otherChange{in one interface cell}, when the change of~$c_\Sigma^+$ between $j+1$ and $j$ 
drops below a user-defined tolerance~\otherChange{$\tau_{c_\Sigma}$}:
\otherChange{${ | c_{\Sigma}^{+,j+1} - c_{\Sigma}^{+,j} | \leq \tau_{c_\Sigma} }$}.
By the Henry jump, $c_\Sigma^+$ and $c_\Sigma^-$ are coupled, so \otherChange{if} one \otherChange{is} converged\otherChange{, then} the other is too.


This method couples two one-sided SGS models~\otherChange{(\cref{eq:6x6a,eq:6x6b,eq:6x6c,eq:6x6d})} via the interfacial species jump condition and species flux continuity~(\cref{eq:6x6_HJ,eq:6x6_MF}) at the interface and iterates until the interfacial concentrations~$c_\Sigma^\pm$ converge, ensuring consistency with all equations in equation system~(\ref{eq:6x6-eqs}).
\otherChange{It is given in the \textcolor{blue}{blue} part of \cref{alg:tosfsgs}.}
\otherChange{We denote this iterative solution strategy for the nonlinear two-sided SGS problem as the \textit{\textbf{T}wo \textbf{O}ne-\textbf{S}ided \textbf{F}itted \textbf{SGS}}~(TOS-F-SGS) algorithm.}

\subsubsection{\otherChange{Segregated} Dirichlet-Dirichlet coupling strategy with SGS modeling}
\label{sss:DD_coupling_w_SGS}

\begin{algorithm}
\caption{\otherChange{TOS-F-SGS s}olution algorithm: \textcolor{brown}{\textit{Solve bulk equations}} \textcolor{black}{and} \textcolor{blue}{\textit{Satisfy interface conditions}}}\label{alg:tosfsgs}
\begin{algorithmic}[1]
\For{time step \textbf{in} time steps}
    \State \textcolor{brown}{\texttt{...}}
    \For{\textcolor{brown}{outer iteration $o$} \textbf{in} \textcolor{brown}{outer iterations}}
        \State \textcolor{blue}{\textbf{Solve eqs.~\ref{eq:6x6-eqs}} for $c_{\Sigma}^\pm$ and $D^{SGS,\pm} $ for all faces~$f$: \Comment{with TOS-F-SGS strategy}}
        \State \textcolor{blue}{set tolerance $\tau_{c_\Sigma}$}
        \State \textcolor{blue}{$c_\Sigma^{\pm,j=0} = c_\Sigma^{\pm,o} $}
        \For{\textcolor{blue}{inner iteration $j$} \textbf{in} \textcolor{blue}{inner iterations}}
            \State \textcolor{blue}{\textbf{Solve eqs.~\ref{eq:6x6-eqs}(a-d):}}
            \State \textcolor{blue}{Fit SGS profile \cref{eq:SGS_anSol} on both sides} using \cref{eq:mass_balance} or~(\ref{eq:mass_balance_2_cells}):
            \State \textcolor{blue}{$\delta^{\pm, j} = \delta^\pm(c_\Sigma^{\pm,j} , c_{1}^{\pm, o} , c_{\inf}^{\pm} )$ or 
            $\delta^{\pm, j} = \delta^\pm(c_\Sigma^{\pm,j} , c_{1}^{\pm, o} , c_{2}^{\pm, o} )$} 
            \State \textcolor{blue}{update $D^{SGS,\pm,j}$ as $D^{SGS,\pm,j} = f(\delta^{\pm, j})$} \Comment{c.f.\ \cref{eq:scaling_diff}}
            \State \textcolor{blue}{\textbf{Solve eqs.~\ref{eq:6x6-eqs}(e+f):}}
            \State \textcolor{blue}{Apply mass flux and Henry-jump (\cref{eq:2sSGS_conti,eq:henry_law}) }
            \State \textcolor{blue}{$c_\Sigma^{\pm,fit}  = f( c_{1}^{\pm, o}, D^{SGS,\pm,j+1} )$ } \Comment{c.f.\ eqs.~(\ref{eq:cSigma_linear}), but with $D^{SGS,\pm}$}
            \State \textcolor{blue}{$c_\Sigma^{\pm,j+1} = ( c_\Sigma^{\pm,fit} + c_\Sigma^{\pm,j} ) / 2 $} 
            \If{\textcolor{blue}{$ \left| c_\Sigma^{\pm,j} - c_\Sigma^{\pm,j} \right| < \tau_{c_\Sigma}$}}
                \State \textbf{break}
            \EndIf
        \EndFor
        \State \textcolor{blue}{update $\delta^{\pm, j+1}$, $D^{SGS,\pm,j+1}$ and $F_f^{SGS,\pm,j+1}$ from last $c_\Sigma^{\pm,j+1}$}
        \State \textcolor{brown}{\textbf{Solve \cref{eq:transport}} for $c_k$ in $\PMDomain$} 
        \State \textcolor{brown}{with fixed $c_\Sigma^{\pm, j+1}$, $D^{SGS, \pm, j+1}$ and $F_f^{SGS,\pm,j+1}$}
        \If{\textcolor{brown}{$c^{\pm, o+1}$ converged acc.\ to \cref{eq:residual}}} 
            \State \textbf{break}
        \EndIf
    \EndFor
    \State \textcolor{brown}{\texttt{...}}
\EndFor
\end{algorithmic}
\end{algorithm}



In the FVM the species concentration is computed by solving \otherChange{the} linearized system~\otherChange{(\ref{eq:clinsystem})}. 
In contrast to the \otherChange{\textit{linear Dirichlet-Dirichlet solution algorithm}} presented in \cref{sss:DD_coupling,alg:linDD}, where an algebraic eqs.~(\ref{eq:cSigma_linear}) suffices \otherChange{to couple the bulk concentrations~$c_k^\pm$ in \cref{eq:transport} in the subdomains~$\PMDomain$}, the introduction of SGS modeling transforms 
coupling \otherChange{the concentrations in both bulk phases} into a highly nonlinear coupl\otherChange{ing} problem\otherChange{.} 
\otherChange{This problem is} represented by equation system~(\ref{eq:6x6-eqs}).
The presence of scaled diffusion coefficients and fluxes near the interface, which reflect the influence of the \otherChange{under-resolved} thin boundary layers, necessitates a modified form of \cref{eq:discr_conc}, which incorporates the updated SGS diffusion coefficients and fluxes.



\otherChange{To solve the nonlinear and coupled equation systems~(\cref{eq:clinsystem,eq:6x6-eqs}), we employ a segregated solution strategy.}
However, \otherChange{in a segregated algorithm each part of the segregated problem is solved by using the values from the previous solution of the other segregated part.} 
\otherChange{Such} inconsistency requires solving the nonlinear, \otherChange{segregated} system iteratively. 
To ensure consistency between the concentration field and the SGS correction, we introduce a nested iteration strategy, \otherChange{iterating between updating the interfacial concentrations, scaled diffusion coefficients and SGS-adjusted concentration fluxes with the (nested) TOS-F-SGS strategy and solving the species transport equation in the bulk.
The details are given in \cref{alg:tosfsgs}.}

    
\noindent\begin{minipage}{\linewidth}
The updated form of the species concentration equation becomes
\begin{equation}
    \left( \frac{c_k^{o+1} - c_k^n}{\Delta t}\right) =
    - \sum_{f \in F_k} {F_f^{\pm, SGS, o, j+1} c_f^{\pm, o+1}} 
    + \sum_{f \in F_k} D_f^{\pm, SGS, o, j+1} (\nabla c)_f^{\pm, o+1} \cdot S_f^\pm
\end{equation}
with
\begin{equation*}
    c_\Sigma^{\pm, o, j+1} \otherChange{=} c_f^{\pm, o+1} \text{ on } \Sigma
    \quad \text{and} \quad F_f^{\pm, SGS, o, j+1} \approx F_f^{\pm, SGS, j}
    \quad \text{and} \quad D_f^{\pm, SGS, o, j+1} \approx D_f^{\pm, SGS, j} \ ,
\end{equation*}
where
\begin{equation*}
    F_f^{\pm, SGS, o, j+1} = f(c_\Sigma^{\pm, o, j}, c_1^{\pm, o}, c_2^{\pm, o}) \quad \text{and} \quad
    D_f^{\pm, SGS, o, j+1} = f(c_\Sigma^{\pm, o, j}, c_1^{\pm, o}, c_2^{\pm, o}) \ .
\end{equation*}
\end{minipage}
\vspace{0.3em}

The outer iterations advance the species transport equation in the bulk in time while incorporating the SGS corrections through Dirichlet-Dirichlet coupling.
The inner iterations of the TOS-F-SGS \otherChange{algorithm} solve the nonlinear equation system~(\ref{eq:6x6-eqs}) until the interfacial boundary layers, concentrations and far-field values are consistent with both the local finite volume values and the interfacial concentration jump and mass flux continuity conditions.
The combined approach ensures that the SGS model remains stable and accurate. 
This nested iterative approach is necessary to capture the non\otherChange{-}linearity of interfacial transport processes with thin boundary layers on one or both interface sides.


\section{Results}
\label{sec:results}

\newcommand{\testFP}{flat plate }

\subsection{(One-sided) \testFP with unknown $c_\infty$}

\begin{figure}[!htbp]
    \centering
    \includegraphics[width=0.9\linewidth]{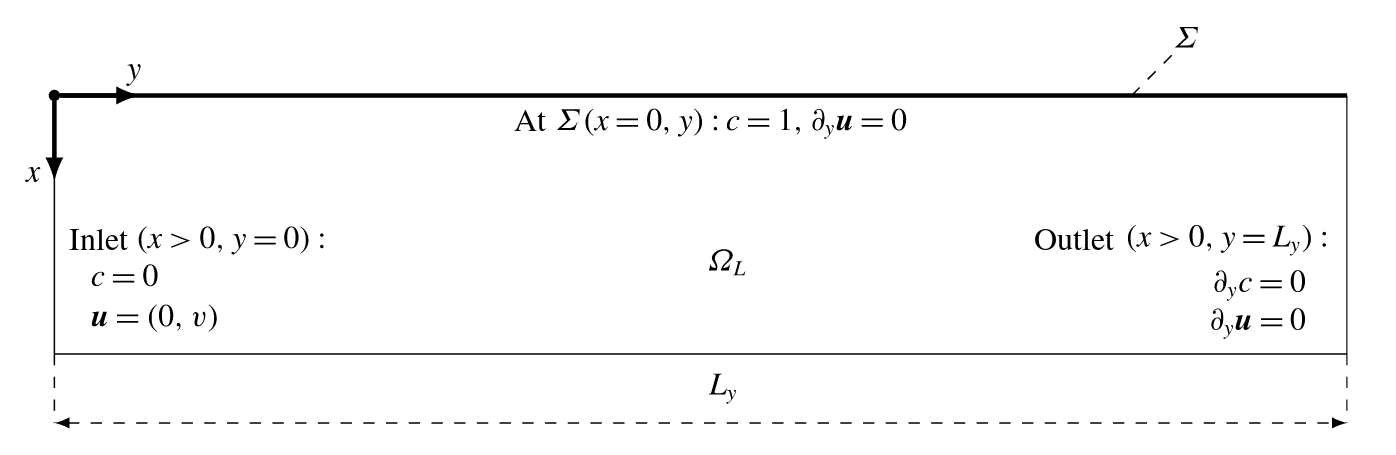}
    \caption{Setup of the \testFP test case, taken from \cite{pesci_SGS_risBubb_surfactants_2018}}
    \label{fig:FP_setup}
\end{figure}

We test the new method to treat the unknown parameter~$c_\infty$ introduced in \cref{sss:free_c_infty} with the \testFP case.
\otherChange{This test case has also been performed by~\mbox{\cite{schwarzmeier_twophaseintertrackfoam_2024,pesci_SGS_risBubb_surfactants_2018,weiner_advanced_2017}}.}
For this test case we can compare the \otherChange{numerical} solution with an analytical solution from~\cref{eq:SGS_anSol}\ReviewerTwo{, which is valid for high \PNs}.
\otherChange{Additionally we run the test case without the SGS model.}

\otherChange{With the \PN~$\Pecl$ characterizing the rate of advective to diffusive transport defined as
\begin{equation} \label{eq:peclet}
    \Pecl = \frac{L\,u}{D} 
\end{equation}
and with taking the length of the plate as reference length~$L$ our test case has \PNs ranging from \SI{10}{} to~$10^7$.} 
The test setup is visualized in \cref{fig:FP_setup}.
The domain has a length of \SI{5}{mm} and height of~\otherChange{\SI{1.5}{mm}}.
Everywhere a velocity of \SI{0.1}{m/s} parallel to the plate is prescibed.
A total time of \otherChange{\SI{0.2}{s}} is simulated \otherChange{with a \TS size of \SI{5e-4}{s}}, what is sufficient to reach a quasi-steady state.
The cell size is varied from \otherChange{\SIrange{7.4e-6}{5.6e-4}{m}} while 
the diffusivities \otherChange{\SIlist{5e-5; 5e-8; 5e-11}{m^2/s}} are \otherChange{used}. 
\otherChange{The coarsest test case has the height resolved with \SI{3}{} finite volume cells and the length with \SI{9}{}, whereas the finest test case is 80 times finer in every spacial coordinate.}

\begin{figure}[!htbp]
    \centering
    \includegraphics[width=1.0\linewidth]{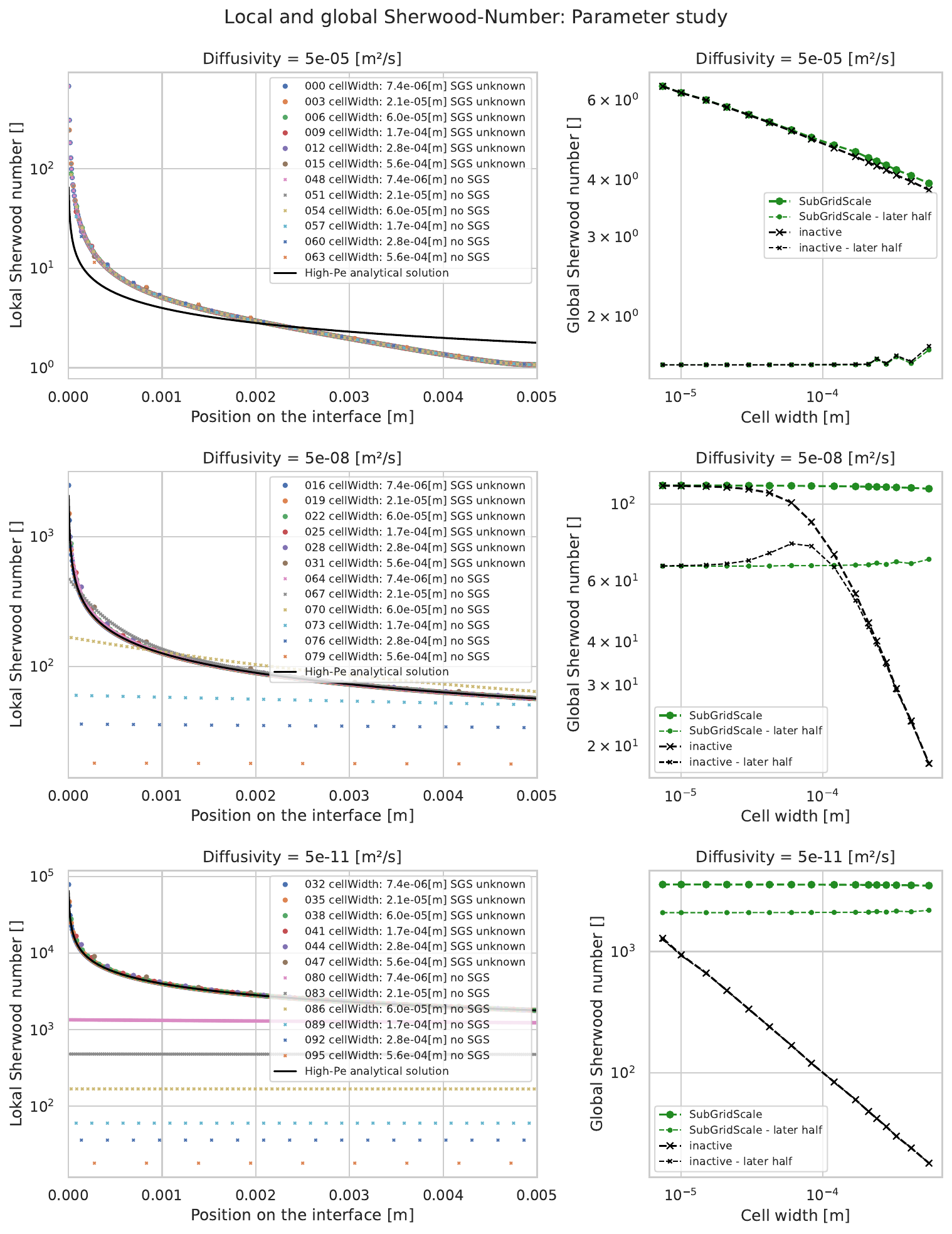}
    \caption{Sherwood numbers \otherChange{for} the plate with the new algorithm with unknown~$c_\infty$}
    \label{fig:FP_cInf_Sh}
\end{figure}

In \cref{fig:FP_cInf_Sh} \otherChange{the local} Sherwood numbers along the interface are displayed, \otherChange{as well as the global Sherwood numbers along the whole plate and within the lower half in downstream direction, that is from \SIrange{2.5}{5}{mm} as position on the plate.} 
We observe excellent agreement \otherChange{of the local Sherwood numbers computed with SGS modeling} for all cell sizes with the analytical solution for small diffusivities.
\otherChange{As a consequence the respective global Sherwood numbers exhibit nearly mesh-independent behaviour, yielding the correct solution.}
\otherChange{The \SNs computed without SGS model are nearly position-independent but are highly mesh-dependent.}

The high diffusivities however violate the model assumption of high \PNs, which lead\otherChange{s} to the simplification of including only diffusive transport normal to the interface \otherChange{in the modeling}.
\otherChange{T}his model\otherChange{ing} assumption does not hold for \otherChange{the} small \PNs \ReviewerTwo{reported in this parameter study}.
Therefore the analytical model cannot predict \otherChange{the} correct solution for small \PNs.
\ReviewerTwo{The SGS assumptions not holding true and consequently not predicting the correct solution} generally poses no problem, since small \PNs usually can be resolved with FVM, \ReviewerTwo{in which case applying the SGS will not affect the solution}.
\otherChange{The global \SN{}s in the whole domain are not mesh-independent but the \SNs computed from the lower half of the domain are and thus we can conclude the solution to be converged in the traditional~FVM.
The results with SGS modeling show a high degree of similarity to those without SGS modeling.}


\otherChange{For the medium \PNs the local and global \SNs computed with SGS model are nearly mesh-independent, as is expected.
The parameter study without SGS modeling over mesh resolution reveals the computation to converge to the correct solution for small cell sizes, but under-resolving the solution for big cell sizes.
In between, as the global \SN is converging, the global \SN in the later half is overpredicted.
}

Overall, we \otherChange{come to the} conclu\otherChange{sion}, the method with a non-fixed~$c_\infty$ to be correctly implemented and working well for this simple test case.

\newcommand{\tsFP}{two-sided flat interface\xspace}

\subsection{Mass transfer in parallel flow along planar interface}
\label{secSub:test_flatPlate}

\subsubsection{Test case setup}


The first test case for the two-sided SGS for model development and implementation is the \tsFP test case.
It consists of a planar interface~$\Sigma$, representing a boundary between two different fluids/phases with different (initial) concentrations~$c^\pm$ and different diffusivities~$D^\pm$.
At the interface~$\Sigma$ the one-sided concentration limits~$c_\Sigma^\pm$ obey Henry's law~\cref{eq:henry_law}.

\begin{figure}[!htbp]
    \centering
    \includegraphics[width=0.6\textwidth]{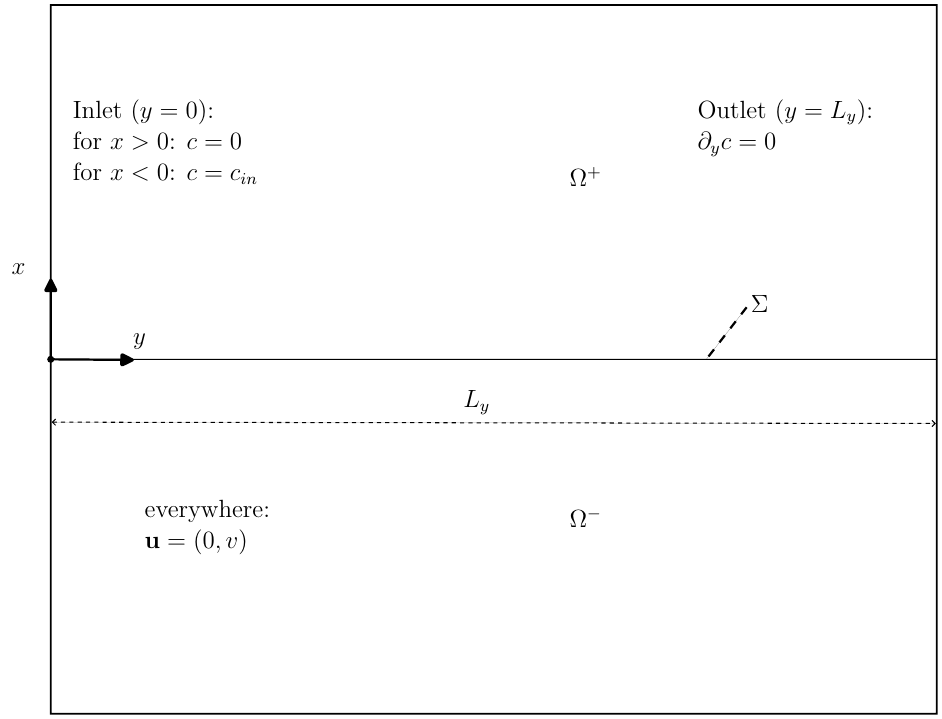}
    \caption{Setup of the \tsFP test case}
    \label{fig:2sFP_setup}
\end{figure}

A flow with the interface-parallel velocity~$\vel$ is present on either side of the interface.
It does not matter, whether one perceives the interface as moving with the flow or as stationary but with perfect slip condition on both sides.
In \cref{fig:2sFP_setup} we visualize the setup of the \tsFP test case.
The problem is a 2D~problem.
For the analytical solution of the problem we introduce a coordinate system, where the point of origin marks the start of the solution domain, $\Omega \ni x>0$.
The interface lies at $y = 0$, therefore we call the "upper" part of the domain, where $y>0$, $\Omega^+$, and the "lower" part~$\Omega^-$.

Initially and in the inflow from the domain boundary at $x=0$, the concentration~$c$ is prescribed.
For $y>0$, following our convention, we write $c^+$ and we initialize it and set the inflow \otherChange{at}~${ \otherChange{(x=0,} \ y>0 \otherChange{)}}$ $c^+_{in}$ to~$0$.
We will treat $c^+_{in}$ as a reference level, which is set to $0$, and will omit it in the following.
The concentration in~$\Omega^-$ and its inflow at~${ (x=0, \ y \otherChange{< 0}) }$ is set to~$c^-_{in}$.
For brevity we will call it $c_{in}$ from here on.

In our test we use the velocity \SI{0.1}{m/s} in a domain that is \SI{5}{mm} long and \SI{4}{mm} high.
The interface cuts the height into two equally sized regions.
The diffusivities used are \SI{2.976e-9}{m^2/s} for $\Omega^+$ and \SI{1.3155e-9}{m^2/s} for $\Omega^-$, representing acetone in water and in toluol, respectively~\cite{wegener_einfluss_2009}.
The Henry coefficient~$H$ is \SI{1.5873}{}~\cite{wegener_einfluss_2009}\footnote{Please note that the reciprocal value is used to the referenced publication.}.
The concentration initially and in the inflow is set to \SI{1}{mol/m^3} for the lower side and \SI{0}{mol/m^3} for the upper side.
The simulation is run for \SI{0.1}{s} with a \TS size of \SI{2e-4}{s}.
The simulation domain is decomposed with a rectangular grid with a cell size varied from \SIrange{1e-5}{2e-4}{m}.

The case is run with the state-of-the art algorithm with prescribed~$c_\infty$ on both sides, 
the $c_\infty$ prescribed on one side only 
and the new algorithm from \cref{sss:free_c_infty} on both sides. 
\ReviewerTwo{The~$+$-side is denoted with \textit{FS} and the~$-$-side with \textit{FSS}.}
If~$c_\infty$ is prescribed, the modifications introduced in \cref{sss:iSGS} are used.

\subsubsection{Analytical solution (steady state)}

For the \tsFP test case we have different conditions, given in the setup visualized in \cref{fig:2sFP_setup}:
\begin{subequations}
    \label{eqs:2sFP_cond}
    \begin{align}
        \rm v \, \partial_\textit{\otherChange{x}} c^\pm &= D^\pm \partial_y^2 c^\pm \qquad \qquad \qquad \text{for } x>0 \ , \\
        c^+(x, y=0) &= c^-(x, y=0) / H  \ , \\
        D^+ \partial_y c^+(x, y=0) &= D^- \partial_y c^-(x, y=0) \ , \\
        c^+(x=0, y) &= 0
        \intertext{and}
        c^-(x=0, y) &= c_{in} \ .
    \end{align}
\end{subequations}
We conduct a Laplace transformation w.r.t.\ $x>0$ with the Laplace variable called~$s>0$:
\begin{align}
    s \, \hat{c}^\pm(s,y) &= c^\pm(x=0,y) = D^\pm / v \partial_y^2 \hat{c}^\pm(s,y) \label{eq:2sFP_laplTransformed}\\
    \intertext{with}
    \hat{c}^+(s, 0) &= \hat{c}^-(s, 0) / H \qquad 
    \intertext{and }
    D^+ \partial_y \hat{c}^+(s,0) &= D^- \partial_y \hat{c}^-(s,0) \ .
\end{align}
\noindent Using the additional BCs at $\infty$, i.e.\ $ c^+(x, + \infty) = 0 $ as well as $ c^-(x, + \infty) = c_{in} \,$, it follows that
\begin{subequations}
    \begin{alignat}{5}
        \hat{c}^+(s,y) &\longrightarrow 0 \qquad &&\text{as } y \longrightarrow &&\infty \label{eq:2sPF_snSol_c+_as_y_inf} \quad
        \intertext{ and}
        \hat{c}^-(s,y) &\longrightarrow c_{in}/s \qquad &&\text{as } y \longrightarrow - &&\infty \ .
    \end{alignat}
\end{subequations}
We can now write the ODEs in $y$ (for all $s>0$) as
\begin{equation} \label{eq:2sFP_anSol_ODE}
    s \hat{c}^-(s,y) - c_{in} = \frac{D^-}{\rm v \partial_y^2 \hat{c}^- (s,y)} \qquad \qquad y < 0 \ .
\end{equation}    
Let $u^-(s,y) = \hat{c}^-(s,y) - c_{in} / s \,$.
Then \cref{eq:2sFP_anSol_ODE} yields
\begin{equation}
    u^-_{yy} = \frac{\rm v s}{D^-} u^- \ .
\end{equation}
Exploiting ${ u^-(s,y) \longrightarrow 0 \text{ as } y \longrightarrow - \infty \, }$,
we obtain
\begin{alignat}{2}
    u^-(s,y) &= &A^-(s) \, e^{\sqrt{\frac{\rm v s}{D^-} y}} \qquad &y > 0 \ \otherChange{,}
    \intertext{hence}
    c^-(s,y) &= \frac{c_{in}}{s} + &A^-(s) \, e^{\sqrt{\frac{\rm v s}{D^-} y}} \qquad &y > 0 \ .
\end{alignat}
Analogously, for $c^+$ we get
\begin{equation}
        \partial_y^2 \hat{c}^+(s,y) = \frac{\rm v s}{D^+} \hat{c}^+(s,y) \ .
\end{equation}
With \cref{eq:2sPF_snSol_c+_as_y_inf} it follows
\begin{equation}
    c^+(s,y) = A^+(s) e^{-\sqrt{\frac{\rm v s}{D^+} y}} \qquad y > 0 \ .
\end{equation}
We evaluate the transmission conditions to get $A^+(s)$ and $ A^-(s) \, $.
Using Henry's law~\cref{eq:henry_law}, we have
\begin{equation} \label{eq:2sPF_anSol_cPlus_s}
        c^+(s,0) = A^+(s) 
        = c^-(s,0) / H 
        = \left( \frac{c_{in}}{s} + A^-(s) \right) / H 
\end{equation}
and, using the mass conservation~\cref{eq:mass_balance},
\begin{equation}
        -D^+ \partial_y \hat{c}(s,0) = A^+(s) \sqrt{\rm v s D^+} \\ 
        = - D^- \partial_y \hat{c}^-(s,0) \\ 
        = -A^-(s) \sqrt{\rm v s D^-} \ .
\end{equation}
The latter yields
\begin{equation}
    A^+(s) \sqrt{D^+} = -A^-(s) \sqrt{D^-} \ .
\end{equation}
Inserting into \cref{eq:2sPF_anSol_cPlus_s}
\begin{equation}
    HA^+(s) = \frac{c_{in}}{s} -A^+(s) \sqrt{\frac{D^+}{D^-}} \ .
\end{equation}
We thus obtain expressions for $A^\pm(s)$ as
\begin{equation}
    A^+(s) = \frac{c_{in}}{H + \sqrt{\frac{D^+}{D^-}}} \frac{1}{s} \qquad \text{and} \qquad
    A^-(s) = -\frac{c_{in} \sqrt{\frac{D^+}{D^-}}}{H + \sqrt{\frac{D^+}{D^-}}} \frac{1}{s} \ .
\end{equation}

Using the fact that $ f(x) = {\rm erfc} \sqrt{\frac{a}{4x}} $ with $ (a > 0)$ has Laplace transform~${ \hat{f}(s) = 1/s \ e^{-\sqrt{as}} \, }$, we hence obtain the steady-state \tsFP concentrations~$c^\pm(x,y)$ to be given as\\
\noindent\begin{minipage}{\textwidth}
\begin{subequations}
    \begin{alignat}{5} \label{eq:2sFP_anaLsg}
        c^+(x,y) 
            &=  &&\frac{c_{in}\sqrt{D^-}}{\sqrt{D^+} \otherChange{+} H \sqrt{D^-}} &&{\rm erfc} \left( \frac{y}{\delta^+(x)} \right) \quad
        &\text{with } \delta^+(x) = \sqrt{4D^+ x/\rm v} & {} \quad x>0, y>0 \\ 
        \intertext{and}
        c^-(x,y) &=  c_{in} - &&\frac{c_{in}\sqrt{D^+}}{\sqrt{D^+} + H \sqrt{D^-}} &&{\rm erfc} \left( \frac{-y}{\delta^-(x)} \right)
        &\text{with } \delta^-(x) = \sqrt{4D^- x/\rm v} & {} \quad x>0, y<0 \ . 
    \end{alignat}
\end{subequations}
\end{minipage}

\ReviewerTwo{
Analogously to \cref{eq:SGS_anSol_grad} from \cref{eq:SGS_anSol} and by inserting the coordinate~${y=0}$, we determine the surface normal gradient at the interface to be\\
\noindent\begin{minipage}{\textwidth}
\begin{subequations}
    \begin{alignat}{1} \label{eq:2s_FP_nSigma}
        \partial_\NSigma c^+(x,y=0) 
            &= \frac{2}{\sqrt{\pi}} \frac{c_{in}\sqrt{D^-}}{\sqrt{D^+} + H \sqrt{D^-}} \frac{1}{\delta^+(x)} \\
        \intertext{and}
        \partial_\NSigma c^-(x,y=0) 
            &=  \frac{2}{\sqrt{\pi}} \frac{c_{in}\sqrt{D^+}}{\sqrt{D^+} + H \sqrt{D^-}} \frac{-1}{\delta^-(x)} \ .
    \end{alignat}
\end{subequations}
\end{minipage}
}

\subsubsection{Results with prescribed $c_\infty$}
\label{sss:2s_FP_results_prescribed}

\begin{figure}[!htbp]
    \begin{subfigure}[t]{.49\textwidth}
        \centering
        \vskip0pt
        \includegraphics[width=0.98\linewidth]{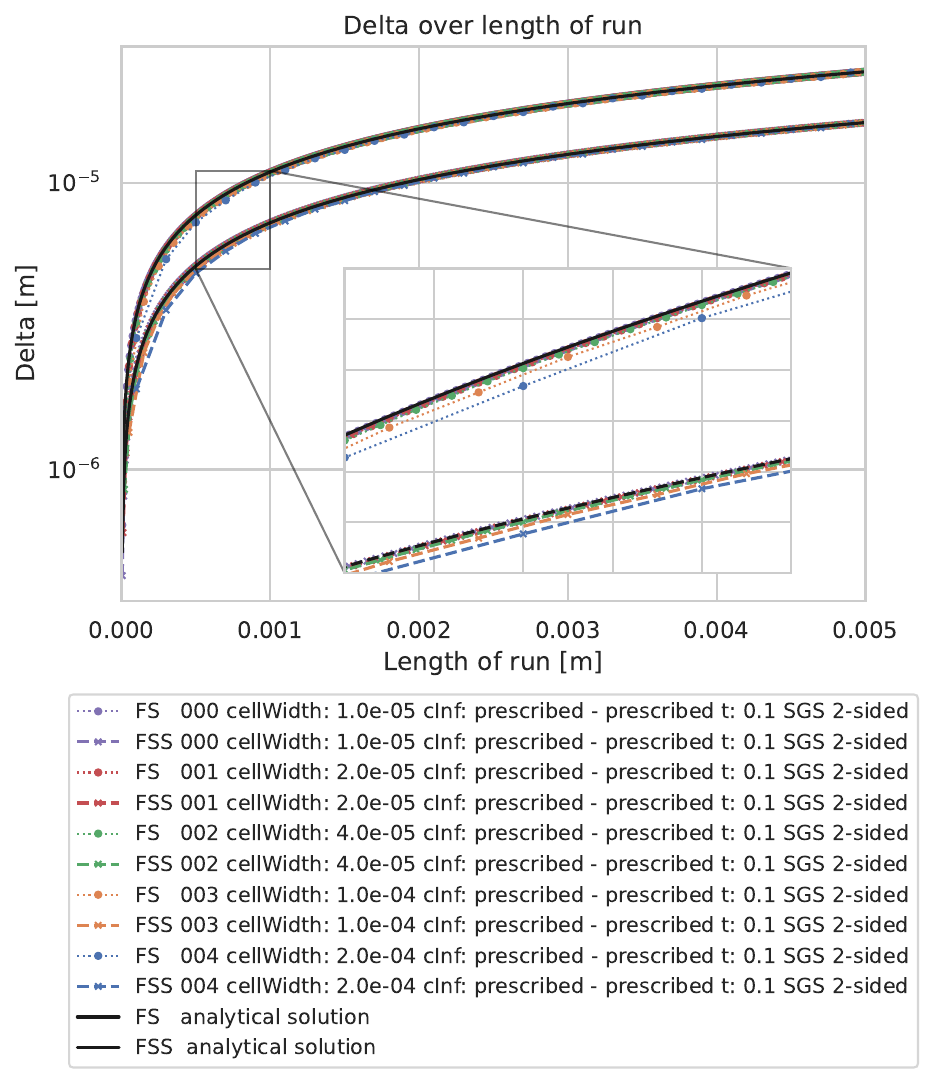}
        \caption{Boundary layer thicknesses}
        \label{fig:2s_FP_prescr_delta}
    \end{subfigure}
    \begin{subfigure}[t]{.49\textwidth}
        \centering
        \vskip0pt
        \includegraphics[width=0.98\linewidth]{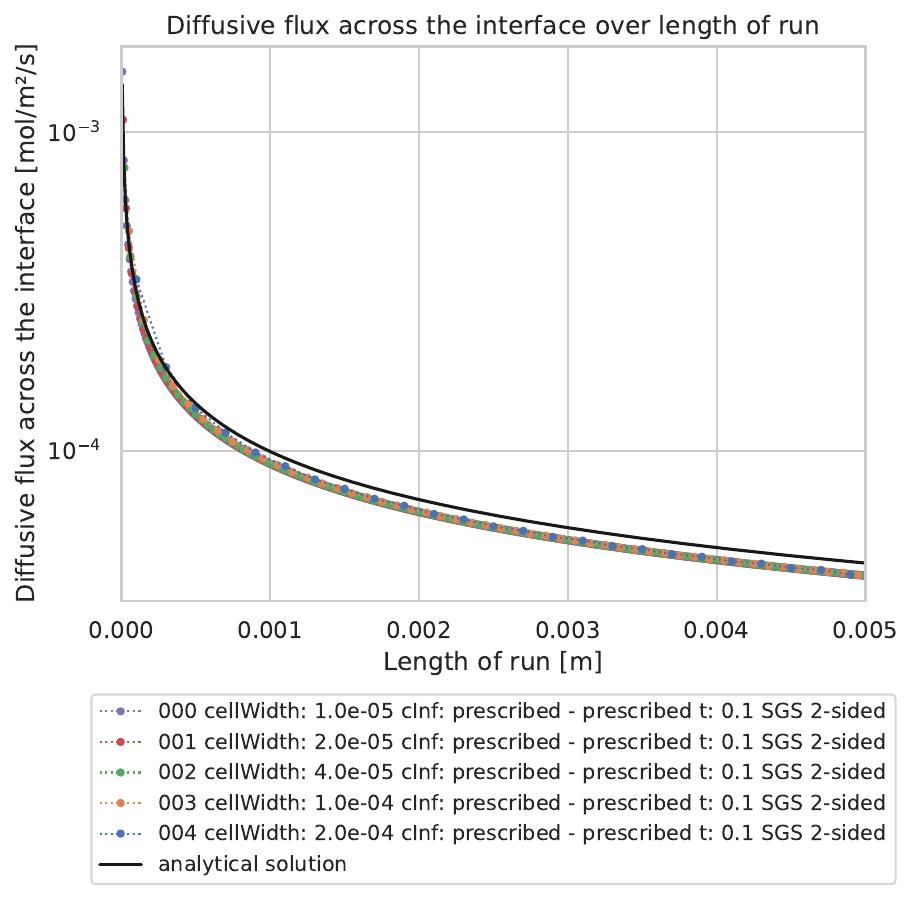}
        \caption{Diffusive fluxes across the interface}
        \label{fig:2s_FP_prescr_diffFlux}
    \end{subfigure}
    \caption{Results for the \tsFP test case with prescribed $c_\infty$}
    \label{fig:2s_FP_prescr_results}
\end{figure}

In our first parameter study, we prescribe the far-field value~$c_\infty$ on both sides.
Therefore, the six equation system~(\ref{eq:6x6-eqs}) simplifies to a four equation system, where the formulas containing $c_2^\pm$~(\cref{eq:6x6b,eq:6x6d}\otherChange{)} are omitted.
In theory, this should make the system easier to solve and is expected to be numerically more stable and the results more precise.

The results for this case are shown in \cref{fig:2s_FP_prescr_results}.
\Cref{fig:2s_FP_prescr_delta} shows the boundary layer thickness on either side of the interface.
The boundary layer thickness\otherChange{es} coincide very well with the analytical solution.
\Cref{fig:2s_FP_prescr_diffFlux} shows the resulting diffusive flux across the interface.
As expected, the flux is very high at the beginning of the domain and diminishes over the length of run.
Similar behaviour has been observed for the one-sided \testFP test case in~\citep{schwarzmeier_twophaseintertrackfoam_2024}.
\otherChange{The flux coincides well with the analytical solution.
It is slightly underestimated in the later part of the domain.}

\subsubsection{Results with prescribed and unknown $c_\infty$}
\label{sss:2s_FP_results_mixed}

\begin{figure}[!htbp]
    \centering
    \begin{subfigure}[t]{.49\textwidth} 
        \centering
        \vskip0pt
        \includegraphics[width=0.98\linewidth]{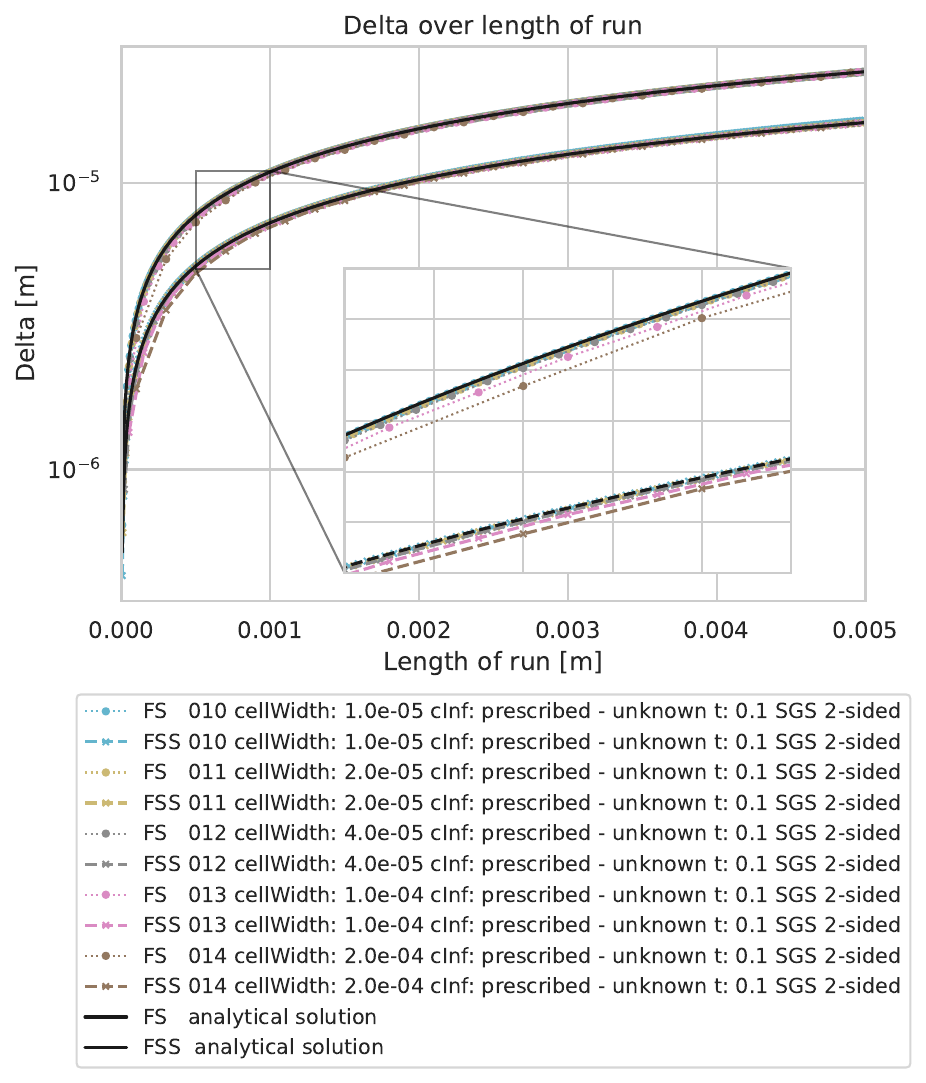}
        \caption{Boundary layer thicknesses}
        \label{fig:2s_FP_mix_delta}
    \end{subfigure}
    \begin{subfigure}[t]{.49\textwidth}
        \centering
        \vskip0pt
        \includegraphics[width=0.98\linewidth]{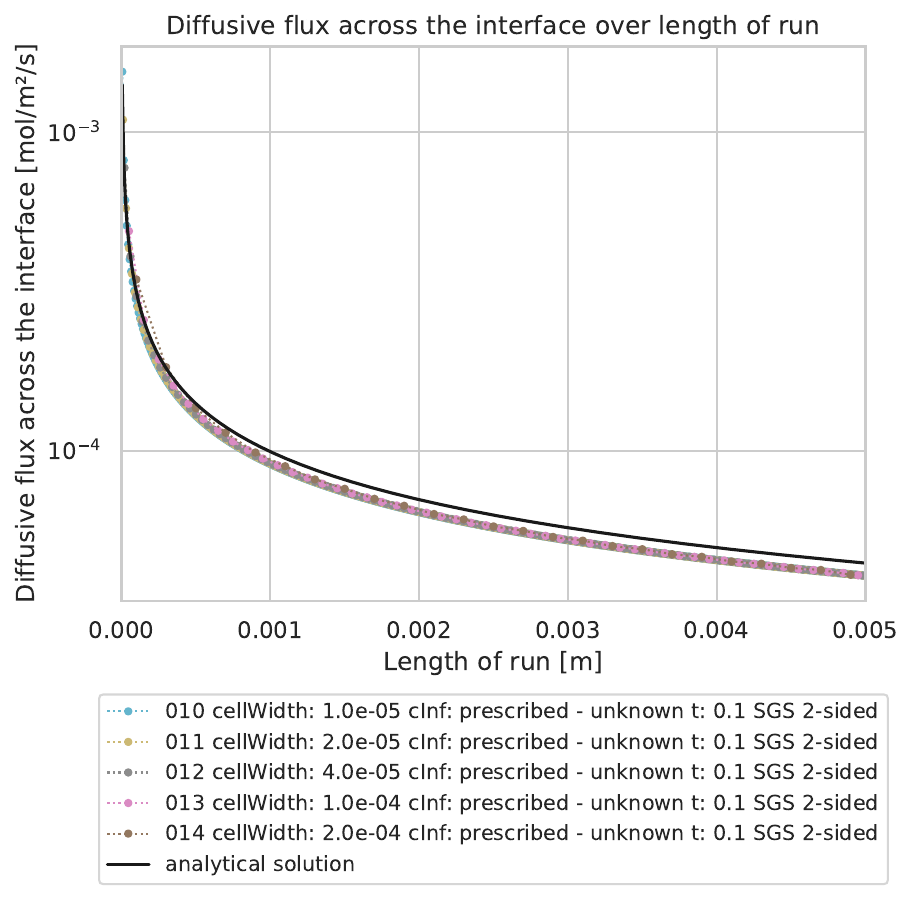}
        \caption{Diffusive fluxes across the interface}
        \label{fig:2s_FP_mix_diffFlux}
    \end{subfigure}
    \caption{Results for the \tsFP test case with mixed computation modes}
    \label{fig:2s_FP_mix_results}
\end{figure}

\Cref{fig:2s_FP_mix_results} shows results if not the same $c_\infty$-algorithm is applied to both sides.
In this setup, on one side, we prescribe $c_\infty$, while on the other we use the algorithm for unknown $c_\infty$ described in \cref{sss:free_c_infty}.
This is equivalent to omitting one equation containing $c_2^\pm$ from the equation system~(\ref{eq:6x6-eqs}). 

In \cref{fig:2s_FP_mix_delta} the \BLT on both sides is displayed.
We observe very good agreement with the analytical solution over the given range of cell sizes.
The boundary layer thickness is of similar size as the cell size for the highest resolution.
\otherChange{T}he fluxes across the interface in \cref{fig:2s_FP_mix_diffFlux} are \otherChange{not only in good agreement with the analytical solution but also} in very good agreement to each other for all meshes.
This is because the SGS has only little effect if the \BLT is bigger than the cell size.

\subsubsection{Results with unknown $c_\infty$}
\label{sss:2s_FP_results_unknown}

In this parameter study we test the ability of the framework to compute both boundary layers adequately without prescribed~$c_\infty$ on either side.
It is the most general case.

\begin{figure}[!htbp]
    \centering
    \includegraphics[width=0.7\textwidth]{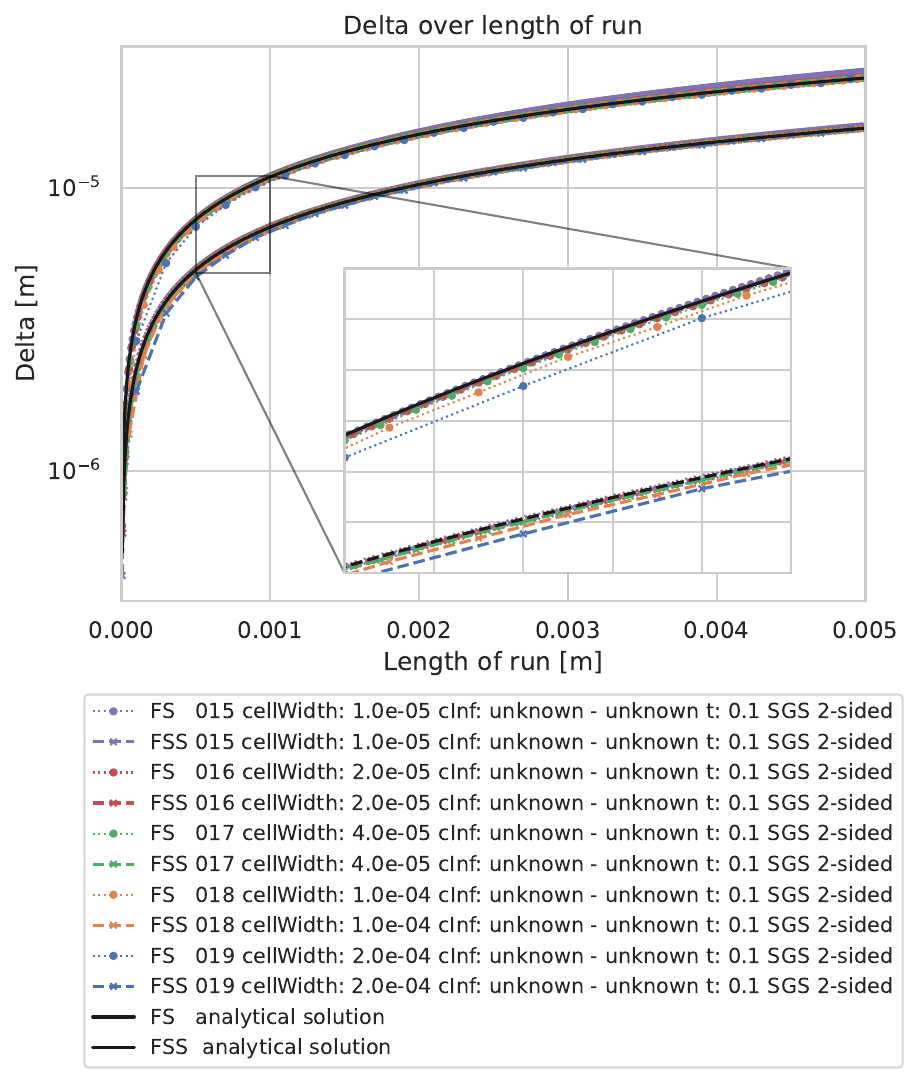}
    \caption{Boundary layer thicknesses for the \tsFP test case with unknown $c_\infty$ on both sides}
    \label{fig:flatPlate_2s_deltas}
\end{figure}

\begin{figure}[!htbp]
    \centering
    \includegraphics[width=0.7\textwidth]{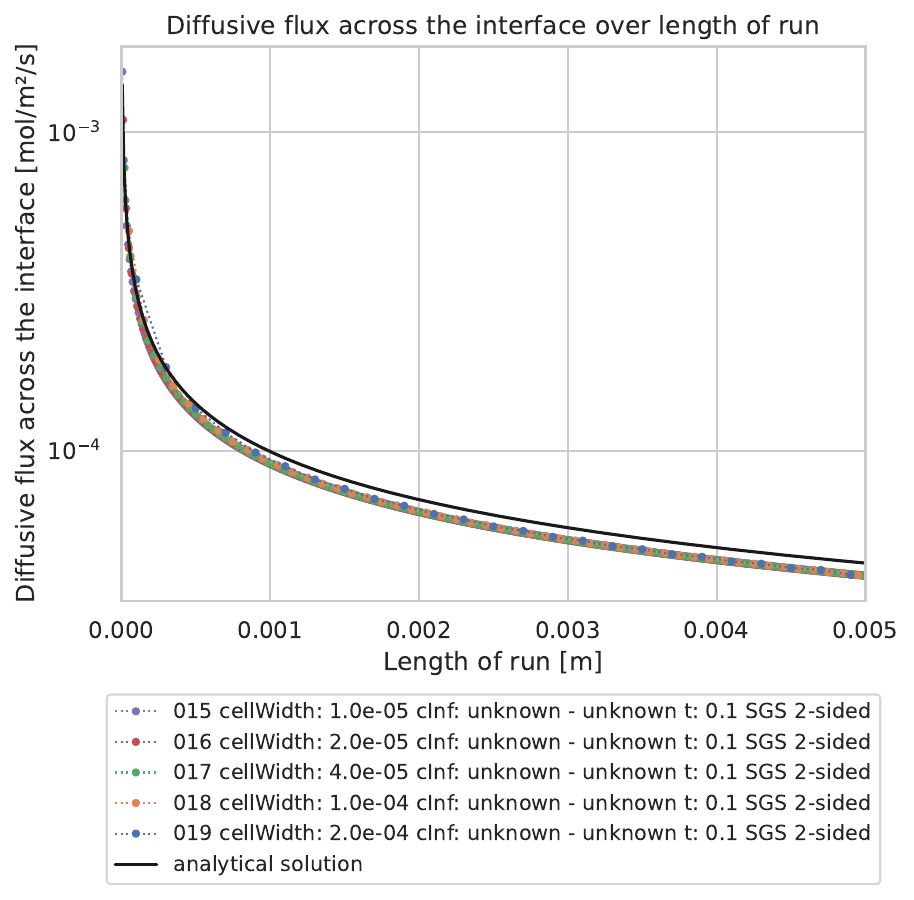}
    \caption{Diffusive fluxes across the interface for the \tsFP test case with unknown $c_\infty$ on both sides}
    \label{fig:flatPlate_2s_diffFluxes}
\end{figure}

In \cref{fig:flatPlate_2s_deltas} the thickness of the boundary layer is displayed over the length of run.
The boundary layer on the interfaces side with higher diffusivity is thicker, as expected from the analytical solution~\cref{eq:2sFP_anaLsg}.
The analytical boundary layer thicknesses are well captured by the numerical results.

\Cref{fig:flatPlate_2s_diffFluxes} shows the (diffusive) fluxes across the interface over the length of run.
The fluxes are \otherChange{close to the analytical solution}, even in the most coarsely meshed cases.

\subsubsection{Conclusion}

With this test case, our method has proven the capability to produce very accurate results, both in terms of the local boundary layer thickness and in terms of the (local) diffusive flux across the interface.
The results have been compared to the analytical solution. 
Tests were performed over widely varying mesh sizes and with three different setups: with prescribed~$c_\infty$ on both sides, with mixed computations and with unknown~$c_\infty$ on both sides.
All cases were able to recover the (local) quantities of boundary layer thickness and mass transfer very well.

\subsection{\Reviewers{Initial mass transfer from a droplet at low Reynolds number}}
\label{ss:SS-droplet}

\subsubsection{Test case setup}
\label{ss:SS-droplet_setup}

We aim to simulate the mass transfer out of or into a moving drop to verify our correct implementation and general applicability of our newly developed two-sided SGS modeling.
Therefore we employ the model to a simple configuration of a moving bubble/drop.
Several publications with and without \SGS modeling have already used the analytical velocity field of Satapathy and Smith given in~\cite{satapathy_motion_1961}, such as~\cite{weber_modeling_2016,weiner_advanced_2017,pesci_SGS_risBubb_surfactants_2018,pesci_computational_thesis_2019,schwarzmeier_twophaseintertrackfoam_2024}.

\subsubsection*{Satapathy-Smith velocity field}

In \cite{satapathy_motion_1961} the authors introduce an analytical solution for the velocity to a spherical drop\footnote{The authors in \cite{satapathy_motion_1961} use the term \textit{drop}, whereas in \cite{weber_modeling_2016} the term \textit{bubble} is used.
We will use the term \textit{drop/droplet} to indicate a non-negligible viscosity within the inner domain in comparison to the viscosity of the outer domain.
\textit{Drops/Droplets} by this definition can have higher or lower density than the surrounding fluid and therefore rise/sink under the influence of gravity and no other influences.}
in low Reynolds number flow inside a spherical domain.
A typographical mistake in the original publication was corrected by \cite{weber_modeling_2016} whilst introducing one by himself in the formula for~$D_b$~\cref{eq:SS_paramDb}.

In the notation used by \cite{weber_modeling_2016}, where $R$ is the domain radius and $\eta$ the ratio of the viscosities~${ \mu_l / \mu_b }$, the velocity field in polar coordinates is given by\\
\noindent\begin{minipage}{\textwidth}
\begin{subequations}
    \label{eqs:SS_vel}
    \begin{alignat}{4}
        v_r &= & 2 \left( \frac{A}{r^3} + \frac{B}{r} + C + Dr^2 \right) &cos(\theta) \\
        \intertext{and}
        v_\theta &= &\left( \frac{A}{r^3} - \frac{B}{r} - 2C - 4Dr^2 \right) &sin(\theta) 
    \end{alignat}
\end{subequations}
\end{minipage}

\noindent with the parameters $A$~to~$D$ given as\\
\noindent\begin{minipage}{\textwidth}
\begin{subequations}
    \label{eqs:SS_params}
    \begin{align}
        B_l &= \frac{3R^6 - 3R + 2R^6\eta + 3R\eta}
            {4R^6 - 9R^5 + 10R^3 - 9R + 4R^6\eta - 6R^5\eta + 6R\eta - 4\eta + 4} \ , \\
        D_l &= \frac{2 B_l R^3 + 4B_l - 6B_l R - 3R}{4R^6 - 10R^3 + 6R} \ , \\
        C_l &= \frac{1}{2} - \frac{2}{3}\frac{B_l}{R} - \frac{5}{3} D_l R^2 \ , \\
        A_l &= -B_l - C_l - D_l \\
        \intertext{and}
        A_b &= B_b = 0 \ , \\
        D_b &= \eta ( A_l + D_l ) \ , \label{eq:SS_paramDb} \\
        C_b &= -D_b 
    \end{align}
\end{subequations}
\end{minipage}
with the index~$l$ denoting the surrounding liquid and the index~$b$ denoting the interior of the bubble.

We prescribe the analytical velocity at all cell centers and faces.
At the interfacial faces we apply $v_r = 0$.
Subsequently we use the Helmholtz decomposition approach to ensure mass flux continuity in all cells.
This is beneficial because the velocity at the faces is set to the value obtained by using the coordinates of the face center, but in FVM context it should represent the integration over the face area.

In our test case we use the moving velocity~\SI{0.37}{m/s}, a bubble diameter of~\SI{0.725}{mm} and an outer radius of~\SI{20}{mm}.
The viscosity ratio is set to~$0.62$.


\subsubsection*{Species parameters}

\begin{table}[!htbp]
    \centering
    \setlength{\tabcolsep}{8pt} 
    \renewcommand{\arraystretch}{1.3} 
    \begin{tabular}{lcc}
         & $D$ [m$^2$/s] & $c \, (t=0)$ [mol] \\
    \hline
        Drop/Bubble & $9\times 10^{-9}$ & $1$ \\ 
    \hline
        Liquid & $4\times 10^{-9}$ & $0$ \\
    \end{tabular}
    \caption{Properties for the mass transfer}
    \label{tab:SS_speciesProp}
\end{table}
In \cref{tab:SS_speciesProp} we show the parameters relevant to setting the mass transfer for this test case.
Additionally we use the same Henry coefficient~$H$ as in \cref{secSub:test_flatPlate} of \SI{1.5873}{} with the definition~${ H = c_b/c_l }$.

\subsubsection*{Calculation settings}

The axisymmetric simulation is run for a physical time of \SI{0.3}{s} with a \TS size of \SI{5e-5}{s}.
Additionally we employ an under-relaxation factor of~$0.8$ to the species equation.

We conduct a parameter study with the two-sided SGS model comprising of four levels of mesh refinement.
For comparison we run a secondary parameter study including the same mesh resolutions as well as two even more refined ones, where we use the standard FVM method, including the approach to compute the fluxes introduced in \cref{sss:thinFilm_detCSigma}.
With the the chosen \TS, under-relaxation and tolerance settings the algorithms usually converge after $6$~iterations.

\subsubsection{Results}
\label{ss:SS-droplet_results}

\begin{figure}[!htbp]
    \centering
    \includegraphics[width=0.75\linewidth]{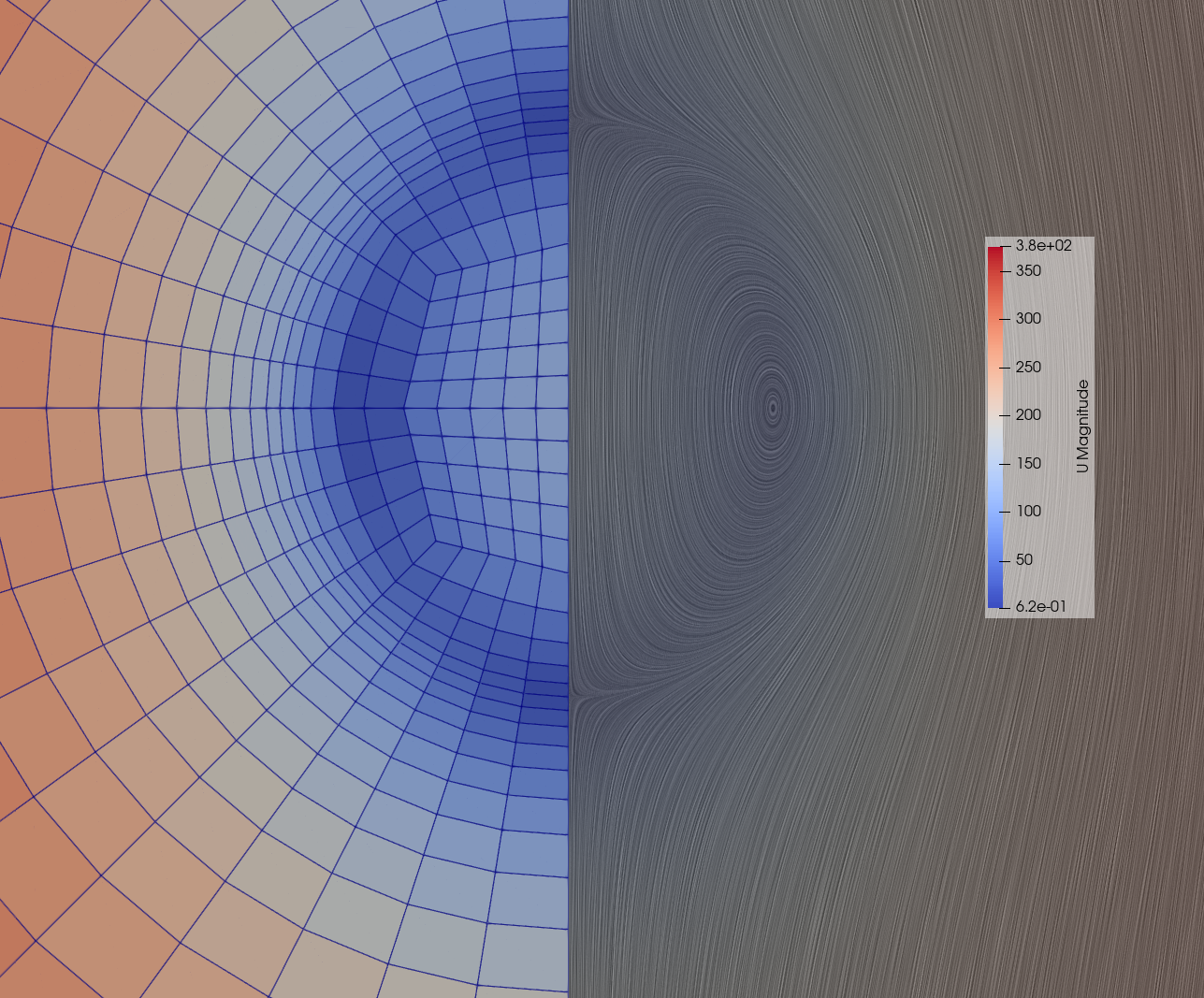}
    \caption{Velocity field: magnitude (coarse mesh) and Surface Line Integral Convolution (fine mesh)}
    \label{fig:SS_vel}
\end{figure}

In \cref{fig:SS_vel} we display the velocity field for the given case.
On the r.h.s. of the figure the magnitude field of the velocity is displayed, where also the coarsest mesh~(\textit{meshN: 5}) is visualized.
The parameter study is such, that the mesh parameter \textit{meshN} controls the mesh size.
Doubling the parameter leads to halving the discretization length in each direction, i.e.\ generating a mesh with 4~times as many cells.
On the l.h.s of the figure the velocity field is visualized with Paraview's \textit{Surface Line Integral Convolution~(LIC)} feature.
The two stagnation points at the rotational axis of the velocity field become visible.
Additionally, it is visible, that the pathlines of a spherical bubble rising at low Reynolds number are closed, thus inhibiting fast mixing inside the bubble.
In contrast, the pathlines outside the bubble are not closed.
Note that for mass transfer on the outside of single rising bubbles the authors in \cite{marschall_development_2021} found that in the area of the wake as a recirculation zone, i.e.\ closed pathlines, chemical reactions can significantly increase local mass transfer, while not affecting the local mass transfer rate in the frontal area of the bubble as much.

\begin{figure}[!htbp]
    \begin{subfigure}[t]{.33\textwidth}
        \centering
        \vskip0pt
        \includegraphics[width=0.95\linewidth]{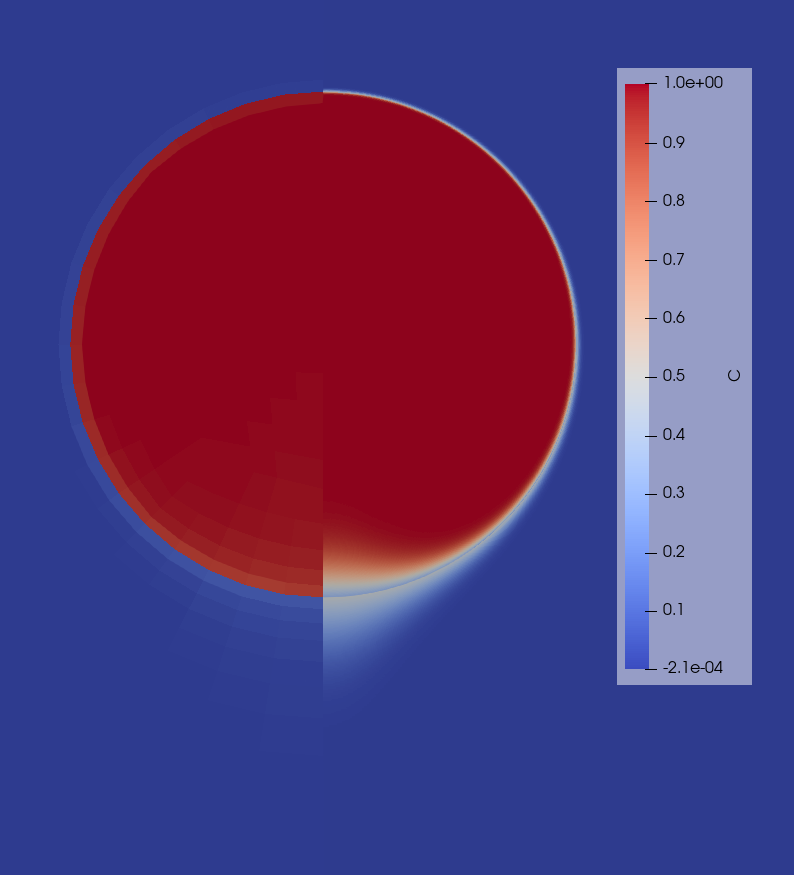}
        \caption{at $t=\SI{0.01}{s}$}
    \end{subfigure}
    \begin{subfigure}[t]{.33\textwidth}
        \centering
        \vskip0pt
        \includegraphics[width=0.95\linewidth]{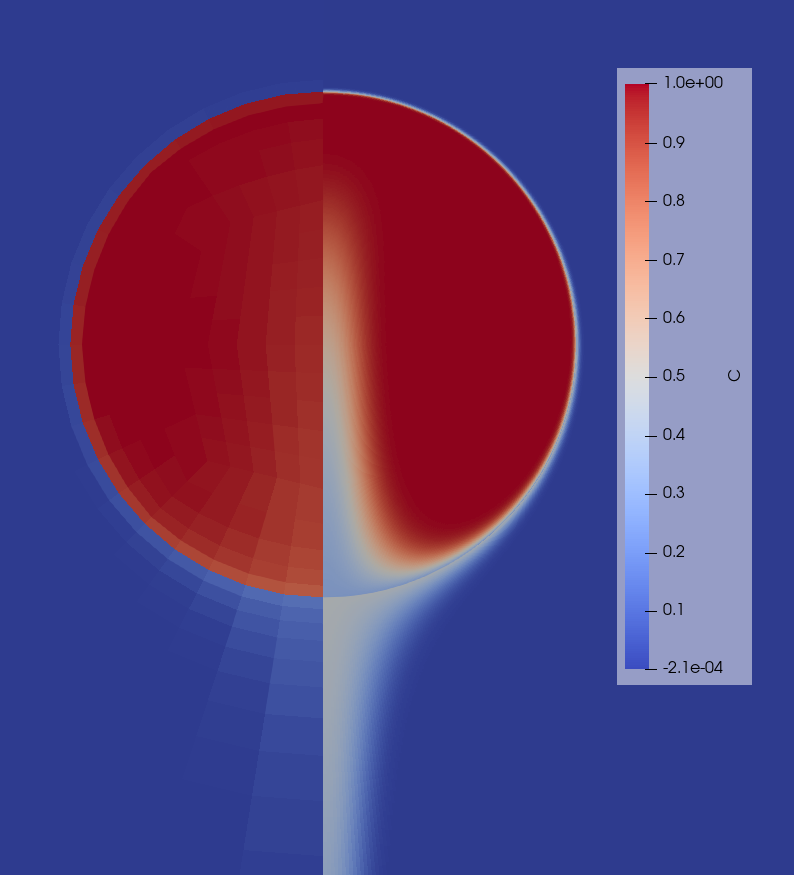}
        \caption{at $t=\SI{0.02}{s}$}
    \end{subfigure}
    \begin{subfigure}[t]{.33\textwidth}
        \centering
        \vskip0pt
        \includegraphics[width=0.95\linewidth]{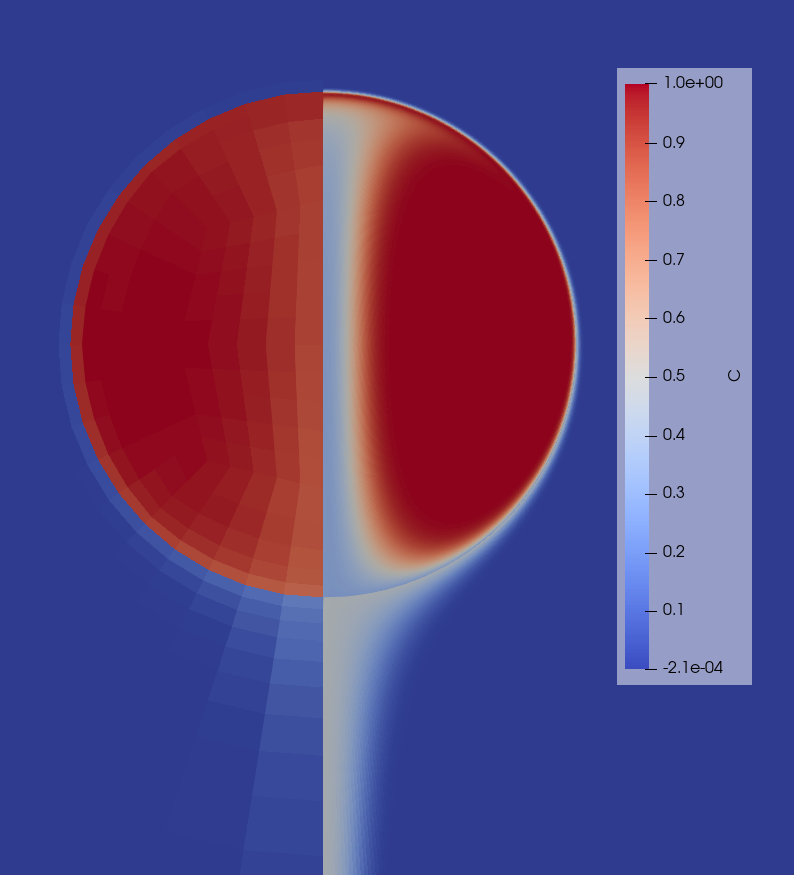}
        \caption{at $t=\SI{0.03}{s}$}
    \end{subfigure}
    \caption{Concentration fields in the coarsest (left) and finest (right) case without SGS modeling}
    \label{fig:SS_C_withoutSGS}
\end{figure}

In \cref{fig:SS_C_withoutSGS} we compare the concentration field at different points in time for the finest and the coarsest case without \SGS modeling.
It becomes obvious that the coarsest resolution is not able to adequately capture the mass transfer.
Note that this is a very coarse resolution in the context of ALE-IT.\\
Another observation from \cref{fig:SS_C_withoutSGS} is the evolution of the concentration over time.
The problem states a uniform concentration of \SI{1}{mol/mm^3} inside the droplet and of \SI{0}{mol/mm^3} as initial condition.
After \SI{0.01}{s} the depletion of concentration has not advanced much into the droplet, except for the rear part, where we observe a depleted area.
At \SI{0.2}{s} the (steady) velocity field carries/"shoots" a depleted jet from the rear pole of the droplet towards its front pole.
We observe visual similarity to the concentration field shown in~\cite{watada_theoretical_1970}, where only mass transfer inside a droplet was considered.
This jet has reached the front stagnation point at \SI{0.3}{s}.
Visible in the fine resolution~(lhs) a thin area between the jet and the surface has developed, where there is still a high concentration.
This can be interpreted as an area, where the species-infested fluid is trapped at low velocity - almost \textit{sitting} stagnant - while from one side the change in concentration approaches via diffusion and from the other side the concentration change approaches, carried by the jet.

\begin{figure}[!htbp]
    \begin{subfigure}[t]{.33\textwidth}
        \centering
        \vskip0pt
        \includegraphics[width=0.95\linewidth]{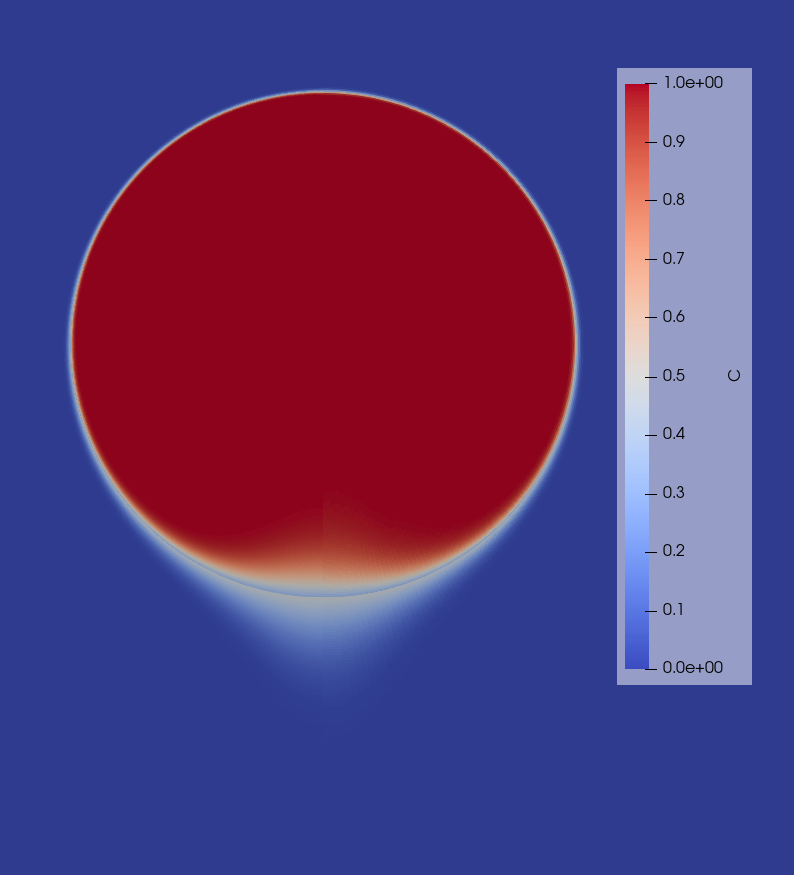}
        \caption{at $t=\SI{0.01}{s}$}
    \end{subfigure}
    \begin{subfigure}[t]{.33\textwidth}
        \centering
        \vskip0pt
        \includegraphics[width=0.95\linewidth]{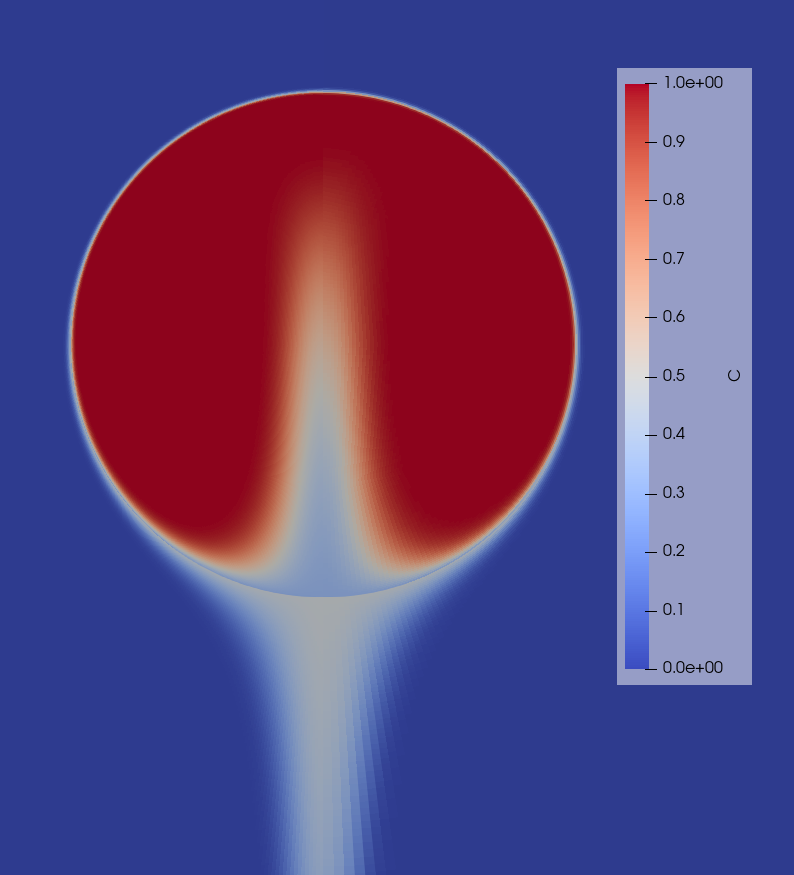}
        \caption{at $t=\SI{0.02}{s}$}
    \end{subfigure}
    \begin{subfigure}[t]{.33\textwidth}
        \centering
        \vskip0pt
        \includegraphics[width=0.95\linewidth]{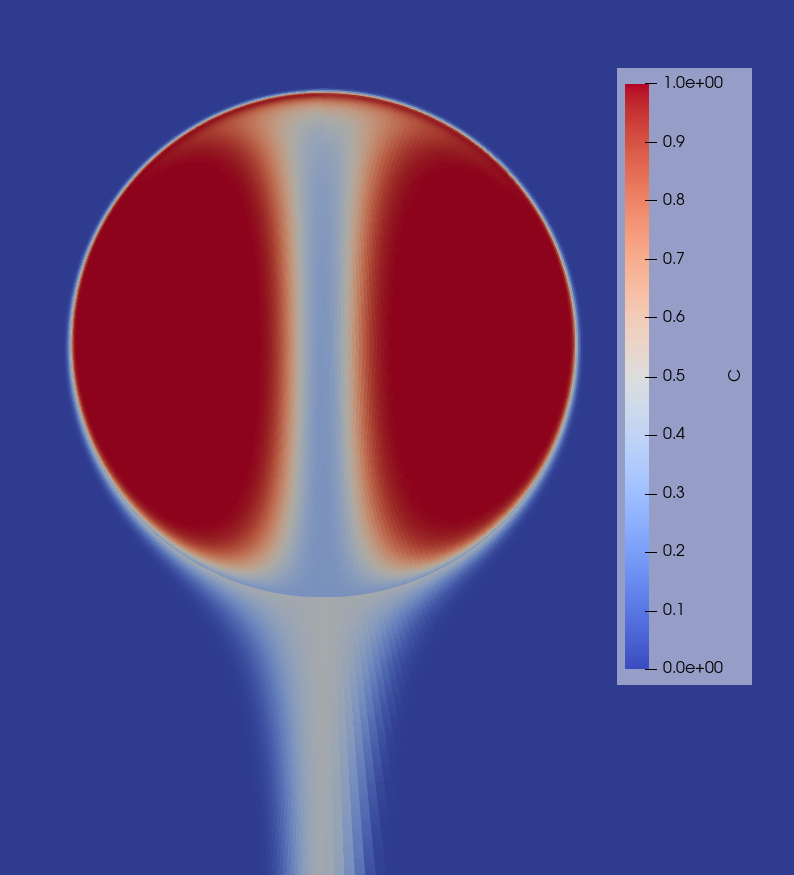}
        \caption{at $t=\SI{0.03}{s}$}
    \end{subfigure}
    \caption{Concentration fields in the respective finest case without (left) and with (right) two-sided SGS modeling}
    \label{fig:SS_C_finest_with-without_SGS}
\end{figure}

We now compare the respective finest resolution concentration fields in \cref{fig:SS_C_finest_with-without_SGS}, where on the lhs two-sided \SGS modeling was used and on the rhs no SGS modeling has been used.
Visually, we hardly see a discrepancy between the two concentration fields.

\begin{figure}[!htbp]
    \begin{subfigure}[t]{.33\textwidth}
        \centering
        \vskip0pt
        \includegraphics[width=0.95\linewidth]{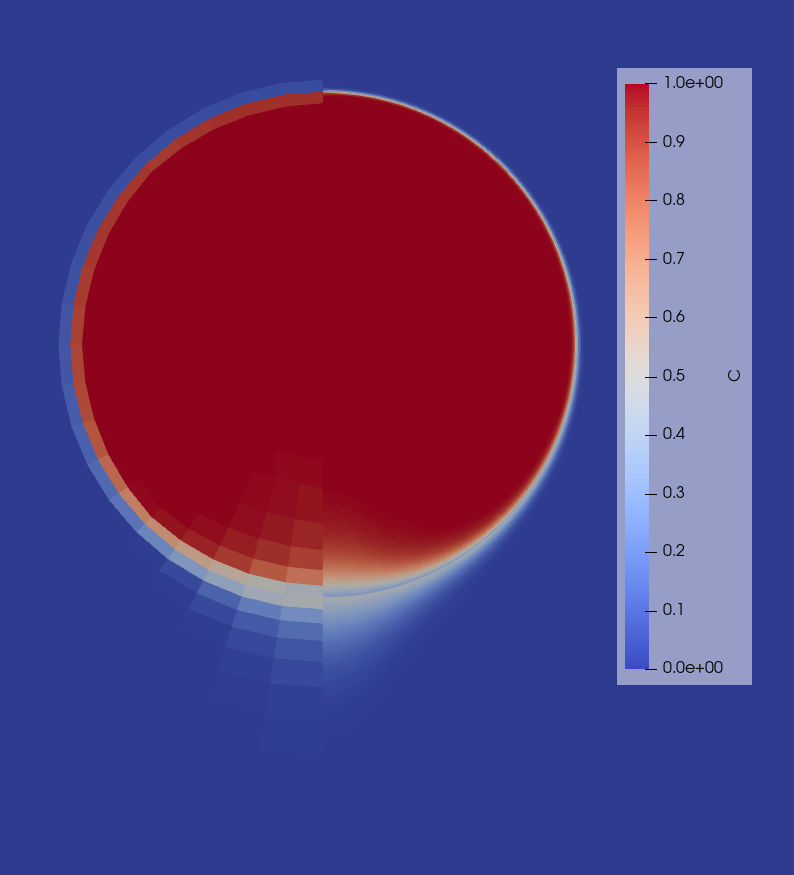}
        \caption{at $t=\SI{0.01}{s}$}
    \end{subfigure}
    \begin{subfigure}[t]{.33\textwidth}
        \centering
        \vskip0pt
        \includegraphics[width=0.95\linewidth]{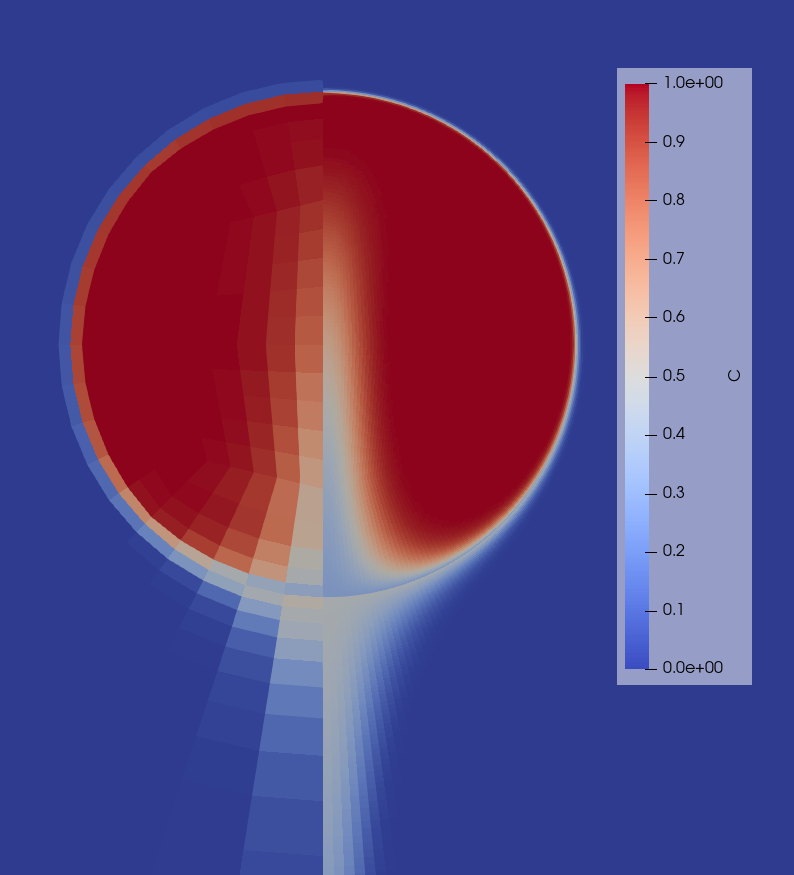}
        \caption{at $t=\SI{0.02}{s}$}
    \end{subfigure}
    \begin{subfigure}[t]{.33\textwidth}
        \centering
        \vskip0pt
        \includegraphics[width=0.95\linewidth]{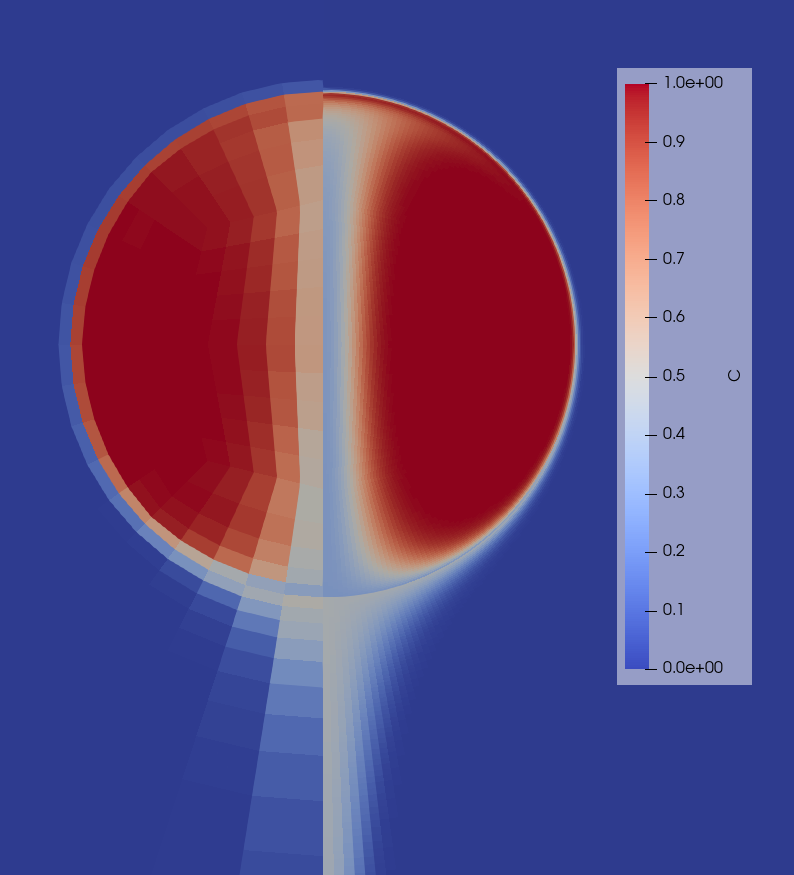}
        \caption{at $t=\SI{0.03}{s}$}
    \end{subfigure}
    \caption{Concentration fields in the coarsest and finest case with SGS modeling}
    \label{fig:SS_C_withSGS}
\end{figure}

In \cref{fig:SS_C_withSGS} we compare the concentration field at different mesh resolution levels in the same manner as in the previous \cref{fig:SS_C_withoutSGS}.
Comparing the respective rhs and lhs sides we observe the concentration field better resolved with the finer mesh.
Apart from the resolution the concentration fields visually appear alike.

\begin{figure}[!htbp]
    \begin{subfigure}[t]{.5\textwidth}
        \centering
        \vskip0pt
        \includegraphics[width=0.99\linewidth]{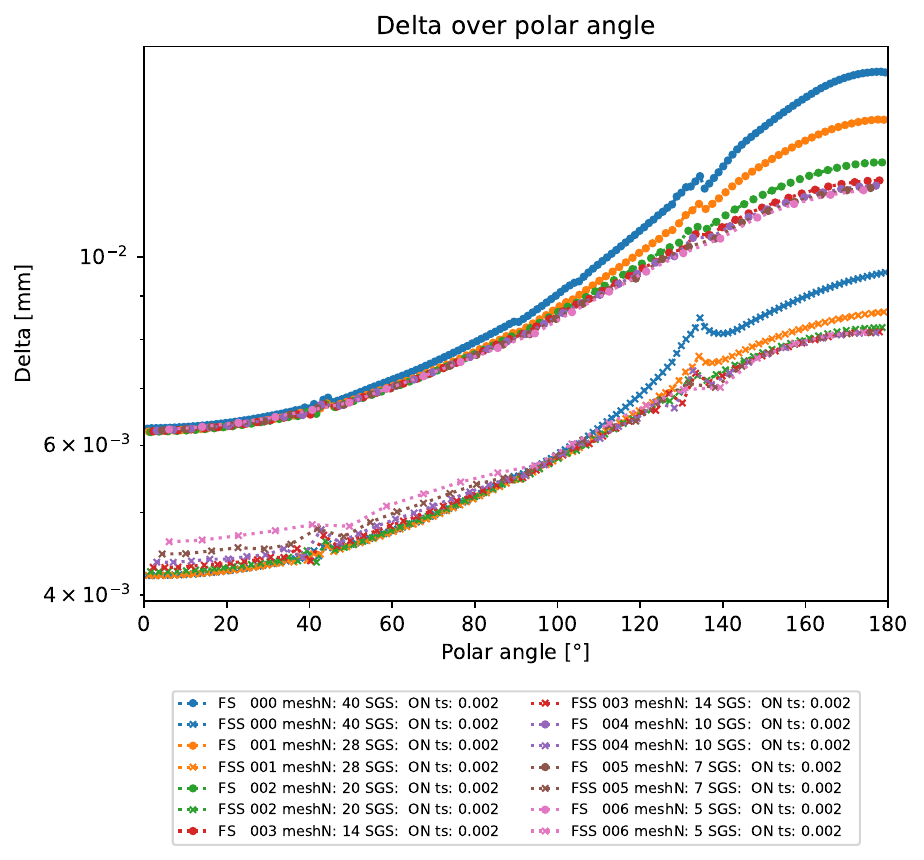}
        \caption{at $t=\SI{0.002}{s}$}
        \label{fig:SS_BLT_over_time_t0-002}
    \end{subfigure}
    \begin{subfigure}[t]{.48\textwidth}
        \centering
        \vskip0pt
        \includegraphics[width=0.99\linewidth]{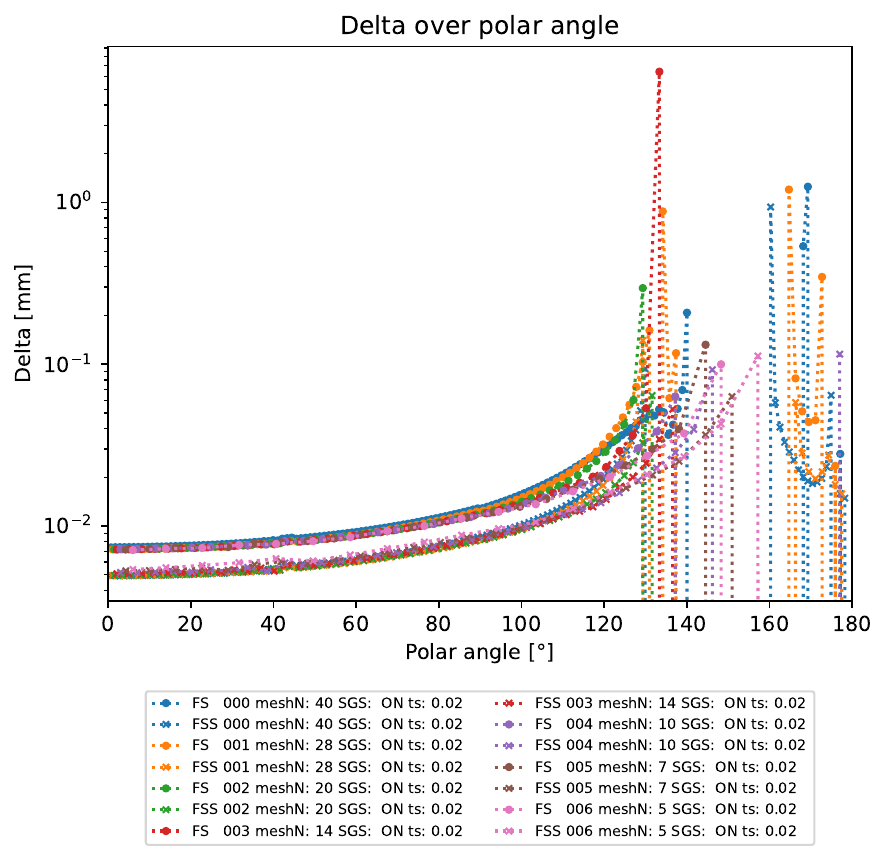}
        \caption{at $t=\SI{0.02}{s}$}
    \end{subfigure}
    \caption{\BLT over the polar angle}
    \label{fig:SS_BLT_over_time}
\end{figure}

\Cref{fig:SS_BLT_over_time} shows the computed \BLT{}es~$\delta^\pm$ over the polar angle, where \SI{0}{\degree} represents the front pole / impingement point at two different points in time.
After \SI{0.002}{s} the \BLT{}es are within approximately \SIlist{4e-3; 9e-3}{mm} or \SIlist{6e-3; 1.4e-2}{mm} respectively, where in this and the subsequent figures the outward pointing side of the droplet is denoted with \textit{FS}, whereas the inner side is represented by~\textit{FSS}.
\Cref{eq:2sSGS_conti} shows that the \BLT{}es are coupled, what is supported by this diagram again.
We see that for the given case, the outer \BLT is approximately $1.5$~times higher than the inner \BLT.\\
After \SI{0.02}{s} the \BLT has grown to at least approx.\ \SI{5e-3}{mm} or \SI{7e-3}{mm}.
In the rear part of the droplet, where the flow detaches from the interface into and away from the droplet respectively, the computed \BLT becomes large before it becomes nearly undeterminably\footnote{This is referring to the coarse resolution in \cref{fig:SS_BLT_over_time} at the later \TS{}.} exactly.\footnote{Generally for the SGS modeling to be valid the local \BLT needs to be small in comparison to the local curvature.
Also, in a physical sense, the \BLT inside a sphere can not be bigger than the sphere.}
In favor to our approach is the fact that \BLT{}es much larger than the cell size have almost no influence on the solution in terms of fluxes or gradient correction anymore, c.f.~\cref{fig:minDeltaViss,fig:limitedSGS}.

\begin{figure}[!htbp]
    \begin{subfigure}[t]{.49\textwidth}
        \centering
        \vskip0pt
        \includegraphics[width=0.95\linewidth]{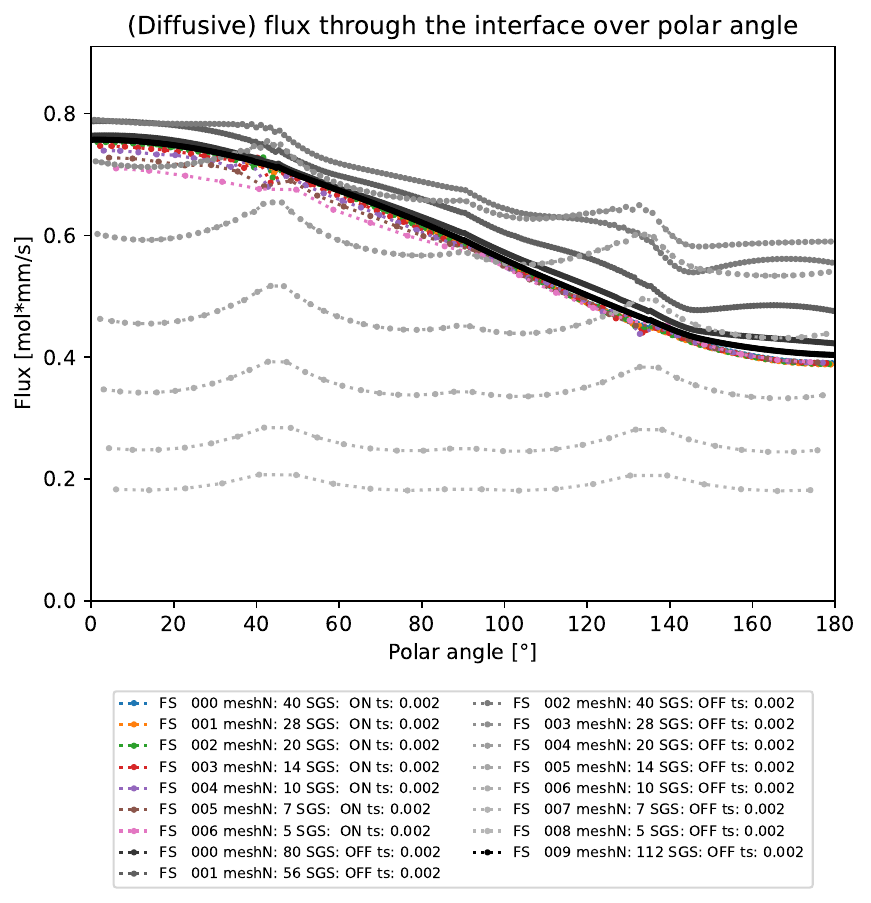}
        \caption{at $t=\SI{0.002}{s}$}
        \label{fig:SS_flux_t0-002}
    \end{subfigure}
    \begin{subfigure}[t]{.49\textwidth}
        \centering
        \vskip0pt
        \includegraphics[width=0.95\linewidth]{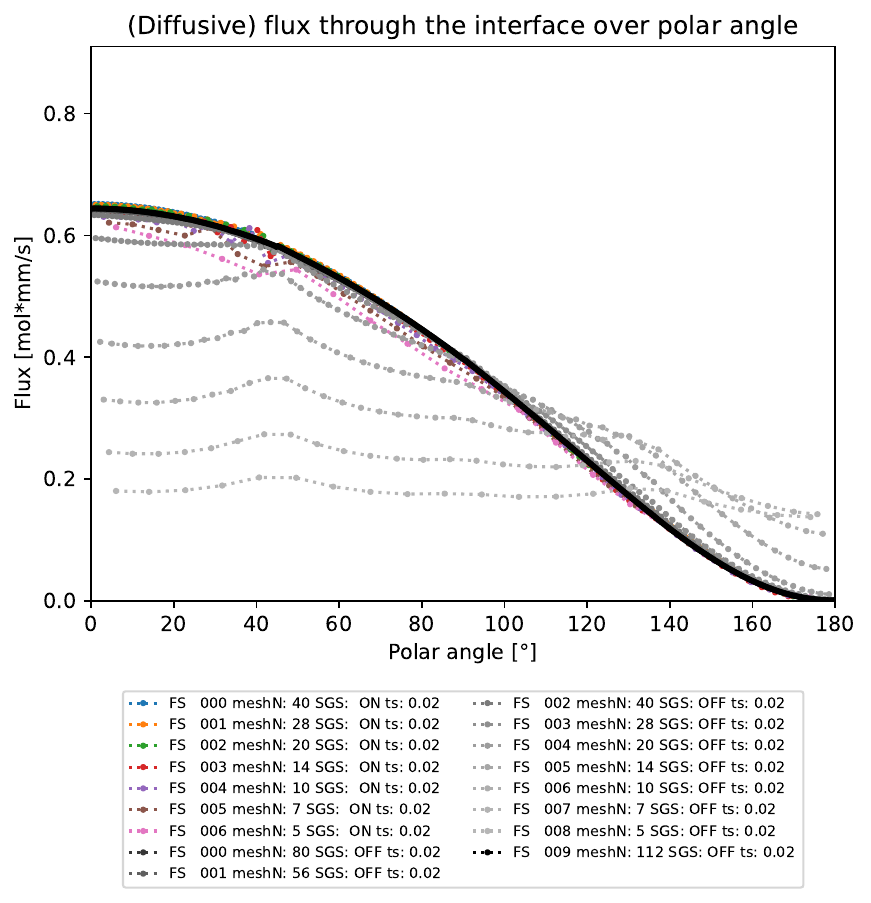}
        \caption{at $t=\SI{0.02}{s}$}
        \label{fig:SS_flux_t0-02}
    \end{subfigure}
    \caption{Flux over the interface over the polar angle}
    \label{fig:SS_flux}
\end{figure}

In \cref{fig:SS_flux} the flux over the interface is depicted over the polar angle.
At the early \TS the flux is high at every point of the interface, yet we already see how the flux diminishes in \cref{fig:SS_flux_t0-002} as the \BLT increases over the polar angle in \cref{fig:SS_BLT_over_time_t0-002}.
The simulation without SGS modeling is not able to resolve the fluxes adequately on the same mesh refinement levels as used for SGS modeling.
As in \cref{secSub:test_flatPlate} we observe that under-resolved gradients upstream can result in over-calculated gradients downstream, an effect already observed by earlier publications on one-sided SGS modeling.\\
The diminishing need for and influence of the SGS model for large boundary layers becomes more obvious in \cref{fig:SS_flux_t0-02}, as at the rear part of the droplet the flux tends to zero, inversely to the computed \BLT and much better resolved by the coarse modeling without SGS than the high fluxes in the front part of the droplet.
The effects described here align very well with our previous observations for species transport to/from an interface with fixed interfacial concentration and SGS modeling with fixed parameter~$c_\infty$, e.g.\ in~\cite[chapter~6.3]{schwarzmeier_twophaseintertrackfoam_2024}.
Note that computing the correct concentration fields and (local and global) mass transfer rates are likely the most interesting quantities for application purposes.

\begin{figure}[!htbp]
    \centering
    \includegraphics[width=0.6\linewidth]{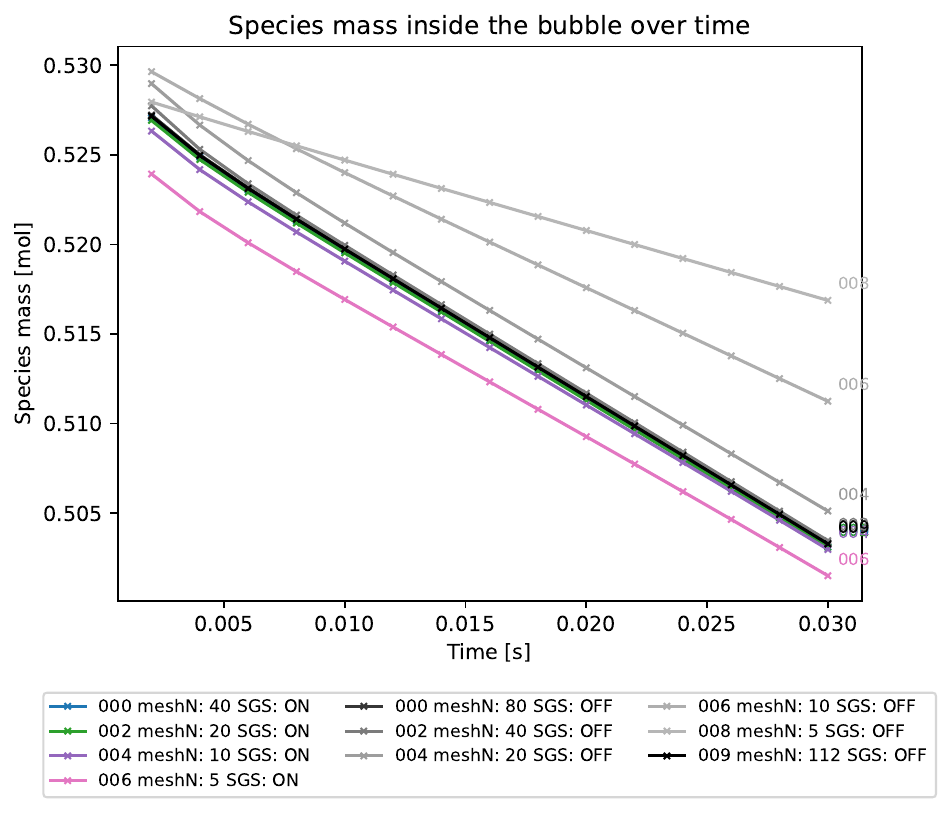}
    \caption{Total species mass over time}
    \label{fig:SS_C_massOverTime}
\end{figure}

In \cref{fig:SS_C_massOverTime} the total mass of species inside the droplet is shown.
Notably the two coarsest resolutions are outliers.
For all other, it can be noted, that the total mass transfer is well captured by all grid resolutions in case our new two-sided \SGS modeling approach is used.
The under- and subsequent overshoot for certain resolutions visible in \cref{fig:SS_flux} leads to seemingly good results even for middle resolutions without SGS modeling.
However, for the coarsest resolution the later overshoot is not sufficient to compensate for the earlier undershoot and the total mass flux is generally underestimated.

\subsubsection{Conclusion}
\label{ss:SS-droplet_conclusion}

With this test case, our method has proven the capability to produce accurate results in the initial phase of mass transfer from a moving droplet.
This instationary case includes a complex flow field with detachment and inner recirculation.
The mass transfer has been quantified with the local and global diffusive flux across the interface and the results have been compared to resolved FVM computations.

As the other test cases, this parameter study has been added to our code repository, the archived repository snapshot and to the data~\cite{GitlabRepositoryV2,tudatalib_code_v2,tudatalib_data_v2}.

\section{Summary}
\label{sec:conclusions}


In the present work, a novel \SGS~(SGS) framework for conjugate mass transfer is developed, which incorporates the \ICs across a fluid interface, addressing the computational challenges posed by bilateral thin concentration boundary layers.

For this purpose, conservative Dirichlet-Dirichlet coupling, derived directly from the \ICs, is introduced in the finite-volume ALE-IT method. 
The existing SGS algorithm is improved for \exSmall boundary layers. 
Moreover, the SGS modeling is extended to handle unknown far-field concentrations. 
Finally, the introduced Dirichlet-Dirichlet coupling is extended by incorporating two one-sided SGS models.
Using an iterative coupling algorithm, the concentrations at the interface \ReviewerTwo{are}  determined \ReviewerTwo{such that they} satisfy the \ICs. 


The verification in the open-source \textit{twoPhaseInterTrackFoam} OpenFOAM module~\cite{schwarzmeier_twophaseintertrackfoam_2024} of the new algorithm for unknown far-field concentrations in the \testFP test case and the subsequent verification of the novel two-sided SGS framework \otherChange{in} the \tsFP test case demonstrate a high degree of accuracy.
\Reviewers{The test case for the initial mass transfer from a droplet also shows how with two-sided SGS modeling the local quantities of mass transfer
can accurately be computed for more complex flows.} 





In a next step the authors aim to apply the developed framework to more challenging cases, such as droplet\otherChange{s} in counter-flow, representing a relevant elementary sub-process of a chemical extraction process.
This case is more complex, involving instationarity, a \otherChange{deforming} interface, an outer flow field \otherChange{and} internal circulation.

\section*{Acknowledgments}

Moritz Schwarzmeier, Tomislav Mari\'{c} and Dieter Bothe would like to thank the Federal Government and the Heads of Government of the Länder, as well as the Joint Science Conference (GWK), for their funding and support within the framework of the NFDI4Ing consortium. Funded by the German Research Foundation (DFG) - project number 442146713.

\medskip

Funded by the German Research Foundation (DFG) – Project-ID 265191195 – SFB 1194.

\medskip

Part of this work was funded by the Hessian Ministry of Higher Education, Research, Science and the Arts - cluster project Clean Circles.

\bibliography{bibliography}

\end{document}